\documentclass[11pt,a4paper]{article}
\usepackage[T1]{fontenc}
\usepackage{lmodern}
\usepackage{amsmath,amssymb,amsthm,mathtools}
\usepackage{geometry}
\usepackage[hidelinks]{hyperref}
\usepackage{microtype}
\usepackage{enumitem}
\usepackage{booktabs,array}
\setlist{nosep}
\newtheorem{theorem}{Theorem}[section]
\newtheorem{lemma}[theorem]{Lemma}
\newtheorem{corollary}[theorem]{Corollary}
\newtheorem{proposition}[theorem]{Proposition}
\theoremstyle{remark}

\theoremstyle{plain}
\theoremstyle{definition}
\newtheorem{definition}[theorem]{Definition}
\newcommand{\Rm}{\operatorname{Remove}}
\newcommand{\Rp}{\operatorname{Rep}}
\newcommand{\Qrat}{\mathbb Q}

\newcommand{\BQP}{\mathbf{BQP}}
\newcommand{\FBQP}{\mathbf{FBQP}}
\newcommand{\ket}[1]{|#1\rangle}
\newcommand{\A}{\mathcal A}
\newcommand{\Q}{\operatorname{QRec}}
\newcommand{\Br}{\operatorname{Br}}
\newcommand{\Sk}{\operatorname{Sk}}
\newcommand{\It}{\operatorname{It}}
\newcommand{\Af}{\operatorname{After}}
\newcommand{\Bu}{\operatorname{Bubble}}
\newcommand{\Cr}{\operatorname{Cross}}
\newcommand{\Rt}{\operatorname{Rot}}
\newcommand{\Fx}{\operatorname{Fix}}
\newcommand{\id}{\mathrm I}
\newcommand{\blank}{\mathsf{blank}}
\makeatletter
\newcommand{\qsrpart}[2]{\par\noindent\textup{(#1)}\ \begingroup\edef\@currentlabel{\@currentlabel(#1)}\phantomsection\label{#2}\endgroup\ignorespaces}
\makeatother
\title{Quantum Step Recursion:\\Complexity Classes, Exact Spectra, and Structural Separations}
\author{Kirill Osipov\thanks{The Hague, The Netherlands. Correspondence: \texttt{d503@acmer.me}.}}
\date{26 November 2026}
\hypersetup{pdftitle={Quantum Step Recursion: Complexity Classes, Exact Spectra, and Structural Separations},pdfauthor={Kirill Osipov}}
\begin{document}
\maketitle
\begin{abstract}
We develop two quantum extensions of Step Recursion and establish a distinction between computational completeness and exact operator expressibility. Fixed-workspace iteration characterizes random-access quantum time polynomial in descent depth under suitable clock assumptions, with a sharp parity threshold; serial doubling characterizes bounded-error quantum space. Shrinking-register recursion captures BQP and FBQP at width one and construction rank at most ten with polynomial padding. Nevertheless, exact operator inclusion is governed by width divisibility, even up to eventual projective agreement. Specified finite gate libraries admit a complete joint basis--width--rank classification.

We establish necessary and sufficient recurrence conditions for determinant, non-torsion rank-one and scalar phase spectra, and prove that every realizable finite-order determinant can be concentrated on a single basis state at finite rank. Intersections of incomparable singleton-width algebras are not width-generated. Register reversal has minimum construction rank two, with a sharp asymptotic operator-norm approximation threshold of $\sqrt{2}$ among fixed terms of rank at most one, even up to global phase. The obstruction to common exact representatives of projectively equivalent families forms an abelian two-primary torsion group whose triviality remains open. Exact alignment is established in several sectors, and arbitrary projective intersections admit faithful interleaved encodings with doubled register length and widths, without increasing construction rank. Finally, no finite term realizes the unbounded quantum Fourier transform exactly on a tail, even up to global phase and with finitely many additional fixed-size gates.
\end{abstract}
\section{Introduction}
Computational complexity classifies computations by their resource requirements. It does not, in general, classify the exact operator families generated by a fixed finite recursion term. Quantum recursion therefore presents two distinct problems: which computational tasks a grammar can perform, and which length-indexed unitary families it can express exactly.

We study these problems using Step Recursion (SR)~\cite{SR}, in which descent is determined by generalized inverses of increasing generators.

We consider two quantum grammars. Fixed-workspace iteration schedules unitary updates by generalized-inverse descent. Shrinking-register recursion~\cite{Yamakami2020} uses predecessor descent with coherent prefix control. The former connects descent geometry to quantum time and space (Theorems~\ref{thm:clock-class}, \ref{thm:clock-space}); the latter permits exact classification by initial basis, recursion width and construction rank.

Our principal results distinguish computational collapse from structural non-collapse. Polynomial padding yields BQP and FBQP even at width one and construction rank at most ten (Theorem~\ref{thm:oneqrec-fbqp}). Coherent width grouping gives forward divisibility for exact operator inclusion (Theorems~\ref{thm:width-order}, \ref{thm:joint-resource}), and incomparable singleton-width algebras have intersections that are not width-generated (Theorem~\ref{thm:operator-meet}). Determinant and non-torsion phase spectra admit necessary and sufficient recurrence conditions (Theorems~\ref{thm:det-full-depth}, \ref{thm:rank-one-spectrum}, \ref{thm:scalar-nontorsion}); finite-order determinants can be concentrated on one basis state (Theorem~\ref{thm:finite-det-concentration}). Register reversal separates construction ranks one and two even under approximation up to global phase (Corollary~\ref{cor:robust-reversal}).

We further study projective intersections. Arbitrary scalar discrepancies reduce to bounded two-primary torsion, yielding an abelian obstruction group (Theorem~\ref{thm:alignment-obstruction-group}). Several sectors admit rank-preserving exact alignment, while the general case admits a faithful interleaved encoding with doubled register length and widths (Theorem~\ref{thm:interleaved-adjoint}).

The two grammars are related by formatted reset-clock transport and a clean polynomial-capacity lift (Proposition~\ref{prop:grammar-interface}). These constructions do not identify their ranks or the SR global pointwise majorant with workspace. The spectral and inclusion results are independent of QTM normalization, used for padded completeness and the internal clean lift. The two grammars thus retain different aspects of SR; neither identifies the classical and quantum hierarchies.

\subsection{Relation to previous work}
\paragraph{Quantum recursion and complexity.} Implicit quantum complexity uses quantum lambda calculi~\cite{DLMZ2010}, coherent recursion~\cite{Ying2014}, schematic function algebras~\cite{Yamakami2020,Yamakami2024}, and restricted recursive procedures (PFOQ)~\cite{HPS2023,HPS2025}. PFOQ compilation counts calls per branch~\cite{SG2026}; we count removed qubits. Random-access polylogarithmic time~\cite{Ferrari2025}, logspace~\cite{Yamakami2025QFO}, and resource analysis for richer recursive languages~\cite{CHPV2026,Barsse2026} address different models. QSynth~\cite{Deng2024} synthesizes inductive unitary families by program logic and SMT; we characterize exact membership at prescribed widths and rank. We vary generalized-inverse descent and answer Yamakami's one-qubit question (\cite[Sec.~6.1]{Yamakami2020}) with a sufficient rank and corrected readout, not a first schematic characterization of BQP. The compiler uses QTM simulation and measurement elimination~\cite{BV1997,Yao1993,FeffermanRemscrim2021,GRZ2021}.

\paragraph{Recurrences and finite-state representations.} Weighted automata and matrix-product representations are standard~\cite{CrosswhiteBacon2008,Droste2009}. Barloy et al.~\cite{Barloy2020} study rational recurrences whose characteristic roots are roots of rational numbers. Our determinant classification specifies the allowed factors $X^s-c$, their width-dependent multiplicities, and converse synthesis. Chuklanov~\cite{Chuklanov2026} uses uniform matrix-product operators for ladder-generator equivalence; we derive transfer representations for shrinking-register terms to bound minimum rank and approximation.

\paragraph{Exact and projective synthesis.} Clifford+$T$ and Clifford-cyclotomic synthesis~\cite{GilesSelinger2013,Amy2024} concern individual matrices and ancillary resources; we classify entire families denoted by one term without allocated qubits. The finite libraries include a non-torsion phase. The QFT field argument~\cite{FreedmanWang2007}, adjoint representation and topological phase-selection obstruction~\cite{Gavorova2024} use established methods. Our contributions are the finite-term obstruction, the native rank-nonincreasing translation and the scope of projective decoding; the general alignment problem remains open.

\section{Step Recursion and quantum descent}\label{sec:bridge}
Write $\mathbb N=\{0,1,\ldots\}$ and $\mathbb N_{>0}=\mathbb N\setminus\{0\}$. For a map $F$, $F^{[j]}$ denotes its $j$-fold iterate ($F^{[0]}=\mathrm{id}$). Write $I_m$ for the identity on $\mathbb C^m$, $|x\rangle$ for the computational-basis vector indexed by a binary word $x$, and $\oplus$ for bitwise XOR on equal-length words (or direct sum for operators). Fix a strictly increasing generator $G:\mathbb N\to\mathbb N$ with $G(0)\ge1$. For stride $s\ge1$ the step function is $\Phi_{G,s}=G^{[s]}$. Its generalized inverse and descent depth are
\[
 \rho_{G,s}(y)=\min\{z:G^{[s]}(z)\ge y\},\qquad
 D_{G,s}(y)=\min\{j:\rho_{G,s}^{[j]}(y)=0\}.
\]
Both vanish at zero. In bounded SR, each recursion requires a global pointwise majorant given by an earlier term; unbounded SR omits this requirement. Workspace bounds storage instead. Record basis $\mathcal B$, generator $G$, strides $S$, rank $d$, workspace $w$, grammar and readout. Rank is zero on initial operations, the maximum under composition and branching, and one plus the operand maximum at recursion. We also call it depth, distinct from descent depth and minimum realization rank.

\begin{lemma}[Clock geometry]\label{lem:clock-geometry}
For $s\ge1$ and $y\ge0$,
\[
 \rho_{G,s}=\rho_{G,1}^{[s]},\qquad D_{G,s}(y)=\lceil D_{G,1}(y)/s\rceil.
\]
\end{lemma}
\begin{proof}
Compose the adjunction $\rho_{G,1}(y)\le z\iff y\le G(z)$; sample the descent every $s$ steps. Zero is absorbing and positive arguments decrease since $G(z)\ge z+1$.
\end{proof}
On a fixed workspace define $F(0)=I$, $F(y)=V_{\rho_{G,s}(y)}F(\rho_{G,s}(y))$ for unitary updates $V_z$. Then $F$ is unitary, with $F(y)=V^{D_{G,s}(y)}$ for constant $V_z=V$. For permutation updates this is the usual permutation representation of the reversible fold, since permutation matrices preserve composition. Noninjective updates need not admit this lift.

For $g_0(z)=z+1$ and $g_1(z)=2z+1$, the inverses are $\max(0,y-s)$ and $\lfloor y/2^s\rfloor$, respectively. In particular,
\[
 D_{g_0,1}(n)=n,\qquad D_{g_1,1}(n)=\lceil\log_2(n+1)\rceil.
\]
These are the predecessor and halving clocks.

\subsection{Typed initial basis and descent-controlled iteration}
Let $x=x_0\cdots x_{n-1}\in\{0,1\}^n$, $n\ge1$. Length $n$ is read-only in binary; $O_x|i,b\rangle=|i,b\oplus x_i\rangle$ with $x_i=0$ for $i\ge n$ and $\ell=\lceil\log_2(n+1)\rceil$ address bits. Instructions are local deterministic controller transitions on finitely many work tapes, addressed quantum gates, or input queries (including indexed bit reads). Controller branches inspect only scanned symbols and finite control. Controller tapes and quantum registers both count as workspace. The fixed quantum library is $H,T=\operatorname{diag}(1,e^{i\pi/4})$, CNOT and inverses; $H|0\rangle=|+\rangle$, $H|1\rangle=|-\rangle$, $|\pm\rangle=(|0\rangle\pm|1\rangle)/\sqrt2$. Write $X=HT^4H$ for NOT; SWAP exchanges two bits, CNOT maps $|a,b\rangle$ to $|a,a\oplus b\rangle$. For exact Toffoli use $H$ on the target and the phase $(-1)^{abc}$ via $4abc=a+b+c-(a\oplus b)-(a\oplus c)-(b\oplus c)+(a\oplus b\oplus c)$, computing and uncomputing the parities with CNOT and using $T^{\pm1}$. Controlled SWAP is Toffoli between two CNOTs; conjugation by $X$ gives zero controls. Each program has a fixed instruction table and bounded-error decision readout.

A source term is generated by elementary instructions, finite composition, finite branching on controller data, and the following descent constructor. With root $n$, a clock constructor for an earlier program $U$ is
\[
 I_{G,s}[U](0)=\mathrm{id},\qquad
 I_{G,s}[U](y)=U\circ I_{G,s}[U](\rho_{G,s}(y))\quad(y>0).
\]
The operand uses the unchanged workspace, root metadata and mutable controller state. Each clock has a private descent counter; inner clocks reset to $n$, not the current shortened argument. Halted controllers and ill-formed gate addresses execute identity. For fixed binary input and initialized controller, the quantum workspace evolves unitarily; the controller itself need not be reversible. Rank counts nested constructors, taking maxima under composition and branching. Each controller transition costs one slot; clock arithmetic and routing are charged separately.

\begin{lemma}[Coherent controlled input access]\label{lem:controlled-query}
The controlled bit oracle $|c,i,b\rangle\mapsto|c,i,b\oplus c x_i\rangle$ is realized by one ordinary call to $O_x$, two controlled SWAPs, and one clean dummy response in state $|+\rangle$ (returned unchanged). This works for coherent controls and out-of-range addresses under the zero convention.
\end{lemma}
\begin{proof}
On the zero-control branch, the oracle sees the swapped-in response $|+\rangle$, fixed by both $I$ and $X$; on the one-control branch the real response receives the oracle. Swapping back restores the dummy on both branches, also coherently. Undoing its initial Hadamard restores zero. Route any selected address to the oracle port and back.
\end{proof}

\begin{lemma}[Instruction budget]\label{thm:clock-budget}
For any finite set of strides $S$, a fixed clock term of depth at most $d$ uses at most $C(1+D_{G,1}(n))^d$ elementary instructions, hence at most that many input queries, for a term-dependent constant $C$. If the inverse maps are computable on binary clock values in polynomial logarithmic time and $O(\log n)$ space, interpretation takes $O((1+D_{G,1}(n))^d\operatorname{polylog}(n+2))$ time before gate routing and $O(\log n)$ extra space (for fixed syntax). With $O(\log n)$ controller and quantum work, addressed gates likewise have polynomial logarithmic overhead. More generally, with polynomial-in-$D_{G,1}(n)+\ell$ workspace, addressing overhead is polynomial in $D_{G,1}(n)+\ell$.
\end{lemma}
\begin{proof}
Induct on syntax: composition and branching add bounds; a clock multiplies the operand bound by $D_{G,s}(n)\le D_{G,1}(n)$ (Lemma~\ref{lem:clock-geometry}). Use one binary counter per level, charging one slot even for an empty iteration. Clock arithmetic uses no queries; scans and controlled swaps give the routing overhead. The estimates hold on every controller path.
\end{proof}

\subsection{Descent-depth calibration and stride collapse}
Write BQP for bounded-error polynomial-time quantum decision and FBQP for its binary-output function class (success probability at least $2/3$ on every input). BQL is bounded-error quantum logarithmic space with polynomial time; BQPOLYLOGTIME uses time polynomial in $\log(n+2)$. BQSPACE$(w)$ uses $O(w)$ space and $2^{O(w)}$ time. Let $D(n)=D_{G,1}(n)$ and suppose $\ell=O(D(n))$, $D(n)\le n$, and each inverse map for the fixed permitted strides has the efficiency stated in Lemma~\ref{thm:clock-budget}. Write $\mathsf{BQTIME}_{\mathrm{ra}}(D^{O(1)})$ for bounded-error length-aware random-access quantum computation in time polynomial in $D(n)$, with a finite universal effective gate library. The uniform controller receives binary length metadata, may read indexed input bits, and emits addressed gates and coherent oracle calls; its local transitions, gate generation and address routing count toward time. Work is bounded by time. For the space-bounded assertions we use the standard circuit model with read-only binary input of~\cite{Watrous1999,FeffermanRemscrim2021}: the uniform controller is counted in the space bound, and read-only input access is available.

\begin{theorem}[Clocked quantum SR captures polynomial closure of descent time]\label{thm:clock-class}
For any nonempty set of permitted fixed strides $S$, the clocked quantum SR language class with work bounded by a program-dependent polynomial in $D(n)$ is exactly
\[
 \mathsf{BQTIME}_{\mathrm{ra}}(D^{O(1)}).
\]
In particular, it is independent of the choice of nonempty $S$. The predecessor clock captures $\mathsf{BQP}$ (and its prefix-free functional version captures $\mathsf{FBQP}$), while the halving clock captures length-aware $\mathsf{BQPOLYLOGTIME}$. With $O(\log n)$ work, the predecessor clock captures $\mathsf{BQL}$.
\end{theorem}
\begin{proof}
Soundness follows from Lemma~\ref{thm:clock-budget} and $\ell=O(D)$. Every term uses finitely many widths. For predecessor the oracle on a binary word can be simulated by polynomially many ordinary gates: scan the read-only input, compare a deterministic scan counter to the coherent address, conditionally flip the target and uncompute the comparison. This uses logarithmic auxiliary space and polynomial time.

For completeness, amplify the computation and simulate its uniform controller by local transitions. For a QTM presentation, apply uniform simulation~\cite{BV1997,Yao1993} to the visited work cells, retaining oracle instructions as addressed calls implemented using Lemma~\ref{lem:controlled-query}. For time $t$, the work-cell encoding, gate generation and routing cost polynomial resources in $t+\ell$; no input copy or oracle table is needed. Retain measurement and halt flags and approximate gates within the error margin. A deterministic controller counts descent, computes a time bound and emits the circuit within $CD(n)^k$ local steps. For any $s\in S$, depth-$k$ reset-clock nests supply $\lceil D(n)/s\rceil^k$ slots; a constant number of copies covers the bound and short lengths. Idle after halt. Work is polynomial in $D(n)$; for functions use polynomial-capacity prefix-free output.

For logarithmic work with predecessor, measurement elimination~\cite{FeffermanRemscrim2021} supplies logspace-uniform unitary circuits of polynomial size and logarithmic width. Their halting logspace generator may read the binary input, and runs in polynomial time by the configuration bound (write-only output). The same clock supplies its slots, including gate emission, with logarithmic total work.
\end{proof}

\begin{corollary}[Clock separation and sharp parity threshold]
\qsrpart{a}{cor:clock-separation}
At $O(\log n)$ work the halving-clock language class is strictly contained in its predecessor-clock counterpart, even allowing all fixed strides of halving. More generally, if parity is recognized by a depth-$d$ clock program, then $D(n)^d=\Omega(n)$; no fixed-depth program with a subpolynomial descent depth can recognize parity.
\qsrpart{b}{cor:sharp-parity}
Use the indexed bit-read instruction of the instruction basis, with a binary address stored on the controller work tape. For any fixed $d\ge1$ and nonempty width set $S$, exact parity on all input lengths is achievable at nesting depth at most $d$ with $O(\log n)$ work if and only if $D(n)^d=\Omega(n)$.
\end{corollary}
\begin{proof}
\par\noindent\textup{(a)} The halving fragment lies in the predecessor class, which includes parity. The $n/2$ bounded-error parity query lower bound~\cite{BBCMW2001} applies also to indexed reads and deferred controller branches; Lemma~\ref{thm:clock-budget} gives $D(n)^d=\Omega(n)$, excluding fixed-depth halving. OR gives the weaker $\Omega(\sqrt n)$ bound; see also~\cite{Yamakami2024,Ferrari2025}.
\par\noindent\textup{(b)} Necessity is Corollary~\ref{cor:clock-separation}. For sufficiency a controller reads consecutive addresses, toggling parity on a one. Maintain an $\ell$-bit address and remaining-count register, with least significant bits near the head origins. Ripple scans for increment/decrement visit position $j$ at most $O(1+n/2^j)$ times, hence cost $O(n+\ell)$ including returns. To test zero without a full scan maintain the number of ones in unary on an $O(\ell)$-cell track with its head at the boundary; each changed bit adds or removes a mark, and an end marker detects emptiness. Initialize from length metadata in $O(\ell)$ slots and conditionally apply $X$ to write the result. Thus $O(n+\ell)$ slots and $O(\ell)$ work suffice (routing and arithmetic separately charged). A constant number of depth-$d$ reset nests of width $s\in S$ supplies enough slots by the assumed bound, including short lengths; idle after halt. This uses indexed reads at one slot each.
\end{proof}

\noindent\textit{Scope of the time characterization.} These are resource characterizations for the typed grammar with length metadata and random access, not for a minimal initial basis or untagged elementary quantum recursion schemes (EQS). Related polylogarithmic-time characterizations are known~\cite{Ferrari2025}; the parameter here is generalized-inverse descent.

\subsection{Workspace-restricted operator recursion}
For a space-constructible $w(n)\ge\log_2(n+2)$, allow $O(w(n))$ total work, with an immutable root clock of length $m=Cw(n)$ for a fixed program constant. Use the serial update $F\mapsto F\circ F$ instead of $F\mapsto U\circ F$:
\[
 J[U](0)=U,\qquad J[U](y+1)=J[U](y)\circ J[U](y).
\]
Thus $J[U](m)=U^{2^m}$. Serial doubling is a separate constructor, not a linear single-use update.
\begin{theorem}[Space calibration]\label{thm:clock-space}
The resulting typed SR model with serial doubling and $O(w(n))$ controller and quantum workspace captures $\mathsf{BQSPACE}(w)$ in its standard exponential-time convention, with a final measurement only.
\end{theorem}
\begin{proof}
Interpret the serial recursion by a binary iteration counter of $m+1=O(w)$ bits; fixed syntax and nested clocks cost $O(w)$ additional space and $2^{O(w)}$ time. The length-aware input oracle can be simulated by a sequential scan of the read-only input using $O(w)$ space, as above. Space constructibility allows the clock length to be obtained in the bound. This proves soundness. Conversely, the equality $\mathsf{BQUSPACE}(w)=\mathsf{BQSPACE}(w)$ of Fefferman--Remscrim~\cite[Theorem 1]{FeffermanRemscrim2021} applies since $w$ is space-constructible and $w=\Omega(\log n)$. For the present time bound $T=2^{O(w)}$, their simulation uses $O(w+\log T)=O(w)$ space and $\operatorname{poly}(T2^w)=2^{O(w)}$ time, with bounded-error decision readout. Use the space-efficient gate-set conversion in~\cite{FeffermanRemscrim2021} for the fixed universal library. Its deterministic $O(w)$-space gate generator, including read-only input access, halts within $2^{O(w)}$ steps by the configuration bound (write-only output cannot influence transitions). Interpret its finite transition rules and gate-emission instructions, with no quantum data measured until the end. The serial clock of sufficiently large constant $C$ covers the generator; identity instructions follow halt. Total space is $O(w)$; the simulated unitary circuit retains its bounded-error guarantee. This does not give extensional Scheme-IV or Scheme-V compilation.
\end{proof}

\subsection{Shrinking-register source grammar}\label{sec:raw-model}
In Yamakami's measurement-free schematic grammar~\cite{Yamakami2020}, width-$s$ recursion reduces length $N$ to $N-s$ until threshold $b\ge s$. Its descent on excess length is $\rho_{g_0,s}(\max(0,N-b))$ and has $\lceil\max(0,N-b)/s\rceil$ steps. The removed $s$-qubit prefix controls the shorter-register operator, a resource absent from fixed-workspace iteration.

Write $U(m)$ for the unitary group on $\mathbb C^m$ and $SU(m)$ for its determinant-one subgroup. On one qubit, NOT is $X=\left(\begin{smallmatrix}0&1\\1&0\end{smallmatrix}\right)$, the phase gate is $\operatorname{PHASE}_\theta=\operatorname{diag}(1,e^{i\theta})$, and rotation has matrix $\left(\begin{smallmatrix}\cos\theta&-\sin\theta\\\sin\theta&\cos\theta\end{smallmatrix}\right)$. SWAP exchanges the first two bits. These gates act as identity on the other bits. Let $S\subseteq\mathbb N_{>0}$ be nonempty. Write $\mathcal A_S$ for the shrinking-register unitary Step Recursion algebra: the operator families denoted by finite terms over the original identity, first-qubit phase, rotation, NOT, and first-two-qubit SWAP gates, closed under composition, binary branching, and the following recursion. For a width $s\in S$, threshold $b\ge s$, active set $C\subseteq\{0,1\}^s$, and earlier terms $g,h,p$, set
\[
 F_n=\begin{cases}g_n&n\le b,\\ h_nD_C(F)_n p_n&n>b,\end{cases}\qquad
 D_C(F)_n=\bigoplus_{u\in\{0,1\}^s}\begin{cases}F_{n-s}&u\in C,\\ I_{2^{n-s}}&u\notin C.\end{cases}
\]
Binary branching $\Br[U,V]$ is identity at length one and $U_{n-1}\oplus V_{n-1}$ otherwise. SWAP is identity at length one. Products act right to left. The threshold $b\ge s$ ensures positive recursive suffixes. Gate constants have polynomial-time approximable entries (equivalently effective phase and rotation angles). Induction on syntax and length gives unitarity; inversion reverses composition, inverts branches and replaces $(g,h,p)$ by $(g^{-1},p^{-1},h^{-1})$. Rank counts nested recursion nodes only; composition and branching take maxima. Write $\mathcal A_S^{[d]}$ for denotations of rank at most $d$ (also called depth). Eventual projective agreement means $G_n=\lambda_n U_n$ on a tail, $|\lambda_n|=1$, without a definability condition on $\lambda_n$.

Write $\mathbb T=\{z\in\mathbb C:|z|=1\}$, $\mathbb T_{\rm eff}$ for its polynomial-time approximable phases, $(Ea)_n=a_{n+1}$ for the forward shift, and $\downarrow S=\{k\in\mathbb N_{>0}:k\mid s\text{ for some }s\in S\}$. A non-torsion phase has infinite multiplicative order. For infinite width sets, polynomial-lcm criteria require a finite permitted subset. Eventual identities hold on a tail, so finite forward shifts may be removed.

\begin{lemma}[Fixed-length evaluation and tables]
\qsrpart{a}{lem:guarded-eval}
Let $V$ be the matrix of a term at fixed positive length $j$. It has a recursion-free extension $L_j(V)$, identity below $j$ and $V\otimes I_{2^{n-j}}$ at length $n\ge j$, over the same initial library (including projections if present). Assume the library contains SWAP and admits branching, composition and fixed-arity gates acting trivially below their arity, as in each library considered here. In an inverse-closed unitary library, $V$ can be localized to length $j$ at rank zero. In any library admitting an inactive recursion of width $a\le j$, it can be localized at rank at most one.

\qsrpart{b}{lem:width-table}
Given a finite list of unitary matrices $V_j\in U(2^j)$, $1\le j\le B$, with polynomial-time approximable entries, there is a term using only the original unitary initial functions, composition and branching that equals $V_j$ at length $j\le B$ and identity at every length greater than $B$.
\end{lemma}
\begin{proof}
\noindent\textup{(a)} Unroll at length $j$ into controlled initial gates. For each arity-$a$ gate route targets to the last $a$ of the first $j$ wires by adjacent swaps (identical-branch wrappers); a branch tree of height $j-a$ implements its controls. The tree or leaf is inactive below $j$ and extends the action above $j$. Compose to obtain $L_j(V)$. For unitary $V$, $L_j(V)L_{j+1}(V^{-1}\otimes I_2)$ localizes at $j$. Otherwise a permitted inactive recursion of width $a\le j$, threshold $j$, base $L_j(V)$ and identity exterior localizes at rank one.
\par\noindent\textup{(b)} For a nonzero column pair $(a,b)$, the two-row gate
\[
 \frac{1}{\sqrt{|a|^2+|b|^2}}\begin{pmatrix}\bar a&\bar b\\-b&a\end{pmatrix}
\]
eliminates $b$. Repeated elimination gives a diagonal unitary. Undo the eliminations and use coordinate phases. Two-level gates are permutation-conjugated controlled one-qubit gates; hypercube edge transpositions (controlled NOTs) generate all permutations. Row and column phases reduce a one-qubit unitary to a real rotation; common phases use $X\operatorname{PHASE}_\theta X\operatorname{PHASE}_\theta$. Fixed nonzero pivots, square roots and local arguments give effective angles; skip zero pivots, without a uniform zero test. This is two-level elimination~\cite{Barenco1995}. Localize by Lemma~\ref{lem:guarded-eval} and compose.
\end{proof}

\begin{lemma}[Uniform circuit expansion with clean controls]\label{lem:circuit-expansion}
A fixed measurement-free term $F$ of construction rank $d$ has an exact uniformly generated circuit on $n$ data wires and $O(n)$ reusable clean control wires. The circuit uses $O((n+1)^d)$ controlled local gate instructions, each conditioned on a conjunction of fixed-width prefix predicates on disjoint wires outside its targets. Expanded into bounded-arity gates from a fixed term-dependent library this costs $O((n+1)^{d+1})$ gates. The controls are restored for every input, including an input entangled with a reference. Over a standard finite universal basis the circuit can be approximated, up to one overall phase, to inverse-polynomial operator-norm error in polynomial time.
\end{lemma}
\begin{proof}
Expand on a wire interval with inherited prefix predicates. Composition concatenates; branching emits the two suffix lists with their bit predicates. Above a recursion threshold emit $p$, one suffix expansion conditioned on $u\in C$, then $h$; otherwise emit $g$. Controls lie outside the controlled interval and the removed prefix is untouched by the suffix. This implements the direct sum even when $p,h$ mix branches, without $|C|$ copies.

Counts add under composition and branching; recursion satisfies $N_F(n)\le N_p(n)+N_h(n)+N_F(n-s)$ above threshold, with base $N_g(n)$. Summation gives $O((n+1)^d)$ from constant rank-zero counts. Disjoint fixed-width control blocks total at most $n$ bits. Compute flags reversibly, combine by a Toffoli chain, apply the gate and uncompute, using $O(n)$ gates and reusable clean bits per instruction. This is polynomial-time uniform and restores controls on entangled inputs. Approximate each whole controlled gate (not its phase-ambiguous uncontrolled version) up to phase with error $\eta/L$ for $L$ gates via Solovay--Kitaev~\cite{DawsonNielsen2006}; telescoping gives error $\eta$ up to overall phase.
\end{proof}

\section{Width-one schematic recursion: characterization of \texorpdfstring{$\FBQP$ and $\BQP$}{FBQP and BQP}}\label{sec:oneqrec}
\subsection{Statement and conventions}
Let $\A=\mathcal A_{\{1\}}$. Write $F=\Q_t[g,h,p\mid F_0,F_1]$ for the width-one constructor with threshold $t$, where $F_i=F$ on active branches and $F_i=\id$ otherwise. All recursions in this section have $t=1$ unless stated otherwise. A fixed-size gate is guarded by its required length and is identity on shorter registers (Lemma~\ref{lem:width-table}). Use input-length-dependent padding and prefix-free output: a raw prefix can encode several distinct answers.

Write $H_j=0^j1$ for $j\ge0$ and $\phi^p(x)=H_nH_{p(n)}H_{11p(n)+6}x$ for $n=|x|$ and a nonnegative integer-valued polynomial $p$. For $s=s_1\cdots s_j$, let $\widetilde s=0s_1\cdots 0s_j11$. These codes are prefix-free on aligned two-bit boundaries. For a binary word $z$ and register length $N\ge1$ define the prefix projector
\[
 \Pi_z^{(N)}=\begin{cases}|z\rangle\langle z|\otimes I_{2^{N-|z|}},&|z|\le N,\\0,&|z|>N.\end{cases}
\]
The empty prefix gives identity. Terms allocate no qubits.

\begin{theorem}[Original-padding completeness]\label{thm:oneqrec-fbqp}
For every polynomially bounded string function $f$, the following are equivalent:
\begin{enumerate}
\item $f\in\FBQP$.
\item For every constant $0<\epsilon<1/2$, some $g\in\A$ and polynomial $p$ satisfy
\[\|\Pi_{\widetilde{f(x)}}^{(N)}g_N\ket{\phi^p(x)}\|^2\ge1-\epsilon,\qquad N=|\phi^p(x)|,\quad x\in\{0,1\}^*.\]
\end{enumerate}
In (2), the term may be chosen of construction rank at most ten. For characteristic functions, the output can instead be a single leading qubit with direct basis-state readout, giving $\BQP$ with precisely the original initial unitary functions. Empty inputs are allowed. The converse uses uniform circuit expansion, not postselection.
\end{theorem}

The proof is in Appendix~\ref{app:compiler}.

\subsection{Derived operations in the original basis}
The rotation, inverse rotation, unary-header skipping and counted-iteration ideas already occur in~\cite[Sec.~3 and Sec.~5.4]{Yamakami2020}; the following explicit versions are used to fix the exact conventions of the compiler. The new task is to compile the machine using only width-one recursions. For $w\ge1$, $\Rm_w$ cyclically moves the first $w$ bits to the end (on lengths at least $w$); $\Rp_w=\Rm_w^{-1}$. In particular $\Rm_1=\Q_1[\id,\id,\mathrm{SWAP}\mid\Rm_1,\Rm_1]$, and fixed powers give the other widths. Short inputs are handled by the grammar. Define $\Fx_w[V]$ by $w$ nested $\Br[V,V]$, i.e. act after $w$ fixed-position bits; $\Fx_0[V]=V$. Likewise finite branching on a fixed $w$-bit word is obtained by a binary tree of height $w$ (construction rank zero). The basic skip and counted iteration are
\begin{align*}
\Sk[V]&=\Q_1[\id,\Br[\id,V],\id\mid \Sk[V],\id],\\
\It[V]&=\Q_1[\id,\Br[\Sk[V],\id],\id\mid \It[V],\id].
\end{align*}
They satisfy $\Sk[V](H_jW)=H_jV(W)$ and $\It[V](H_jW)=H_jV^j(W)$ on nonempty payloads. Induction on $j$ proves both identities. Let $D=\Q_1[\id,\mathrm{SWAP},\id\mid D,\id]$, so $D(H_j0W)=H_{j+1}W$ and $D^{-1}(H_{j+1}W)=H_j0W$.

\paragraph{Compiler and clean transport.} Appendix~\ref{app:compiler} proves Theorem~\ref{thm:oneqrec-fbqp}: the native machine interface (Theorem~\ref{thm:native-machine-interface}) uses ranks four, eight and ten for the stopped transition, extraction and complete simulator. The clean SR lift and coherent descent fold require ranks at most twelve and thirteen (Proposition~\ref{prop:clean-sr-lift}, Corollary~\ref{cor:coherent-clock-fold}). Rank ten bounds complexity in padded length, not a uniform exponent in input length. Appendix~\ref{sec:historical-model} treats the readout and guard corrections.

\section{Determinant spectra at fixed construction rank}\label{sec:determinants}

In addition to the rotation $\Rm_1$ and its inverse $\Rp_1$ from Section~\ref{sec:oneqrec}, the cyclic increment $P$ on little-endian rank $r(x)=\sum_{i=1}^n2^{i-1}x_i$ has base NOT, pre-operation $I$, post-operation NOT on the first bit, and only branch $1$ active. Induction gives $P|r\rangle=|r+1\bmod 2^n\rangle$. Rotation and increment have depth one.

For $Q(X)=\sum_jq_jX^j\in\mathbb Z[X]$, set $(Q(E)\delta)_n=\prod_j\delta_{n+j}^{q_j}$ and $(Q(E)a)_n=\sum_jq_ja_{n+j}$ for phase and additive sequences, respectively. On eventual sequences $E$ is invertible. Over a field the monic generator of a nonzero annihilator ideal is the minimal eventual annihilator. Write $\mathbb F_p=\mathbb Z/p\mathbb Z$ and $\bar Q$ for reduction modulo the indicated prime. Determinants multiply under composition; $c$ active blocks contribute a $c$th power. Effective phases admit fixed-order roots by local arguments. Polynomial addition multiplies phase sequences; multiplication composes their actions.
\begin{lemma}[Primary synthesis reduction for phase recurrences]\label{lem:primary-synthesis}
Let $P\in\mathbb Z[X]$ be monic, with characteristic-zero factorization $P=\prod_i q_i^{e_i}$ into distinct monic irreducibles. Let $\Lambda$ be a multiplicative subgroup of unit phases admitting roots of every fixed positive integer order within $\Lambda$. If $P(E)\delta=1$ eventually and $\delta_n\in\Lambda$, then on a tail
\[
 \delta_n=\prod_{i,j}\gamma_{ij}^{b_{ij}(n)},\qquad
 \gamma_{ij}\in\Lambda,\quad b_{ij}(n)\in\mathbb Z,\quad q_i^{e_i}(E)b_{ij}=0\ \text{eventually},
\]
with finitely many factors. This applies to effective unit phases and to the groups of all odd-order or all $2$-power-order roots of unity. For $P=1$ the tail is identically one. The decomposition is performed in characteristic zero before specializing the phases; no coprimality of reduced factors modulo the phase order is required.
\end{lemma}
\begin{proof}
By monicity, on a tail $\delta_n=\prod_j\beta_j^{u_j(n)}$ for initial phases $\beta_j$ and integral recurrence solutions $u_j$ with Kronecker initial data. Rational CRT projectors $B_i$ satisfy $\sum_iB_i=1\bmod P$ and $q_i^{e_i}B_i=0\bmod P$. Clear denominators by $D>0$: $b_{ij}=DB_i(E)u_j$ are integral with the required annihilators and sum $Du_j$. Choose $\gamma_{ij}^D=\beta_j$ in $\Lambda$. This step needs divisibility of $\Lambda$, not divisibility of the cyclic group generated by $\beta_j$. For effective phases use a local argument. For a phase of $p$-power order $p^a$, writing $D=p^hD_0$ with $p\nmid D_0$, take a root of order dividing $p^{a+h}$ and invert $D_0$ modulo that order. Apply this prime by prime for odd-order phases. Thus primary order type is preserved, but a specified finite order need not be.
\end{proof}

\begin{lemma}[Phase synthesis by differences]\label{lem:phase-difference}
Let $\delta$ be a sequence of effective unit phases. The following table specifies a base polynomial $B$, the allowed factors $F_s$ (with $s\in S$), and the realization requested:
\[
\begin{array}{c|c|c}
 B & F_s & \text{realization at length }n\\ \hline
 X-2 & X^s-c,\ 1\le c\le 2^s & \det U_n=\delta_n\\
 1 & X^s-1 & I+(\delta_n-1)|0^n\rangle\langle0^n|\\
 X-1 & X^s-1 & \delta_n I_{2^n}
\end{array}
\]
If $B(E)\prod_{i=1}^r F_{s_i}(E)\delta=1$ eventually, the indicated realization has depth at most $r$. For finite $S$, let $L$ be the monic lcm of the factors in the corresponding row. The same conclusion with depth $d$ holds if $(BL^d)(E)\delta=1$ eventually. In the rank-one row, product synthesis for phases in the cyclic group of a supplied phase gate uses only that gate, NOT, swaps, branching and composition besides the indicated recursions.
\end{lemma}
\begin{proof}
For product synthesis with $r=0$, the tails are respectively $\delta_{N+i}=\delta_N^{2^i}$, identity, and constant. Realize the first by a guarded unitary extension at $N$ of determinant $\delta_N$, the second by identity, and the third by a constant one-qubit multiple of identity. Correct all earlier lengths by finite tables.

For the last factor $X^s-c$ (take $c=1$ in the last two rows), set $\eta_n=\delta_n\delta_{n-s}^{-c}$ for $n>s$, with arbitrary initial values. The remaining annihilator kills $\eta$ on a tail. In a width-$s$ recursion, take its earlier realization as post-operator, identity as pre-operator, and a finite base matching $\delta$ up to $s$. In the first row use $c$ active prefixes; determinants obey $\delta_n=\eta_n\delta_{n-s}^c$. In the second use only $0^s$; in the third use all prefixes. In these rows the exceptional phase or common phase is multiplied by $\eta_n$, yielding $\delta_n$. Induction proves the bound. For a supplied phase gate the fixed rank-one bases and their corrections use a controlled integral power of that gate, localized by Lemma~\ref{lem:guarded-eval}, without arbitrary angles.

For the lcm assertion, apply the separate root-extraction step of Lemma~\ref{lem:primary-synthesis}. Each allowed factor is squarefree in characteristic zero; every primary factor of $BL^d$ divides a product of $B$ and at most $d$ allowed factors (repeat one containing it). Apply product synthesis to each component, compose, and correct finite exceptions. The roots are effective; the lcm step may enlarge finite phase order as in Lemma~\ref{lem:primary-synthesis}.
\end{proof}

\begin{lemma}[Lifting annihilators modulo prime powers]\label{lem:lift-annihilator}
Let $a$ be an eventual sequence modulo $p^m$, for a prime $p$ and $m\ge1$. If $P_i(E)a=0$ for finitely many integer polynomials $P_i$ and $\bar H\in(\bar P_i)_i\mathbb F_p[X,X^{-1}]$, then $H(E)^m a=0$ eventually.
\end{lemma}
\begin{proof}
In the Laurent ring lift the ideal identity as $H=\sum_i A_iP_i+pZ$. On eventual sequences the shift is invertible. Thus $H(E)a=pZ(E)a$ and, since these operators commute, $H(E)^m a=p^mZ(E)^m a=0$.
\end{proof}

\begin{theorem}[Complete fixed-depth determinant image]\label{thm:det-full-depth}
For a finite nonempty width set $S$, define the monic integer polynomial
\[
 L_S(X)=\operatorname{lcm}_{s\in S,\ 1\le c\le2^s}(X^s-c),\qquad P_{S,d}(X)=(X-2)L_S(X)^d.
\]
A sequence $\delta_n\in\mathbb T_{\rm eff}$ is the exact determinant sequence of a term of depth at most $d$ in $\mathcal A_S$ if and only if $P_{S,d}(E)\delta=1$ eventually. For infinite $S$ the criterion is existence of a finite nonempty subset with this property. In particular, the least admissible $d$ is the exact minimum determinant-realization depth, not necessarily the minimum depth of any specified operator having that determinant. Necessity is valid even for arbitrary fixed gate constants; sufficiency as stated uses admissible angles.
\end{theorem}
\begin{proof}
All integer-polynomial lcms are monic over $\mathbb Q[X]$ and integral by Gauss's lemma. For necessity, a recursion-free initial gate has eventual determinant $\beta^{2^{n-r}}$ with fixed $r$ and phase $\beta$. Its annihilator is $X-2$. Composition multiplies determinants and branching multiplies shifted determinants, so both preserve a common multiplicative annihilator. At depth at most $d\ge1$, a width-$s$ recursion with $c$ active branches has
\[
 \delta_n=\eta_n\delta_{n-s}^c
\]
eventually, with $\eta$ a product of earlier operand determinants of depth at most $d-1$. Thus $(X^s-c)(E)\delta=E^s\eta$ on a tail. If $c>0$, the factor divides $L_S$, and $P_{S,d}$ annihilates $\delta$ by induction. If $c=0$, the shift $E^s$ can be removed on tails and the earlier annihilator suffices.

For sufficiency, apply the determinant row of Lemma~\ref{lem:phase-difference} with $B=X-2$ and $L=L_S$. For infinite $S$ each term uses finitely many widths, giving the stated convention.
\end{proof}

\begin{corollary}[Finite-order and non-torsion determinant exponents]
\qsrpart{a}{cor:torsion-depth}
Let $m\ge2$ and $\zeta_m=\exp(2\pi i/m)$. A sequence $\zeta_m^{a_n}$, with $a_n\in\mathbb Z/m\mathbb Z$, is a determinant of depth at most $d$ over finite $S$ if and only if $P_{S,d}(E)a=0$ eventually in $\mathbb Z/m\mathbb Z$. For prime $m=p$, let $q_a$ be the monic minimal eventual annihilator in $\mathbb F_p[X]$, if one exists; otherwise no realization exists. The exact minimum depth is the least $d\ge0$ such that
\[
 q_a\mid (X-\bar 2)\overline{L_S}^{\,d}\quad\text{in }\mathbb F_p[X].
\]
In particular, ignoring depth, membership is equivalent to every irreducible factor of $q_a$ dividing $\overline{L_S}$. For the nonzero exponential sequence $a_n=\lambda^n$ with $\lambda\in\mathbb F_p^\times$, this is equivalent to $\lambda^s=\bar c$ for some $s\in S$ and integer $1\le c\le2^s$.
\qsrpart{b}{thm:det-image}
For any effective phase $\beta$ of infinite order, $\beta^{a_n}$ with integer exponents is a determinant sequence exactly when $a$ has an eventual annihilator that is a finite product of $X^s-c$ with $s\in S$ and $1\le c\le 2^s$. At fixed depth $d$ the exact criterion is that its minimal eventual monic rational annihilator divide $(X-2)L_S(X)^d$ (using a finite subset when necessary).
\end{corollary}
\begin{proof}
\par\noindent\textup{(a)} For a primitive $m$th root, multiplicative annihilation is exactly additive annihilation modulo $m$. Apply Theorem~\ref{thm:det-full-depth}. Over $\mathbb F_p$, annihilators form a polynomial ideal; $X-\bar2$ divides $\overline{L_S}$ since $X-2$ divides $X^s-2^s$. For $a_n=\lambda^n\ne0$ the minimal polynomial is $X-\lambda$. Compute the lcm before reduction, since distinct factors can merge modulo $p$.
\par\noindent\textup{(b)} Infinite order converts the determinant recurrence in Theorem~\ref{thm:det-full-depth} to an additive recurrence. The factor $X-2$ divides every $X^s-2^s$, and finite products divide some $L_{S_0}^d$ for finite $S_0\subseteq S$. Since $L_S$ is squarefree in characteristic zero, minimum depth is the largest multiplicity of an allowed factor in the minimal annihilator, discounting one copy of $X-2$ and taking a nonnegative maximum.
\end{proof}
\section{Non-torsion phase spectra: rank-one and scalar operators}\label{subsec:rank-one-spectrum}
Determinant exponents can grow exponentially; individual phase ratios obey a polynomial bound.
\begin{lemma}[Specialization and integral phase coordinates]\label{lem:phase-specialization}
Let $K\subseteq\mathbb C$ be a finitely generated field containing a non-torsion phase $\beta$, and let $\Gamma\subset K^\times$ be a finitely generated multiplicative subgroup containing $\beta$. Let $H=\{x\in K^\times:x^m\in\Gamma\text{ for some }m\ge1\}$. Given a finite list of elements of $K$ (which may include conjugates of the gate constants), there exist a number field $F$, an integer $D\ge1$, a representation of $K$ by $D\times D$ matrices over a rational function field $L=F(t_1,\ldots,t_r)$, and a rational specialization of the $t_i$ such that:
\begin{enumerate}
\item the matrices for the finite list have no poles at the specialization, and their entrywise evaluation defines a ring homomorphism on the ring generated by that list and $\mathbb Z$;
\item evaluation of the field norm gives a multiplicative homomorphism $\nu:H\to F^\times$, with $\nu(\beta)$ non-torsion, and its image is finitely generated;
\item when $x\in H$ is in the ring in (1), $\nu(x)$ is the determinant of its specialized multiplication matrix. In particular there is an integer-valued homomorphism $v:H\to\mathbb Z$ with $v(\beta)>0$.
\end{enumerate}
\end{lemma}
\begin{proof}
For algebraic $\beta$ take $F=\mathbb Q(\beta)$; otherwise take $F=\mathbb Q$ and extend $t_1=\beta$ to a transcendence basis of $K/F$. Put $D=[K:L]$ and use multiplication matrices $M_x$ in an $L$-basis, with $\det M_x=N_{K/L}(x)$. Include $\beta^{\pm1}$ in the list. Choose rational $t_i$ avoiding all matrix denominator zeros and numerator and denominator zeros of generator norms. In the transcendental case impose $t_1>1$. A finite product of nonzero polynomials cannot vanish throughout a rational open box.

For $x\in H$, $N_{K/L}(x)^m$ is a product of powers of generator norms. Unique factorization in $F[t_1,\ldots,t_r]$ implies every numerator or denominator factor already occurs among those norms, so evaluation is defined and nonzero. The resulting norm homomorphism $\nu$ sends $\beta$ to $\beta^D$ or $t_1^D>1$, both non-torsion. Since some power of each $\nu(x)$ lies in $\nu(\Gamma)$, valuations outside a fixed finite set vanish. This group is finitely generated: its valuation image is a subgroup of a finite-rank free abelian group and its kernel is a subgroup of the ordinary unit group (Dirichlet~\cite{MilneANT}). Thus $\nu(H)$ is finitely generated. A free integer coordinate nonzero on $\nu(\beta)$, with sign chosen positive, gives $v$ by composition. Matrix evaluation on the listed generated ring respects every polynomial identity between its elements and commutes with determinants. No specialization of the conjugation operation is required.
\end{proof}

\begin{lemma}[Finite saturation of fixed phase generators]\label{lem:finite-phase-saturation}
Let $K\subseteq\mathbb C$ be finitely generated over $\mathbb Q$, and let $\Gamma\subseteq K^\times$ be a finitely generated group of unit phases. Its saturation $H=\{x\in K^\times:x^m\in\Gamma\text{ for some }m\ge1\}$ is a finitely generated group of unit phases. If the generators of $K$ are effective, every element of $H$ is effective. In particular $H$ splits as $H_{\rm free}\times H_o\times H_2$, where the last two factors are finite cyclic groups of odd and power-of-two order respectively.
\end{lemma}
\begin{proof}
The torsion subgroup of $K^\times$ is finite: if $[K:\mathbb Q(t)]=D$ for a transcendence basis $t$, then a primitive $m$th root in $K$ has $\varphi(m)\le D$. Indeed adjoining the independent variables $t$ preserves the degree of the cyclotomic field. The formula for $\varphi$ implies $\varphi(m)^2\ge m/2$, by checking each prime power, so only finitely many orders occur. Every element of $H$ is a unit phase by taking absolute values of its defining power identity.

Let $r$ be the free rank of $\Gamma$. The integer homomorphisms of Lemma~\ref{lem:phase-specialization} separate $(\Gamma/\operatorname{Tor}\Gamma)\otimes\mathbb Q$: clear denominators and apply the lemma to the resulting non-torsion element. Choose at most $r$ by successive kernel reduction. Their joint map $H\to\mathbb Z^r$ has torsion kernel, since a power of each kernel element lies in $\Gamma$ with all coordinates zero. The kernel is finite, the image finitely generated; lift generators and adjoin the kernel. This includes $r=0$. The structure theorem gives the splitting because finite subgroups of $\mathbb C^\times$ are cyclic. Fixed field elements are rational expressions in the generators, with fixed nonzero denominators, hence effective if the generators are. No uniform computation of the splitting is asserted.
\end{proof}

\begin{lemma}[Polynomial growth of individual phase ratios]\label{lem:entry-growth}
Let $G$ be a finite unitary term of recursion depth $d$, and suppose that for every sufficiently large $n$ two of its matrix entries $u_n,w_n$ have modulus one and satisfy $u_n/w_n=\beta^{a_n}$ with $a_n\in\mathbb Z$ and fixed non-torsion $\beta\in\mathbb T$. Then $|a_n|=O((n+1)^d)$. The entries may be chosen at different positions at different lengths.
\end{lemma}
\begin{proof}
Apply Lemma~\ref{lem:phase-specialization} to the field of gate entries, conjugates, $\beta$ and any short exceptional matrices, listing these entries and taking $\Gamma$ containing $\beta$. The specialized multiplication representation respects products and direct sums; local gates become, up to block permutation, sums of identical fixed blocks. Choose a place of $F$ with $|\nu(\beta)|\ne1$: otherwise $\nu(\beta)$ is an algebraic unit with all conjugates of modulus one, hence torsion by Kronecker's theorem. Use the complex Euclidean operator norm at an infinite place, and the maximum entry norm at a finite place (submultiplicative by ultrametricity). Direct sums take maxima; permutation norms are one.

For a term $V$ let $B_V(n)$ be the logarithm of the maximum of one and the norm of its specialized matrix. Initial gates have $B_V(n)=O(1)$, composition adds bounds, and branching takes a maximum of shifted bounds. A width-$s$ recursion satisfies
\[
 B_V(n)\le B_h(n)+B_p(n)+B_V(n-s)\qquad(n>b),
\]
where $b$ is its base threshold; an inactive branch contributes identity. By induction on syntax, then by summing this bound along at most $n$ descent steps, $B_V(n)=O((n+1)^{\operatorname{depth}(V)})$. The same proof applies to the entrywise complex-conjugated syntax, which has the same depth and uses the included conjugate constants. Short lengths contribute fixed bounds. The specialized matrices need not be unitary.

The identities $u_n\bar u_n=w_n\bar w_n=1$ and $u_n\bar w_n=\beta^{a_n}$ hold in $K$. Include $\beta^{\pm1}$ and the conjugate gate entries in the listed ring, which contains all these products at every length. Its ring homomorphism preserves the identities even if it does not preserve conjugation. The multiplication matrices for $u_n,w_n$ are blocks of specialized $G_n$, and their inverses are the corresponding blocks of specialized $\overline{G_n}$. Thus the norm and inverse norm of specialized $M_{u_n/w_n}$ are $\le\exp(O((n+1)^d))$. At either place, determinants of $D\times D$ matrices are bounded by a fixed constant times the $D$th power of the norm. Applying this to the matrix and inverse gives $|a_n\log|\nu(\beta)||=O((n+1)^d)$, since its determinant is $\nu(\beta)^{a_n}$.
\end{proof}

\begin{lemma}[Slow part of the recurrence]\label{ws:slow}
Suppose an integer sequence $d_n$ is eventually annihilated by a product of factors $X-2$ and $X^s-c$ with $1\le c\le2^s$. Then eventually
\[
d_n=a_n+2^n b_n,
\]
where $a_n$ is rational, has an annihilator involving only $X^s-c$ with $c<2^s$, and $a_n=o(2^n)$. The sequence $b_n$ is rational and is a polynomial in $n$ on each residue class modulo a fixed positive integer. There is an integer $D>0$ such that $Db_n\in\mathbb Z$ on a tail.
\end{lemma}
\begin{proof}
Factor the annihilator into the product $P_< P_=$ of the factors whose roots have modulus strictly less than two and exactly two, respectively. The latter consists of powers of $X-2$ and $X^s-2^s$; the former of powers of $X^s-c$ with $c<2^s$. They are coprime over $\Qrat$. A B\'ezout identity $A P_<+B P_=1$ in $\Qrat[X]$ splits $d$ into $a=B(E)P_=(E)d$ and $u=A(E)P_<(E)d$ on a sufficiently late tail. The former is killed by $P_<(E)$ and the latter by $P_=(E)$. Both are rational sequences. If either factor is constant its corresponding component is zero. The standard finite linear recurrence solution over $\mathbb C$ expresses $a_n$ as a finite sum of polynomial multiples of $\alpha^n$, $|\alpha|<2$, proving $a_n=o(2^n)$.

Set $b_n=2^{-n}u_n$. Its characteristic roots are roots of unity. Take $M$ divisible by all widths appearing in $P_=$ (or $M=1$ if there are none). For sufficiently large $r$, $(E^M-1)^r b=0$ eventually. On each residue class, the $r$th forward difference of the rational sequence vanishes. Its values are therefore a rational polynomial in the progression index, equivalently in $n$, by the binomial-coefficient formula for finite differences. The finitely many polynomial coefficients have a common denominator $D$.
\end{proof}

For finite $S$, write $M_S(X)=\operatorname{lcm}_{s\in S}(X^s-1)$ and let $\Phi_k$ denote the $k$th cyclotomic polynomial. For a phase $\beta$ and integer exponents $a=(a_n)_{n\ge1}$ (or exponents modulo its finite order), put
\[
 R(\beta,a)_n=I+(\beta^{a_n}-1)|0^n\rangle\langle0^n|.
\]
\begin{theorem}[Complete projective and exact rank-one non-torsion spectrum]\label{thm:rank-one-spectrum}
Let $\beta\in\mathbb T_{\rm eff}$ be non-torsion, $a_n\in\mathbb Z$, and $d\ge0$. The following are equivalent:
\begin{enumerate}
\item $R(\beta,a)$ is realized at every positive length by a term in $\mathcal A_S$ of recursion depth at most $d$.
\item A term of that depth in $\mathcal A_S$ agrees with $R(\beta,a)$ up to arbitrary length-dependent global phases at all sufficiently large lengths.
\item $M_S(E)^d a=0$ eventually (with the finite-subset convention for infinite $S$).
\end{enumerate}
For $d=0$ the last condition means $a$ is eventually zero. Equivalently, when $a$ is not eventually zero, its minimal eventual rational recurrence polynomial must be a product of cyclotomic polynomials $\Phi_k$, every occurring $k$ must divide some allowed width, and each multiplicity must be at most $d$. The largest multiplicity is the minimum depth, both for exact and eventual projective realization. The necessity holds even if the candidate term uses arbitrary fixed gate constants rather than effective angles.
\end{theorem}
\begin{proof}
(1) implies (2). For (2), write $G_n=\lambda_nR(\beta,a)_n$ on a tail. Let $K$ be generated by $\beta$, gate entries and conjugates. The entries at $0^n,1^n$ have ratio $\beta^{a_n}$ and modulus one; Lemma~\ref{lem:entry-growth} gives $|a_n|=O((n+1)^d)=o(2^n)$. The second entry is $\lambda_n\in K$. Let $\Gamma$ be generated by $\beta$ and local (including exceptional short) determinants. All term determinants belong to $\Gamma$ by induction, and $\lambda_n^{2^n}=\det G_n\beta^{-a_n}\in\Gamma$. Thus $\lambda_n$ belongs to its saturation $H$. Lemma~\ref{lem:phase-specialization} gives $v:H\to\mathbb Z$ with $e=v(\beta)>0$. For finite $S_0\subseteq S$ containing the used widths, determinant necessity yields annihilator $(X-2)L_{S_0}^d$ for $v(\det G_n)$; roots of modulus less than two have multiplicity at most $d$.

Write $v(\det G_n)=u_n+2^n b_n$ by Lemma~\ref{ws:slow}, where $u_n=o(2^n)$ and $b_n$ is a rational quasipolynomial with bounded denominators. Choose a fixed positive integer $D_0$ clearing the denominators of $b_n$. The determinant identity gives
\[
 D_0(u_n-ea_n)=2^nD_0(v(\lambda_n)-b_n)\in 2^n\mathbb Z
\]
The left side is $o(2^n)$, hence eventually zero; $u_n=ea_n$ on a tail. The slow component inherits the original annihilator (it is a polynomial in $E$ applied to the determinant sequence). Allowed slow roots have modulus at least one. Polynomial growth excludes roots of modulus $R>1$: dividing by the largest $R^n n^k$ would give a nonzero finite sum of distinct unit-circle exponentials tending to zero, whereas its Ces\`aro mean square tends to the sum of squared coefficients. Thus only roots of $X^s-1$ remain, with multiplicity at most $d$, giving $M_{S_0}(E)^da=0$.

For (3) implies (1), apply the rank-one row of Lemma~\ref{lem:phase-difference} to $\delta_n=\beta^{a_n}$ with $L=M_{S_0}$. Its factorization $X^s-1=\prod_{k\mid s}\Phi_k$ gives the cyclotomic criterion and the minimum depth.
\end{proof}

\begin{corollary}[Non-torsion rank-one meets and concentration obstruction]
\qsrpart{a}{cor:rank-one-meet}
For a fixed non-torsion admissible phase $\beta$, let $\mathcal R_{S,d}(\beta)$ be the set of rank-one families $R(\beta,a)$ that have a realization of depth at most $d$ in $\mathcal A_S$. The same set is obtained using eventual projective realization. For any nonempty $S,T$ and $d,e\ge0$, with $J=\downarrow S\cap\downarrow T$,
\[
 \mathcal R_{S,d}(\beta)\cap\mathcal R_{T,e}(\beta)
 =\mathcal R_{J,\min(d,e)}(\beta).
\]
Taking the union over depths gives the unrestricted-rank identity, including singleton widths.
\qsrpart{b}{cor:non-torsion-no-concentrate}
For any admissible non-torsion $\beta$, the determinant sequence $\beta^{2^n}$ is realizable without recursion, but the family $R(\beta,(2^n)_n)$ has no finite realization using \emph{any} set of widths, even allowing arbitrary fixed gate constants, arbitrary length-dependent global phases and finitely many exceptional lengths. More generally every superpolynomial integer exponent sequence is excluded on a rank-one phase by Lemma~\ref{lem:entry-growth}, although the determinant image contains many such sequences. Thus determinant realizability is not equivalent to rank-one phase realizability; the finite-order concentration theorem below does not extend to arbitrary non-torsion phases.
\end{corollary}
\begin{proof}
\par\noindent\textup{(a)} Theorem~\ref{thm:rank-one-spectrum} requires cyclotomic orders in the respective downsets and multiplicities bounded by the respective ranks. Intersect these conditions; the finitely many occurring orders suffice even for infinite width sets.
\par\noindent\textup{(b)} Apply a one-qubit phase of value $\beta^2$ to the first qubit. At length $n$ its determinant is $(\beta^2)^{2^{n-1}}=\beta^{2^n}$. The rank-one family violates the polynomial entry-growth lemma for every finite depth. The same argument excludes the general superpolynomial case.
\end{proof}

\begin{theorem}[Non-torsion scalar families: exact depth and width meets]\label{thm:scalar-nontorsion}
Let $\beta\in\mathbb T_{\rm eff}$ have infinite order and let $a_n\in\mathbb Z$. For finite nonempty $S$ the family $\beta^{a_n}I_{2^n}$ belongs to $\mathcal A_S$ at depth at most $d$ if and only if
\[
 (E-1)M_S(E)^d a=0\quad\text{eventually},\qquad M_S(X)=\operatorname{lcm}_{s\in S}(X^s-1).
\]
The same criterion holds for eventual exact agreement; for infinite $S$ use a finite subset. Thus if the minimal eventual rational annihilator is $\prod_k\Phi_k^{m_k}$, membership requires that every occurring order $k$ divide some allowed width, and the minimum depth is $\max(0,m_1-1,\max_{k>1}m_k)$ (a missing multiplicity is zero). For arbitrary nonempty $S,T$, the scalar-sector intersection at depths $d,e$ is exactly the scalar sector at common downset $\downarrow S\cap\downarrow T$ and depth $\min(d,e)$. This theorem concerns exact scalars, not projective equality, which would make every scalar equivalent to identity.
\end{theorem}
\begin{proof}
The determinant is $\beta^{2^na_n}$, so Theorem~\ref{thm:det-full-depth} gives $P_{S,d}(2E)a=0$. Rational CRT splits roots of modulus one from roots of modulus less than one. Both components have bounded denominators as rational combinations of shifts of integer $a$; the decaying component is eventually zero. The unit-circle factors of $P_{S,d}(2X)$ give $(X-1)M_S(X)^d$.

For sufficiency, apply the last row of Lemma~\ref{lem:phase-difference} with $L=M_S$. Comparing cyclotomic multiplicities gives the minimum depth and the intersection criterion; only finitely many orders occur in any minimal polynomial, so the finite-subset convention applies to infinite sets as well.
\end{proof}

\section{Mixed-width inclusion and determinant intersections}\label{subsec:width-order}
\begin{corollary}[Projective width witness]\label{ws:main}
For $k\ge2$ and an effective non-torsion phase $\beta$, the family $R(\beta,\mathbf1_{n\equiv1\pmod k})$ belongs to $\mathcal A_S$ exactly when $k\mid s$ for some $s\in S$. The same criterion holds for eventual projective realizability.
\end{corollary}
\begin{proof}
The periodic indicator has minimal rational annihilator $X^k-1$ (its Fourier coefficient at every $k$th root is nonzero). Apply Theorem~\ref{thm:rank-one-spectrum}, which also gives depth one in the positive case.
\end{proof}
\begin{theorem}[Exact divisibility inclusion classification]\label{thm:width-order}\label{cor:width-lattice}
For nonempty sets $S,T\subseteq\mathbb N_{>0}$, the following are equivalent:
\[
 \mathcal A_S\subseteq\mathcal A_T
 \quad\Longleftrightarrow\quad
 \forall s\in S\ \exists t\in T:\ s\mid t.
\]
The same condition characterizes inclusion up to arbitrary length-dependent global phases and eventual agreement. The exact translation does not increase depth, with no padding, approximation, extra initial function, or change of domain.
\end{theorem}
\begin{proof}
Necessity follows from Corollary~\ref{ws:main} for $s\ge2$; $1$ divides every width. For sufficiency induct on syntax. Only recursion needs translation. Let $F$ have width $s$, threshold $b\ge s$, earlier translated operands $g,h,p$ and active set $C$. Choose $t=ms\in T$. The finite binary branch tree $D_C(V)$ is multiplicative on positive suffixes, with $D_C^j(V)$ controlling on $C^j$ and $D_C^0(V)=V$.

At length $n>b+(m-1)s$, all of the first $m$ levels of the original recursion are above its base threshold. Its exact expansion is
\[
 F_n=\left(\prod_{j=0}^{m-1} D_C^j(h)_n\right)
 D_C^m(F)_n
 \left(\prod_{j=m-1}^{0} D_C^j(p)_n\right),
\]
The left factors are in increasing $j$ and the right factors in decreasing $j$; multiplicativity gives the expansion without commuting any factors, even if $p$ mixes branches. Set $B=b+(m-1)s\ge t$. Lemma~\ref{lem:width-table} supplies the rank-zero base $g'$ equal to $F_j$ for $1\le j\le B$ (these finite evaluations have effective entries). Use width $t$, threshold $B$, the two exterior products and active set $C^m$. Induction on length proves equality, since $n-t\ge1$ for $n>B$. The translated operands have no greater rank, so neither does the recursion.
\end{proof}

\begin{corollary}[Padded computational collapse versus exact width separation]\label{cor:padded-collapse}
For every nonempty $S$, the languages and binary string functions with prefix-free readout realized by $\mathcal A_S$ on the padding $\phi^p$ are exactly $\mathsf{BQP}$ and $\mathsf{FBQP}$, respectively. The same holds for $\mathcal A_S^{[d]}$ at every $d\ge10$. In particular, the choice of an individual permitted width does not change either padded class, although distinct single widths have distinct exact operator algebras.
\end{corollary}
\begin{proof}
Soundness is width-independent by Lemma~\ref{lem:circuit-expansion}. Theorem~\ref{thm:width-order} embeds width one at unchanged rank and padding, so Theorem~\ref{thm:oneqrec-fbqp} gives completeness at rank ten; the same width theorem gives strict structural distinction.
\end{proof}
This answers the padded decision and prefix-free output question in~\cite[Sec.~6.1]{Yamakami2020}, not unpadded equality, length-independent clean output or a coherent universal normal form.

\subsection{Initial-basis strength: the joint basis--width--rank order}\label{sec:finite-basis}
To vary initial-basis strength, use finite gate libraries; the arbitrary-angle length-table lemma no longer applies directly.
For $j\ge3$ let $K_j=\mathbb Q(\zeta_{2^j})$, $\gamma=(3+4i)/5$, and let $\mathcal B_j$ consist of identity, first-bit NOT and Hadamard, first-two-bit SWAP, and first-bit phase gates of values $\gamma,\zeta_{2^j}$ and their inverses. Fixed gates of arity two act trivially on one bit. Write $\mathcal A_{j,S}^{[d]}$ for the schematic algebra with this finite library, allowed widths $S$, and nesting at most $d$. It is a subalgebra of $\mathcal A_S^{[d]}$.

\begin{lemma}[Finite tables relative to the gate library]\label{lem:relative-table}
For any finite list of fixed lengths $1\le r\le B$ and matrices $V_r$ obtainable at length $r$ from terms over $\mathcal B_j$, the family equal to $V_r$ at those lengths and identity above $B$ has a recursion-free realization over the same library. The width grouping construction of Theorem~\ref{thm:width-order} therefore works without enlarging this library.
\end{lemma}
\begin{proof}
Localize each evaluation by the rank-zero unitary rule of Lemma~\ref{lem:guarded-eval} and compose. Width grouping uses only such fixed evaluations.
\end{proof}

\begin{theorem}[Joint initial-basis, width and nesting inclusion]\label{thm:joint-resource}
For $j,k\ge3$, nonempty width sets $S,T$ and finite $d,e\ge0$,
\[
 \mathcal A_{j,S}^{[d]}\subseteq\mathcal A_{k,T}^{[e]}
 \quad\Longleftrightarrow\quad
 j\le k,\quad d\le e,\quad
 \bigl(d=0\ \text{or}\ \downarrow S\subseteq\downarrow T\bigr).
\]
The same criterion holds for inclusion up to eventual projective agreement. Thus for positive depth the three parameters form an exact product order (identifying width sets with their divisibility downsets); depth zero has no width parameter. The unbounded-depth version omits $d\le e$ and requires the width condition. No arbitrary-angle gate or workspace padding is used in this classification.
\end{theorem}
\begin{proof}
For sufficiency, $\zeta_{2^j}=\zeta_{2^k}^{2^{k-j}}$ for $j\le k$, so replace each phase gate by a fixed power in the larger library, at depth zero. For positive depth, group each recursion into an available multiple width by Theorem~\ref{thm:width-order} and Lemma~\ref{lem:relative-table}. This does not increase depth. For $d=0$ no recursion is present.

For necessity, all entries of a target term lie in $K_k$: $i,\sqrt2\in K_k$, every initial gate has entries in $K_k$, and finite matrix operations preserve the field. The first-bit phase of value $\zeta_{2^j}$ has relative diagonal entry $\zeta_{2^j}$ at every positive length. Eventual projective agreement would place that ratio in $K_k$. Since $[K_j:\mathbb Q]=2^{j-1}$, this is impossible for $j>k$.

The trace of $\gamma$ over $\mathbb Q$ is $6/5$, so $\gamma$ is not an algebraic integer and has infinite order. Fixed-length rank-one phases $\gamma^a$ use an all-zero branch prefix and $X\operatorname{diag}(1,\gamma)^aX$, with relative tables. For each $s\in S$, recursion on the single active branch $0^s$ with identity exterior realizes $a_n=\mathbf1_{s\mid n}$ at depth one. Its minimal polynomial contains $\Phi_s$, so Theorem~\ref{thm:rank-one-spectrum} forces $s\mid t$ for some $t\in T$ if $d>0$. For $d>e$, take $a_n=n^{d-1}$; the rank-one row of Lemma~\ref{lem:phase-difference} uses only powers of $\gamma$ at depth $d$ since $(E^s-1)^d a=0$. Its minimal polynomial $(X-1)^d$ excludes projective depth $e$. The same width witnesses give the unbounded-depth criterion.
\end{proof}
The finite libraries are not identified with minimal SR bases, and the arbitrary-angle compiler need not be exact over them.

By Theorem~\ref{thm:width-order}, the width-generated poset is isomorphic to the nonempty divisibility downsets: $\mathcal A_S\subseteq\mathcal A_T$ iff $\downarrow S\subseteq\downarrow T$, and every nonempty downset generates itself. Their joins and meets \emph{within this poset} correspond to union and intersection. For singleton widths these are $\mathcal A_{\{s,t\}}$ and $\mathcal A_{\{\gcd(s,t)\}}$; the join need not be $\mathcal A_{\{\operatorname{lcm}(s,t)\}}$, as the period-six witness of Corollary~\ref{ws:main} for $s=2,t=3$ shows.

\begin{lemma}[Eventual recurrence kernels]\label{lem:recurrence-kernel}
Write $K_A(P)$ for the kernel of $P(E)$ on eventual additive sequences in an abelian group $A$. Let $P,Q$ be monic integer polynomials, $R=\gcd_{\mathbb Q[X]}(P,Q)$ monic, $P=RU$, $Q=RV$, and $\mathcal L=\mathbb Z[X,X^{-1}]$.
\begin{enumerate}
\item If $A$ is torsion-free, then $K_A(P)\cap K_A(Q)=K_A(R)$. For arbitrary $A$ the equality follows from $R\in(P,Q)\mathcal L$. For $A$ of exponent dividing $q$ it follows from $R\in(P,Q,q)\mathcal L$.
\item The first ideal condition holds iff $\gcd(\bar U,\bar V)$ is a power of $X$ at every prime; the second iff this test holds at primes dividing $q$.
\item If $h\ne X$ is monic irreducible over $\mathbb F_p$ and $h^k$ divides both $\bar P,\bar Q$ but not $\bar B$ for an integer polynomial $B$, there is a periodic sequence modulo $p$ killed by $P(E),Q(E)$ but not by $B(E)$. Conversely, if every nonzero root of $\gcd(\bar P,\bar Q)$ is a root of $\bar B$, each common kernel sequence modulo $p^m$ is killed by a power of $B(E)$.
\end{enumerate}
\end{lemma}
\begin{proof}
The reverse inclusion in (1) is immediate. Polynomial B\'ezout over $\mathbb Q$, after clearing denominators, gives a nonzero integer multiple of $R$ in $(P,Q)$; torsion-freeness allows its cancellation. In the other cases apply the stated ideal identity, using invertibility of $E$ on eventual sequences.

For (2), $U,V$ are coprime over $\mathbb Q$. The ideal $(U,V)$ contains the nonzero resultant $C$. At a prime $p$ where the test holds, lifting finite-field B\'ezout gives $X^j=DU+FV+pZ$. Thus $p$ is a unit in $\mathcal L/(U,V)$, and the same holds in its quotient modulo $q$. Testing all primes dividing $C$ (or all dividing $q$ in the second case) makes the quotient zero. Multiplication by $R$ gives the required ideal identity, without cancelling modulo $q$. Necessity follows from (3) with $B=R$ and $k$ one greater than the multiplicity of $h$ in $\bar R$: an ideal identity would contradict the witness, also modulo $q$ if $p\mid q$.

For (3), in $W=\mathbb F_p[X]/(h^k)$ choose a linear functional $\ell$ with $\ell(\bar B)\ne0$. Multiplication by $X$ is a permutation of $W$. The sequence $a_n=\ell(X^n)$ is periodic, is killed by $P(E),Q(E)$, and $B(E)a$ is nonzero whenever $X^n=1$ in $W$. For the converse a sufficiently large power of $\bar B$ belongs to $(\bar P,\bar Q)$ in the Laurent ring, which discards the factor $X$. Apply Lemma~\ref{lem:lift-annihilator}.
\end{proof}

For non-torsion $\beta$, let $\mathcal D_S$ denote the integer exponents of realizable determinant sequences $\beta^{a_n}$. Write $\Delta_S$ for the full determinant image (without a restriction to powers of one phase).
\begin{corollary}[Intersection of non-torsion determinant exponent spectra]\label{cor:det-meet}
For nonempty $S,T$,
\[
 \boxed{\mathcal D_S\cap\mathcal D_T=\mathcal D_{\downarrow S\cap\downarrow T}.}
\]
In particular, $\mathcal D_{\{s\}}\cap\mathcal D_{\{t\}}=\mathcal D_{\{\gcd(s,t)\}}$.
\end{corollary}
\begin{proof}
For finite $S,T$ and $J=\downarrow S\cap\downarrow T$, we first have $\gcd(L_S,L_T)=L_J$. Every common complex root $\alpha$ satisfies $\alpha^s=c$ and $\alpha^t=c'$ for some $s\in S,t\in T$ and positive integers $c\le2^s,c'\le2^t$. With $g=\gcd(s,t)$, B\'ezout on the exponents gives $\alpha^g=c^u (c')^v$ for integers $u,v$ with $us+vt=g$. This is a positive rational number, and it is an integer because its $(s/g)$th power is the integer $c$. It is at most $2^g$ by absolute values. Thus $\alpha$ is a root of an allowed factor $X^g-e$ in the common downset. Conversely each factor at a common divisor width divides an allowed factor at a multiple width, since $X^g-e$ divides $X^{kg}-e^k$ and $e^k\le2^{kg}$. The polynomials are squarefree, proving the gcd identity. In particular, for depths $d,e$ the characteristic-zero gcd of $P_{S,d},P_{T,e}$ is $P_{J,\min(d,e)}$. Apply Lemma~\ref{lem:recurrence-kernel}(1) to integer exponents. Corollary~\ref{thm:det-image} proves the forward inclusion, also at fixed depths $d,e$. For infinite sets use finite subsets of the original widths and then embed their common downset in $J$. The reverse follows from Theorem~\ref{thm:width-order}.
\end{proof}

\begin{theorem}[Exact finite-depth determinant intersection test]\label{thm:fixed-depth-meet}
For finite nonempty width sets $S,T$ and depths $d,e\ge0$, put $J=\downarrow S\cap\downarrow T$, $f=\min(d,e)$, and
\[
 P=P_{S,d},\qquad Q=P_{T,e},\qquad R=P_{J,f},\qquad A=P/R,\quad B=Q/R.
\]
The quotients are monic integer polynomials and coprime over $\mathbb Q$. Write $\Delta_{S,d}$ for the exact determinant image at depth at most $d$. The following are equivalent:
\begin{enumerate}
\item $\Delta_{S,d}\cap\Delta_{T,e}=\Delta_{J,f}$.
\item $R\in(P,Q)\mathbb Z[X,X^{-1}]$.
\item For every prime $p$, $\gcd(\bar A,\bar B)$ is a power of $X$ (possibly $1$) in $\mathbb F_p[X]$.
\end{enumerate}
Only primes dividing the nonzero resultant $\operatorname{Res}(A,B)$ need be tested. For determinant values restricted to $q$th roots of unity, $q\ge2$, replace (2) by membership in $(P,Q,q)\mathbb Z[X,X^{-1}]$ and in (3) test only primes dividing $q$. Every failure admits a periodic witness taking values in the group of $p$th roots for a single obstructing prime. The expected intersection always holds at these depths on any fixed finitely generated torsion-free phase group.
\end{theorem}
\begin{proof}
The preceding gcd identity gives $R=\gcd(P,Q)$. By Theorem~\ref{thm:det-full-depth}, determinant images are the corresponding eventual recurrence kernels in the effective phase group (written multiplicatively). Lemma~\ref{lem:recurrence-kernel}(1)--(2) gives sufficiency of the ideal and prime tests, including the torsion-free and order-$q$ versions. Only primes dividing $\operatorname{Res}(A,B)$ can obstruct the unrestricted test. If the prime test fails at $p$, Lemma~\ref{lem:recurrence-kernel}(3) gives a periodic exponent sequence killed by $\bar P,\bar Q$ but not by $\bar R$; exponentiation with $\zeta_p$ gives a common determinant outside $\Delta_{J,f}$. It belongs to the order-$q$ group when $p\mid q$. Thus the prime test is necessary, completing both equivalences.
\end{proof}

\begin{corollary}[Fixed-depth and unrestricted determinant meet obstructions]
\qsrpart{a}{cor:fixed-depth-torsion}
The sign determinant $\delta_n=(-1)^n$ is in $\Delta_{\{2\},1}\cap\Delta_{\{3\},1}$ but not $\Delta_{\{1\},1}$. Its width-one minimum determinant depth is two. Thus the unrestricted two-primary intersection theorem below does not extend to fixed depth.
\qsrpart{b}{cor:torsion-meet-fail}
One has
\[
 \Delta_{\{1\}}\subsetneq\Delta_{\{2\}}\cap\Delta_{\{3\}}.
\]
In fact $\delta_n=\zeta_5^{3^n}$ lies in the latter intersection, not in the former, even allowing finitely many determinant disagreements.
\end{corollary}
\begin{proof}
\par\noindent\textup{(a)} The exponent $a_n=n\bmod2$ has minimal eventual polynomial $(X-1)^2$ over $\mathbb F_2$. For width two the polynomial $L_{\{2\}}$ has reduction $X^4(X-1)^4$, while width three has reduction $X^{12}(X^3-1)^4$. Both have the required multiplicity at depth one. At width one, $P_{\{1\},d}$ reduces to $X^{d+1}(X-1)^d$, giving exact minimum $d=2$ by Corollary~\ref{cor:torsion-depth}.
\par\noindent\textup{(b)} In $\mathbb F_5$, $3^2=4$ and $3^3=2$. For width two take four active branches and for width three take two active branches; in either case use identity exterior operators and finite base determinants $\delta_j$ for $1\le j\le s$. Their determinant recurrences give exactly $\delta_n$ on all positive lengths. For width one, $L_{\{1\}}=(X-1)(X-2)$; no power of this polynomial, even with the extra $X-2$ factor, vanishes at $3$ in $\mathbb F_5$. Corollary~\ref{cor:torsion-depth} excludes the sequence at every depth. Inclusion of $\Delta_{\{1\}}$ follows from Theorem~\ref{thm:width-order}.
\end{proof}

\begin{theorem}[Complete criterion for unrestricted determinant-image meets]\label{thm:det-meet-primes}
For nonempty width sets $S,T$, put $J=\downarrow S\cap\downarrow T$ and, for each prime $p$, define
\[
 R_p(S)=\{\alpha\in\overline{\mathbb F}_p^\times:\exists s\in S,\ 1\le c\le 2^s,\ \alpha^s=\bar c\}.
\]
Then
\[
 \Delta_S\cap\Delta_T=\Delta_J
 \quad\Longleftrightarrow\quad
 R_p(S)\cap R_p(T)=R_p(J)\ \text{for every prime }p.
\]
For finite $S,T$ the condition is effectively decidable by finitely many polynomial calculations over finite fields. In particular, for all $S,T$ the meet identity holds on determinant sequences whose values have orders that are powers of two; the same holds for sequences taking values in a finitely generated group with no odd-primary torsion.
\end{theorem}
\begin{proof}
All root sets are monotone under the divisibility downset operation: if $a\mid s$ and $\alpha^a=\bar c$ for $1\le c\le2^a$, then $\alpha^s=\overline{c^{s/a}}$, with $c^{s/a}\le2^s$. Thus the inclusion from right to left in the root-set equality is automatic.

If the root-set condition fails at $p$, the irreducible polynomial $h\ne X$ of a root in the difference divides reductions of allowed factors for $S,T$, but none for $J$. The witness in Lemma~\ref{lem:recurrence-kernel}(3) with $k=1$ may be chosen as a nonzero sequence with minimal annihilator $h$ (in the field $\mathbb F_p[X]/(h)$, any nonzero linear functional has this property). Its $p$th-root phase is realized over $S,T$ but no allowed annihilator for $J$ kills it. This proves necessity.

For sufficiency take a common determinant sequence $\delta$. Theorem~\ref{thm:det-full-depth} gives monic annihilators $P,Q$ using finite subsets of $S,T$. Monicity implies that every value on a late tail is in the group generated by finitely many initial values, and adding the finitely many exceptional values gives a finitely generated abelian subgroup $G$ of $\mathbb T_{\rm eff}$ containing the sequence. Use a decomposition of $G$ into a free abelian group and finite cyclic prime-power groups. The generators of these components are elements of $G$, hence admissible phases. In the free coordinates the multiplicative annihilators become ordinary integer recurrences. The characteristic-zero gcd proof of Corollary~\ref{cor:det-meet} provides an allowed annihilator for $J$, and Theorem~\ref{thm:det-full-depth} realizes each component.

On a cyclic component of order $p^e$, represent the sequence by exponents $a_n$ modulo $p^e$. Both $P(E)a$ and $Q(E)a$ vanish on a tail. Put $g=\gcd(\bar P,\bar Q)$ in $\mathbb F_p[X]$. Each nonzero root of $g$ is in $R_p(S)\cap R_p(T)$ (the initial factor $X-2$ causes no extra nonzero root, since it divides each $L_S$). By hypothesis every such root belongs to $R_p(J)$. Because only finitely many irreducibles occur in $g$, choose finitely many allowed factors for $J$ whose product $B$ has reductions containing every nonzero root of $g$. The product may be empty. Lemma~\ref{lem:recurrence-kernel}(3) gives a power of $B(E)$ annihilating $a$. Theorem~\ref{thm:det-full-depth} realizes this component over $J$; compose the finitely many components. For finite $S,T$, take $L_S,L_T$ and their monic characteristic-zero gcd $G_0$. Each root of $G_0$ is in an allowed factor for $J$ by the argument of Corollary~\ref{cor:det-meet}; since the polynomials are squarefree, $G_0$ divides $L_J$ (here $J$ is finite). There is an integer $C_0>0$ and integer polynomials $A,B$ such that $A L_S+B L_T=C_0G_0$, by clearing denominators in a rational B\'ezout identity. For $p\nmid C_0$, every common root of $\overline{L_S},\overline{L_T}$ is a root of $\overline{G_0}$ and hence of $\overline{L_J}$. Only prime divisors of $C_0$ can be exceptional. For each of these primes, factor the reduced polynomials over $\mathbb F_p$ (or compare their squarefree irreducible supports), discarding $X$; the root-set condition is then finite and decidable.

Finally, for $p=2$ every nonzero root of an allowed factor satisfies $\alpha^s=1$. If it is a common root for widths $s,t$, its multiplicative order divides $\gcd(s,t)$, so it lies in $R_2(J)$. Thus the cyclic-component argument works unconditionally for every $2$-power order. A sequence of $2$-power roots with a common monic recurrence takes values in a finitely generated group, so their orders are bounded on the whole sequence. The same proof covers a free part and $2$-primary torsion together. This establishes the final assertions.
\end{proof}

\section{Finite-order rank-one membership spectra}\label{subsec:two-primary-rank-one}
The following arguments distinguish the $2$-primary sector, where a determinant obstruction survives global phase, from the odd-order sector, where it can be removed.
\begin{lemma}[Projective detection of $2$-primary determinant phases]\label{lem:two-primary-projective}
Let $K$ be a finitely generated subfield of $\mathbb C$. Its subgroup $\mu_{2^\infty}(K)$ of roots of unity of $2$-power order has bounded exponent. For a sufficiently large fixed $N$, the quotient map
\[
 K^\times\longrightarrow K^\times/(K^\times)^{2^N}
\]
is injective on $\mu_{2^\infty}(K)$. If a depth-$d$ term over finite widths $S$ satisfies $G_n=\lambda_n U_n$ eventually, where $\det U_n=\zeta_{2^m}^{a_n}$ with $a_n\in\mathbb Z/2^m\mathbb Z$, and $\lambda_n\in K$ on that tail (with all gate constants and $\zeta_{2^m}$ also in $K$), then
\[
 \bigl((X-2)L_S(X)^d\bigr)(E)a=0\quad\text{eventually in }\mathbb Z/2^m\mathbb Z.
\]

\end{lemma}
\begin{proof}
Choose a transcendence basis $t$ of $K/\mathbb Q$, with $[K:\mathbb Q(t)]=D$ finite. For a root of unity $\xi\in K$, the degree $[\mathbb Q(\xi):\mathbb Q]$ is at most $D$: adjoining the algebraically independent $t$ to $\mathbb Q(\xi)$ does not change the finite extension degree over $\mathbb Q(t)$. The degrees of primitive $2^k$th roots are $2^{k-1}$, giving the bounded exponent. Choose $N$ at least its exponent in powers of two. If $x^{2^N}\in\mu_{2^\infty}(K)$, then $x$ itself is a root of unity of $2$-power order and $x^{2^N}=1$. Thus the intersection of the subgroup with the power subgroup is trivial.

For $n\ge N$, the determinant identity gives $[\det G_n]=[\det U_n]$ in the quotient, regardless of the global phases. The determinant necessity in Theorem~\ref{thm:det-full-depth} is an identity in an abelian multiplicative group, so passes to this quotient without division. Apply it on a tail. Since $\zeta_{2^m}$ injects, the exponent recurrence follows.
\end{proof}

\begin{theorem}[Exact and projective rank-one $2$-primary membership]\label{thm:two-primary-rank-one}
Fix $m\ge1$, $\beta=\zeta_{2^m}$ and a sequence $a_n\in\mathbb Z/2^m\mathbb Z$. For nonempty $S$, the following are equivalent:
\begin{enumerate}
\item $R(\beta,a)$ is exactly realizable in $\mathcal A_S$ at every positive length.
\item $R(\beta,a)$ has an eventual projective realization in $\mathcal A_S$.
\item The sequence $a$ is eventually annihilated modulo $2^m$ by a finite product of the polynomials $X^s-1$, $s\in S$ (allowing repetitions).
\item The determinant sequence $\beta^{a_n}$ is realizable by a term in $\mathcal A_S$ at every positive length.
\end{enumerate}
For a realization of depth at most $d$ in (2), the sharper necessary condition of Lemma~\ref{lem:two-primary-projective} holds. A finite product in (3) with $r$ factors gives exact depth at most $r$. More quantitatively, for finite $S$ let
\[
 k_S=\max_{h\ne X}\left\lceil
 \frac{v_h(\overline{L_S})}{v_h(\overline{M_S})}\right\rceil,
\]
where $h$ ranges over the nonzero-root monic irreducible factors of $\overline{L_S}$ in $\mathbb F_2[X]$, and $v_h$ is factor multiplicity. These factors all occur in $\overline{M_S}$. For a realizable sequence, its minimum determinant, eventual projective rank-one, and exact rank-one depths satisfy
\[
 d_{\det}\le d_{\rm proj}\le d_{\rm ex}\le m k_S d_{\det}.
\]
For $S=\{s\}$, $k_S=2^{s-1}$. These are upper and lower bounds, not a general equality of depth minima. For infinite $S$ choose a finite subset realizing the determinant at its minimum depth.
\end{theorem}
\begin{proof}
The implication (1) to (2) is immediate. In (2), the diagonal entry at $1^n$ is $\lambda_n$, so it lies in the finitely generated field of the term constants and $\beta$. Lemma~\ref{lem:two-primary-projective} supplies an eventual exponent annihilator $P=(X-2)L_{S_0}^d$ modulo $2^m$, for a finite nonempty subset $S_0$ of $S$ (take $S_0=S$ when $S$ is finite). In (4) the same annihilator is supplied directly by Theorem~\ref{thm:det-full-depth}. We show that such an annihilator implies (3).

Modulo two, $X^s-c$ reduces to $X^s$ or $X^s-1$, and $X-2$ reduces to $X$. Consequently the nonzero-root factors of $\bar P$ occur in $\overline{M_{S_0}}$. By the definition of $k_{S_0}$, $\overline{M_{S_0}}^{\,k_{S_0}d}$ belongs to $(\bar P)$ in the Laurent ring. Lemma~\ref{lem:lift-annihilator} yields $M_{S_0}(E)^{m k_{S_0} d}a=0$. This polynomial divides a product as in (3). The rank-one row of Lemma~\ref{lem:phase-difference} gives depth at most $m k_{S_0}d$; its product version proves (3) implies (1). Finally (1) implies (4) by the determinant, and Lemma~\ref{lem:two-primary-projective} gives the lower bound on projective depth. For one width the factors are pairwise coprime in characteristic zero and
\[
 \overline{L_{\{s\}}}=X^{s2^{s-1}}(X^s-1)^{2^{s-1}},
\]
so $k_{\{s\}}=2^{s-1}$.
\end{proof}

\begin{corollary}[Two-primary depth bounds, meets and primary splitting]
\qsrpart{a}{cor:additive-two-primary}
For one width $s$, phase order $2^m$, and finite determinant minimum $d\ge1$,
\[
 d_{\mathrm{ex}}\le\min\{m2^{s-1}d,\ 2^{s-1}d+m-1\}.
\]
More generally for finite nonempty $S$, replace $2^{s-1}d$ in the second bound by $r=d\sum_{s\in S}2^{s-1}$ and take the minimum with the earlier bound $mk_Sd$. The additive bound can be sharp for the single-active difference method: at width two and determinant depth one the exponent $a_n=(3^{\lfloor(n-1)/2\rfloor}-1)/2\pmod{2^m}$ requires exactly $m+1$ factors $E^2-1$ to annihilate it. This last assertion is about that synthesis method, not a lower bound of $m+1$ on arbitrary quantum terms.
\qsrpart{b}{cor:two-primary-rank-one-meet}
For fixed $\beta=\zeta_{2^m}$, let $\mathcal R_S(\beta)$ denote the exactly realizable rank-one families with exponents in $\mathbb Z/2^m\mathbb Z$. This class is unchanged if eventual projective realization is used. For all nonempty $S,T$,
\[
 \mathcal R_S(\beta)\cap\mathcal R_T(\beta)=\mathcal R_{\downarrow S\cap\downarrow T}(\beta).
\]
Thus the full rank-one gcd identity holds for this sector too, even projectively and for arbitrary mixed-width sets.
\qsrpart{c}{cor:finite-order-split}
If $\beta$ has finite order $2^m q$ with $q$ odd, write $1=u2^m+v q$ with integers $u,v$. The operator $R(\beta,a)$ belongs to a width algebra exactly when both factors $R(\beta^{u2^m},a)$ and $R(\beta^{vq},a)$ do. The same equivalence holds for eventual projective membership. In both cases the maximum of the two component minimum depths is the minimum depth of the original operator. The trivial order-one component is the identity. Thus only odd-order phases can obstruct the unrestricted-depth rank-one projective meet identity, since the $2$-primary component has the exact meet theorem above.
\end{corollary}
\begin{proof}
\par\noindent\textup{(a)} Work in the module of eventual sequences modulo $2^m$, on which the shift $E$ is invertible. Every even-count factor $E^s-c$ is invertible, by a finite geometric series for $1-cE^{-s}$; the factor $E-2$ is likewise invertible. The determinant polynomial divides the product of all indicated factors with their depth multiplicities, so the remaining product of the $r$ odd-count factors $F_i=X^{s_i}-c_i$ annihilates $a$. Write $M=M_S$. For each $i$, since $M$ is a multiple of $X^{s_i}-1$, we have $M=A_iF_i+2B_i$ over $\mathbb Z[X]$. In the quotient by $(F_1\cdots F_r,2^m)$, expand $M^{r+m-1}$ as follows: at each multiplication use the identity for the first factor index not yet collected in that summand. The first term collects that $F_i$; the second contributes a factor two. After $r+m-1$ multiplications every summand has either collected all $r$ factors or at least $m$ factors of two. Both vanish in the quotient. Thus $M(E)^{r+m-1}a=0$. Lemma~\ref{lem:phase-difference} realizes the rank-one family at depth at most $r+m-1$. For a singleton $r=2^{s-1}d$; the earlier bound is Theorem~\ref{thm:two-primary-rank-one}.

For the stated example put $k=\lfloor(n-1)/2\rfloor$. Its integer lift satisfies $(E^2-1)a=3^k$ and $(E^2-3)3^k=0$. The determinant criterion therefore gives depth one (depth zero has zero tail modulo $2^m$). Induction gives $(E^2-1)^h a=2^{h-1}3^k$ for every $h\ge1$. Since $3^k$ is odd, this vanishes modulo $2^m$ exactly when $h\ge m+1$. The upper bound is attained by this recurrence method. Whether a different raw quantum construction attains the determinant lower bound remains open.
\par\noindent\textup{(b)} Combine the determinant equivalence of Theorem~\ref{thm:two-primary-rank-one} with the two-primary meet in Theorem~\ref{thm:det-meet-primes}.
\par\noindent\textup{(c)} Fixed integer powers preserve rank and projective agreement. Conversely the product of the components recovers the original operator by B\'ezout\'s identity and has rank at most the maximum. Apply the equivalence to each width set and their common downset.
\end{proof}

\subsection{Native synthesis of odd-order projective phases}\label{subsec:odd-periodic}
We construct length and suffix-rank selectors using existing bits and depth-one terms.
\begin{lemma}[Two-level selection by commutators]\label{lem:two-level-selection}
Consider a family of orthogonal blocks with a designated two-dimensional subspace in each. Assume each fixed algebraic $SU(2)$ gate is realizable at depth zero simultaneously on all designated pairs, with identity on their orthogonal complement. Suppose a finite list of realizable unitaries $A_i$ preserves the blocks and these subspaces. On a desired set of blocks their two-dimensional restrictions are fixed nonscalar algebraic matrices (one for each $i$); on every other block at least one restriction is a multiple of identity. Then constant algebraic two-level gates, the $A_i$ and their inverses generate any prescribed algebraic $SU(2)$ gate on the desired subspaces, identity elsewhere. This uses only finite composition and inverses and has depth at most the maximum of zero and the depths of the $A_i$.
\end{lemma}
\begin{proof}
Start with a constant nonscalar algebraic $B\in SU(2)$ on the designated pair, identity on the complement. For each $i$ replace the current gate by $[B,CA_iC^{-1}]$, where $[U,V]=UVU^{-1}V^{-1}$ and $C$ is a constant algebraic special unitary on the pair. It is identity on the complement and on all blocks already eliminated or where $A_i$ is a multiple of identity. On the desired blocks, align the eigenbases of the two restrictions, then tilt the second by a sufficiently small nonzero algebraic rotation. The resulting commutator is not identity (nonscalar two-dimensional unitaries commute only when their eigenbases agree up to permutation and phases), and is close enough to identity not to be $-I$. Thus it is nonscalar and fixed on those blocks. Algebraic eigenbases and tilts exist by algebraic normalization and the rational parametrization of the circle.

For the remaining synthesis, conjugate the final noncentral value of $B$ to $D=\operatorname{diag}(\lambda,\lambda^{-1})$, $\lambda=e^{i\alpha}$, $0<\alpha<\pi$ (the conjugating gate may be chosen algebraic and special unitary). Write
\[
 C_t=\begin{pmatrix}\sqrt{1-t}&-\sqrt t\\\sqrt t&\sqrt{1-t}\end{pmatrix},\qquad 0\le t\le1.
\]
Direct multiplication gives $\tfrac12\operatorname{tr}[D,C_t]=1-2t\sin^2\alpha$. For any algebraic target $T\in SU(2)$, take a fixed sufficiently large $h$ and an algebraic $h$th root $T_0$ with eigenangle $\theta\in[0,2\min(\alpha,\pi-\alpha)]$. Then $t=(1-\cos\theta)/(2\sin^2\alpha)$ is algebraic in $[0,1]$, and the commutator has the same trace as $T_0$. Its two eigenvalues are the roots of the same characteristic polynomial as those of $T_0$; the two normal matrices are conjugate by a special unitary with algebraic entries, using normalized algebraic eigenvectors and adjusting one column phase to determinant one. If the target root is identity take $t=0$. Conjugating and taking the $h$th power gives $T$ on the selected pair and identity elsewhere. All constant conjugations are on the designated subspaces, so they preserve eliminated blocks. An empty list gives the constant target directly.
\end{proof}

\begin{lemma}[Periodic local special unitaries]\label{lem:periodic-local}
Let $k,b\ge1$ be fixed, and let $V_0,\ldots,V_{k-1}$ be algebraic matrices in $SU(2^b)$. There is a width-one term of depth at most one acting at every length $n\ge b$ as $V_{n\bmod k}\otimes I_{2^{n-b}}$. Arbitrary prescribed algebraic unitary matrices at shorter lengths can be supplied without increasing depth.
\end{lemma}
\begin{proof}
First take $b=1$. For $k=1$ use a fixed gate. Otherwise let $D=\operatorname{diag}(1,\exp(2\pi i/k))$. The term $A_n=D^n\otimes I_{2^{n-1}}$ is native at depth one: use base $D$ at length one, both recursive branches active, pre-operator SWAP and post-operator $D\,\mathrm{SWAP}$ (the $D$ here acts on the first bit). The recursive call acts on the original first bit after SWAP, and the second SWAP restores it. Induction proves the displayed action.

For target residue $t$, the gates $AD^{-r}$, $r\ne t$, are identity at residue $r$ and fixed nonscalar matrices at $t$. Lemma~\ref{lem:two-level-selection} gives any algebraic special unitary supported at $t$. Compose over residues to obtain the prescribed schedule.

For $b>1$ decompose each $V_t$ into a finite product of special-unitary two-level gates by algebraic Givens elimination (the final determinant-one diagonal is a product of diagonal two-level special unitaries). Concatenate these finitely many decompositions, activating each constituent only at its assigned residue. Each two-level gate is implemented by a fixed permutation of the first $b$ bits, a binary branch tree controlling a one-bit periodic $SU(2)$ gate, and the inverse permutation. The controlled gate sees suffix length $n-b+1$, so shift its periodic schedule by $b-1$. This acts on all suffixes, not on a clean-workspace sector. The permutations are fixed-block gates as in Lemma~\ref{lem:width-table}; branching and composition do not increase recursion depth. Correct the shorter lengths by a finite table.
\end{proof}

\begin{lemma}[Borrowed-prefix modular selector at depth one]\label{lem:borrowed-prefix}
Fix an odd integer $q\ge3$, put $\omega=\exp(2\pi i/q)$, and choose $b$ with $2^b>q$. At lengths $n\ge b$ write the register as a prefix of $b$ bits followed by a suffix $y$, whose little-endian rank is $r(y)$ (rank zero for an empty suffix). For any fixed $k\ge1$ and algebraic special unitaries $V_{t,j}\in SU(2^b)$ indexed by $t\in\mathbb Z/k\mathbb Z$, $j\in\mathbb Z/q\mathbb Z$, there is a width-one depth-one term preserving the suffix and acting on its prefix by $V_{n\bmod k,r(y)\bmod q}$. Arbitrary fixed algebraic unitary matrices at the shorter positive lengths can be supplied by a finite table.
\end{lemma}
\begin{proof}
Label the $b$-bit prefix codes $0,\ldots,2^b-1$. For $x\in\{0,1\}$ define the fixed permutation $A_x$ on these codes by $u\mapsto 2u+x\pmod q$ for $u<q$ and by identity for $u\ge q$. This is a permutation since $q$ is odd. On a register of length $n>b$ write the input as $(u,x,z)$, with $x$ the first suffix bit. A width-one recursion with threshold $b$, identity base, both branches active, pre-operator moving $x$ to the front of the $b+1$ bits, and post-operator moving that bit back then applying $A_x$ to the prefix, gives
\[
 F_n|u,y\rangle=|f_y(u),y\rangle,\qquad
 f_y(u)=\begin{cases}2^{n-b}u+r(y)\pmod q,&u<q,\\u,&u\ge q.\end{cases}
\]
Indeed recursion first processes the remaining suffix, then applies $A_x$, so the digits enter in reverse order and produce the little-endian rank. The pre- and post-operators are fixed gates on $b+1$ bits, independent of length; the suffix recursion is on a positive length containing the entire prefix state. This is one recursion level. Its base is identity at $n=b$ and all shorter lengths. The equality holds for every initial prefix, not just $u=0$.

Let $D$ be the fixed prefix diagonal matrix $D_{uu}=\omega^u$ for $u<q$ and $D_{uu}=1$ otherwise. It has determinant one, because $q$ is odd. The schedule $D^{2^{n-b}}$ is periodic and special unitary, hence available at depth one by Lemma~\ref{lem:periodic-local}. The composition
\[
 B=F^{-1}(D\otimes I)F(D^{-2^{n-b}}\otimes I)
\]
therefore acts in each suffix block as $\omega^{r(y)}I_q\oplus I_{2^b-q}$. Its determinant in that block is one. In particular the operator retains all invalid prefix codes. All factors have depth at most one.

Choose a prefix pair $u<q$, $v\ge q$. The gates $B(\omega^{-j_1}I_q\oplus I)$ preserve the pair and restrict to $\operatorname{diag}(\omega^{j-j_1},1)$ in a suffix block of residue $j$. For target $(t_0,j_0)$, take those with $j_1\ne j_0$ together with a fixed nonscalar two-level gate selected at length residue $t_0$ by Lemma~\ref{lem:periodic-local}. Lemma~\ref{lem:two-level-selection} gives any algebraic $SU(2)$ action on this pair at the desired residue pair and identity elsewhere. Conjugate by prefix permutations to select other pairs; algebraic two-level $SU(2)$ gates generate $SU(2^b)$ by Givens elimination. Compose the prescribed factors and correct shorter lengths by finite tables. No operation increases depth beyond one or uses an initialized prefix.
\end{proof}

\begin{lemma}[Modular adjacent-pair swaps]\label{lem:rank-pair-swap}
For odd $q\ge3$, let $T_j$ exchange each adjacent pair $(2y,2y+1)$ of full register ranks with multiplier $i$ when $y\equiv j\pmod q$, and fix all other ranks. Each $T_j$ is a width-one depth-one term.
\end{lemma}
\begin{proof}
Use the prefix length $b$ of Lemma~\ref{lem:borrowed-prefix}, label its codes in little-endian order, and let $v=r(y)\bmod q$ be the rank residue of the suffix after the first $b$ bits. The full basis rank is $u+2^b r(y)$. For each even prefix code $u$, the adjacent pair $(u,u+1)$ has rank pair $(2h,2h+1)$ with $h\equiv u/2+2^{b-1}v\pmod q$. In each suffix block apply $iX$ on the disjoint pairs for which this expression equals $j$, and identity on the remaining pairs. This is an algebraic $SU(2^b)$ matrix, since every $iX$ block has determinant one. Lemma~\ref{lem:borrowed-prefix} realizes the finite family indexed by $v$ at depth one. At $n<b$ the desired finitely many matrices are supplied directly. Thus the modular rank selection is exact on every basis state, with no workspace.
\end{proof}

\begin{lemma}[Cyclic concentration]\label{lem:cyclic-concentration}
Fix odd $q\ge3$, $\omega=\zeta_q$, and any $a_n\in\mathbb Z/q\mathbb Z$. Suppose the local first-bit gate with exponents $\pm a_n$ has a width-one realization of depth $d$. Then the normalized rank-one family
\[
 G_n=\omega^{-2^{-n}a_n}R(\omega,a)_n
\]
has a width-one realization of depth at most $\max(1,d)$.
\end{lemma}
\begin{proof}
Put $b_n=-2^{-n}\pmod q$, a periodic sequence. For each value $a$, multiply the given local gate by the fixed gate with exponents $\mp j$ for each $j\ne a$. At $a_n=j$ the product is identity, at $a_n=a$ it is nonscalar since $q$ is odd. Include a periodic noncentral gate supported where $b_n$ has the desired value. Lemma~\ref{lem:two-level-selection}, followed by composition over value pairs, gives the local gate $C_n$ with exponents $\pm a_nb_n/2$ at depth at most $\max(1,d)$.

Let $E_j=CT_jC^{-1}T_j^{-1}$, using Lemma~\ref{lem:rank-pair-swap}. On a selected pair it gives exponents $(a_nb_n,-a_nb_n)$, and elsewhere identity. With the cyclic increment $P$ defined in Section~\ref{sec:determinants}, form the fixed product
\[
 G=\prod_{j=0}^{q-1} E_j^{2j}(PE_jP^{-1})^{2j+1}.
\]
It assigns to each cyclic edge $r\to r+1\pmod{2^n}$ the exponent $\alpha_r=a_nb_nr$ at the source and $-\alpha_r$ at the destination. The sum at $r>0$ is $\alpha_r-\alpha_{r-1}=a_nb_n$, and at zero it is $a_nb_n(1-2^n)=a_n(b_n+1)$. Thus the product is the required diagonal operator on the entire basis. Its determinant is one and the depth bound follows from the depth-one pair swaps and increment.
\end{proof}
\subsection{Transfer invariant at construction rank one}\label{subsec:nonperiodic-projective}
\begin{theorem}[Finite transfer representation at depth one]\label{thm:depth-one-transfer}
Let $U$ be a measurement-free term of rank at most one, using any finite selection of widths and arbitrary fixed gate constants. There are a finite dimension $D$, a row $\ell$, a column $v$, and four $D\times D$ matrices $M_{ab}$ ($a,b\in\{0,1\}$), over the field generated by the gate constants, such that for binary words $x,y$ of equal length $n\ge1$,
\[
 \langle x|U_n|y\rangle=\ell M_{x_1y_1}\cdots M_{x_ny_n}v.
\]
In particular the operator Schmidt rank (minimum number of tensor-product summands) across every contiguous cut of the register is bounded by $D$, independently of $n$. This bound persists up to arbitrary length-dependent global phases.
\end{theorem}
\begin{proof}
Use weighted automata~\cite{CrosswhiteBacon2008,Droste2009}. A rank-zero term is eventually a fixed prefix gate tensored with identity (induction on composition and branching). Store a bounded prefix of the paired word, weight it by the gate entry, then check equality of remaining pairs; a finite table handles short exceptions. Weighted automata on $\{0,1\}^2$ give the stated row--matrix--column form by summing parallel weights.

For a width-$s$ recursion $F=hD_C(F)p$ with depth-zero exterior and base terms, choose $B>s$ at least the exterior prefix supports and strictly larger than the base threshold. At all lengths $n\ge B$, its matrix entry is the sum over $u,v\in\{0,1\}^B$ of
\[
 h_{x_{1:B},u}p_{v,y_{1:B}}\,\mathbf1_{u_{1:s}=v_{1:s}}
 \begin{cases}
 \langle u_{s+1:B}x_{B+1:n}|F_{n-s}|v_{s+1:B}y_{B+1:n}\rangle,&u_{1:s}\in C,\\
 \mathbf1_{u_{s+1:B}x_{B+1:n}=v_{s+1:B}y_{B+1:n}},&u_{1:s}\notin C.
 \end{cases}
\]
Extend the exterior gates to $B$ bits. An automaton buffers at most $B$ pairs; when full, apply the displayed weighted transition. Active transitions retain the new $B-s\ge1$ pairs; inactive ones check equality of residual and subsequent pairs. At end use the exact short-length table of $F$ (or equality acceptance). Fold each transition into the symbol filling the buffer. Induction on remaining length proves the representation, including short lengths.

For composition use matrices $\sum_{z=0}^1M^U_{az}\otimes M^V_{zb}$ and tensor boundary vectors (sum over the intermediate word). Branching checks first-bit agreement and runs the selected suffix automaton, with a short table. Tables and sums use direct sums. Induction on syntax completes the construction.

Across a cut after $k$ bits, insert the $D$ coordinate states between the two products of transfer matrices. Each summand separates the paired indices of the first $k$ bits from those of the last $n-k$ bits; equivalently the operator is a sum of at most $D$ tensor products. Multiplication by a nonzero scalar does not change this rank.
\end{proof}

\begin{corollary}[Finite-valued relative phases are regular at depth one]\label{cor:depth-one-periodic}
If a depth-one term agrees projectively on a tail with $R(\omega,a)$, for a fixed primitive $q$th root $\omega$ of any finite order, then $a_n\in\mathbb Z/q\mathbb Z$ is eventually periodic. The conclusion holds for all allowed widths and arbitrary gate constants. More generally, if a depth-one term is diagonal and its diagonal entries relative to the all-zero entry take values in a fixed finite set at all lengths, the words attaining each value form a regular language.
\end{corollary}
\begin{proof}
In the rank-one case put $u_n=\langle0^n|U_n|0^n\rangle$ and $w_n=\langle1^n|U_n|1^n\rangle$. Both have modulus one on the tail, and $u_n\overline{w_n}=\omega^{a_n}$. By Theorem~\ref{thm:depth-one-transfer} this product is a matrix coefficient of powers of $M_{00}\otimes\overline{M_{11}}$. Cayley--Hamilton gives a fixed monic linear recurrence. Its tail values lie in a finite set, hence the vector of consecutive values of length the recurrence order evolves deterministically in a finite set. It is eventually periodic; injectivity of $a\mapsto\omega^a$ proves the claim.

For the more general statement the relative diagonal value on $x$ is $\langle x|U_n|x\rangle\overline{\langle0^n|U_n|0^n\rangle}$. The transfer matrices are $M_{aa}\otimes\overline{M_{00}}$. Any finite-valued recognizable word series $f(x)=\ell M_x v$ has finitely many distinct residual functions $f_x(y)=f(xy)$: choose finitely many suffixes $y_1,\ldots,y_r$ such that $M_{y_i}v$ span all $M_yv$. The tuple $(f(xy_i))_{i=1}^r$ determines the residual and has only finitely many possible values. These residuals are states of a deterministic finite automaton, with transitions $f_x\mapsto f_{xa}$ and outputs $f_x(\epsilon)$. The empty word can be assigned an arbitrary value by a finite correction. This proves regularity of every fiber.
\end{proof}

\begin{theorem}[Periodic odd-order projective phases]\label{thm:odd-depth-one-sharp}
For odd $q\ge3$, eventual projective realization of $R(\zeta_q,a)$ at depth at most one in any nonempty width algebra is equivalent to eventual periodicity of $a$. In the positive case the normalized family is realized at all lengths at depth at most one. Minimum projective depth is zero exactly for eventually zero $a$, and one for every other eventually periodic $a$.
\end{theorem}
\begin{proof}
Necessity is Corollary~\ref{cor:depth-one-periodic}. For sufficiency Lemma~\ref{lem:periodic-local} realizes the balanced local gate with exponents $\pm a_n$ at depth one on the periodic tail, and a finite table supplies earlier values. Apply Lemma~\ref{lem:cyclic-concentration} and width grouping, which preserves depth. A depth-zero term is eventually $V\otimes I$ for a fixed prefix gate $V$; compare its entries at $0^n$ and the word differing only in the last bit to see that a rank-one projective realization forces $a_n=0$ eventually. Conversely finite exceptions are realized by a depth-zero table.
\end{proof}

\begin{corollary}[Exact and approximate reversal]
\qsrpart{a}{cor:reversal-depth}
Reversal of all register wires has minimum recursion depth two in every nonempty width algebra, even projectively on a tail.
\qsrpart{b}{cor:robust-reversal}
For any fixed rank-at-most-one unitary term $U$, let $D$ be a transfer dimension as in Theorem~\ref{thm:depth-one-transfer}, and let $W_{2m}$ reverse $2m$ wires. With the normalized Hilbert--Schmidt norm $\|A\|_{2,\mathrm{av}}=2^{-m}\sqrt{\operatorname{tr}(A^*A)}$ on this register, for every $|\lambda|=1$,
\[
 \|U_{2m}-\lambda W_{2m}\|_{\rm op}
 \ \ge\ \|U_{2m}-\lambda W_{2m}\|_{2,\mathrm{av}}
 \ \ge\ \sqrt{2-2\min(1,\sqrt D/2^m)}.
\]
In particular, even approximation with eventual uniform operator-norm error bounded by a constant $\epsilon<\sqrt2$ is impossible at rank one, regardless of length-dependent phases or the fixed finite selection of widths. At a single length, error at most $\epsilon<\sqrt2$ requires $D\ge (1-\epsilon^2/2)^2 4^m$. The asymptotic projective threshold $\sqrt2$ is attained by the rank-zero identity.
\end{corollary}
\begin{proof}
\par\noindent\textup{(a)} For the known reversal construction of~\cite[Sec.~3]{Yamakami2020}, with base $I$, both branches active, pre-operation $I$, and post-operation $\Rm_1$, width-one recursion reverses the suffix and moves the first bit to the end. This is a depth-two term; width grouping preserves depth. At length $2m$ the middle-cut operator Schmidt rank is $2^{2m}$: reversal differs from the swap of the two $m$-qubit registers by local permutations, and the swap has a full matrix-unit decomposition. Theorem~\ref{thm:depth-one-transfer} excludes depth one.
\par\noindent\textup{(b)} Put $k=2^m$. The middle-cut Schmidt coefficients of reversal are those of swap, $\sum_{i,j=1}^k E_{ij}\otimes E_{ji}$: all $k^2$ equal one. Realignment of $W_{2m}/k$ therefore has operator norm $1/k$ and Frobenius norm one; realignment of $U_{2m}/k$ has rank at most $D$ and Frobenius norm one. SVD and Cauchy--Schwarz bound their inner product by $\min(1,\sqrt D/k)$. Expanding the squared normalized Hilbert--Schmidt distance gives the bound, which is at most the operator norm distance. Sharpness follows from $(I-iW)^*(I-iW)=2I$, since $W=W^*$ and $W^2=I$.
\end{proof}

\begin{lemma}[A nonperiodic native local length selector]\label{lem:josephus-selector}
For every fixed algebraic $V\in SU(2)$ there is a width-one depth-two term acting as $V$ on the first qubit and identity on the rest when $n$ is a power of two, and as identity otherwise. It can be restricted further to any fixed residue class of $n$ without increasing depth.
\end{lemma}
\begin{proof}
Use the left cyclic rotation $\Rm_1$, which has depth one. Define $J$ by width-one recursion with threshold one, identity base and post-operation, pre-operation $\Rm_1$, and both branches active. The first output wire is the second input wire, then the same rule is applied to the remaining circular list, with the original first wire moved to the end. More precisely, rotation sends the list $(1,2,\ldots,n)$ to $(2,\ldots,n,1)$; the wire $2$ is then fixed as the first output and the suffix is recursively permuted. Thus the output wire order is Josephus elimination with step two. The last output wire is the survivor. With zero-based input indices its index satisfies $j(1)=0$ and $j(n)=(j(n-1)+2)\bmod n$, because the remaining circular list starts at input index $2$ modulo $n$. Induction gives\[j(n)=2(n-2^{\lfloor\log_2 n\rfloor}).\]It increases by two between consecutive powers of two and is zero at powers of two. In particular the last wire is the original first wire exactly at powers of two. All actions are permutations of wire positions, on the entire register.

Let $A=iX$ and $B=\operatorname{diag}(e^{i\pi/6},e^{-i\pi/6})$ act on the first bit. The term $J^{-1}\Rm_1 B\Rp_1 J$ acts as $B$ on the survivor input wire. Its commutator with $A$ is identity when the wires are different and is the fixed noncentral special unitary $B^{-2}$ on the first bit when they coincide (including $n=1$). All operators used have depth at most two. The normal-generation argument of Lemma~\ref{lem:two-level-selection} converts this selected noncentral value to any prescribed algebraic $V$, retaining identity on the other lengths. To restrict to a residue, take a commutator with a periodically selected noncentral first-bit special unitary from Lemma~\ref{lem:periodic-local}; a fixed algebraic conjugation can make the value noncentral on the intersection of the two selected sets, by the tilt argument of Lemma~\ref{lem:two-level-selection}. Normal generation then gives the required target value. No depth increases, since only finite products, inverses and fixed gates are used outside $J$.
\end{proof}

\begin{corollary}[Nonperiodic projective rank-one phases]\label{thm:nonperiodic-projective}
For odd $q\ge3$ and $a_n=\mathbf1_{\{n\text{ is a power of two}\}}$, $R(\zeta_q,a)$ has minimum projective depth two at every nonempty width set, but no exact realization at any widths, even eventually.
\end{corollary}
\begin{proof}
The local selector of Lemma~\ref{lem:josephus-selector}, followed by Lemma~\ref{lem:cyclic-concentration}, realizes the normalized family at depth two. The exponent has infinitely many ones with unbounded gaps, so the transfer invariant excludes depth one. An exact realization would give a monic determinant recurrence over the finite exponent ring, forcing eventual periodicity by finiteness of its recurrence states. This is impossible.
\end{proof}
\subsection{Exact finite-order synthesis and intersections}
\begin{lemma}[Odd-order determinant normalization]\label{lem:odd-normalization}
Fix odd $q$, $\omega=\zeta_q$, and $a_n\in\mathbb Z/q\mathbb Z$. If $a$ has eventual annihilator $(X-2)\prod_{i=1}^r(X^{s_i}-c_i)$ with allowed widths and positive branch counts, then a diagonal term $U$ of depth at most $r$ has determinant $\omega^{a_n}$, and its normalization $V_n=\omega^{-2^{-n}a_n}U_n$ has depth at most $r+1$ (zero when $r=0$). In particular the compensating scalar $UV^{-1}$ has this latter bound. All nonzero entries of these product-synthesis diagonal realizations are powers of $\omega$; in the lcm construction auxiliary phases may have larger odd order, but their product has the required compensating phase. If $P_{S,d}(E)a=0$ eventually, the compensating scalar has depth at most $d+1$ (zero when $d=0$).
\end{lemma}
\begin{proof}
The base $r=0$ has $a_{N+i}=2^i a_N$ on a tail. Use a guarded diagonal extension of determinant $\omega^{a_N}$ and multiply by the constant normalization $\omega^{-2^{-N}a_N}I$, correcting the finite initial matrices. Both terms have depth zero. If the last factor is $X^s-c$, let $h_n=a_n-ca_{n-s}$ for $n>s$ and $b_n=2^{-n}a_n$ modulo $q$. By induction obtain the diagonal determinant realization $H$ for $h$ and its normalization $H'_n=\omega^{-2^{-n}h_n}H_n$. Choose $c$ active prefixes $C$. For $n>s$ put
\[
 U_n=H_nD_C(U)_n,\qquad V_n=H'_nL_nD_C(V)_n,
\]
with diagonal finite bases of the desired determinants and their normalizations. Here $L_n$ is the local diagonal gate on the first $s$ bits whose exponent on prefix $u$ is
\[
 b_{n-s}(\mathbf1_{u\in C}-c2^{-s})\pmod q.
\]
The sum of these exponents is zero. The sequence $b_n$ is eventually periodic (a monic recurrence over a finite ring, multiplied by the periodic sequence $2^{-n}$). Lemma~\ref{lem:periodic-local} and width grouping give $L$ at depth at most one, with finite corrections. The determinant recursion for $U$ gives exponent $a_n$. On an active block, the quotient $V_nU_n^{-1}$ has exponent $-2^{-n}h_n+b_{n-s}(1-c2^{-s})-b_{n-s}$; on an inactive block it has exponent $-2^{-n}h_n-c2^{-s}b_{n-s}$. Both equal $-b_n$. The normalized finite bases start the induction, establishing $V_n=\omega^{-b_n}U_n$ at every length. The exterior terms have depths at most $r-1$ for $U$ and at most $r$ for $V$; hence the claimed bounds.

For the last assertion, use Lemma~\ref{lem:primary-synthesis} on $\omega^{a_n}$ with $P_{S,d}$ in the group of all odd-order phases. Each integer root taken there has finite odd order; a single finite common odd order suffices for the finitely many components. Every component is killed by $(X-2)$ times at most $d$ allowed factors, as in Lemma~\ref{lem:phase-difference}. Apply the preceding construction modulo the order of each component and compose the compensating scalars. Raising to the inverse of $2^n$ is a homomorphism on the odd-order phase group. Correct finite exceptions by tables. For $d=0$ the base case suffices.
\end{proof}

\begin{theorem}[Exact concentration of all finite-order determinants]\label{thm:finite-det-concentration}
Let $q\ge2$, $\omega=\exp(2\pi i/q)$, $S$ nonempty, and $a_n\in\mathbb Z/q\mathbb Z$. At unrestricted finite recursion depth the following are equivalent:
\begin{enumerate}
\item $R(\omega,a)\in\mathcal A_S$;
\item the determinant sequence $\omega^{a_n}$ is realizable in $\mathcal A_S$;
\item $a$ has an eventual annihilator modulo $q$ that is a finite product of $X^s-c$ with $s\in S$ and $1\le c\le2^s$ (including the empty product).
\end{enumerate}
For odd $q$, a product with $r$ factors gives a rank-one realization of depth at most $r+1$, and if $r=0$ depth zero suffices. The determinant and rank-one minima are not asserted to coincide.
\end{theorem}
\begin{proof}
(1) implies (2) by determinants; (2) implies (3) by Theorem~\ref{thm:det-full-depth}, since $X-2$ divides an allowed factor. Conversely for odd $q$ apply Lemma~\ref{lem:odd-normalization} to obtain $\omega^{2^{-n}a_n}I$ at depth at most $r+1$. The exponent is eventually periodic, so Theorem~\ref{thm:odd-depth-one-sharp} gives the normalized rank-one operator at depth at most one. Their product realizes $R(\omega,a)$; for $r=0$ a finite table suffices. For even $q$ split into odd and $2$-primary components (Corollary~\ref{cor:finite-order-split}), apply this result and Theorem~\ref{thm:two-primary-rank-one}, and compose.
\end{proof}

\begin{theorem}[Singleton-width operator intersections]\label{thm:operator-meet}
For positive $s,t$ and $g=\gcd(s,t)$,
\[
 \mathcal A_{\{s\}}\cap\mathcal A_{\{t\}}=\mathcal A_{\{g\}}
 \quad\Longleftrightarrow\quad s\mid t\ \text{or}\ t\mid s.
\]
In the incomparable case the intersection is not any width-generated algebra. A rank-one witness is realizable at depth at most two in both larger algebras, and cannot be realized exactly on a tail at width $g$.
\end{theorem}
\begin{proof}
Comparable widths follow from Theorem~\ref{thm:width-order}. Otherwise $\min(s,t)\ge2g$. Bertrand's postulate gives an odd prime $p$ with $2^g+1<p<2(2^g+1)$, hence $2^g<p-1\le2^{g+1}\le2^{\min(s,t)}$. Set $a_n=(-1)^{\lfloor(n-1)/g\rfloor}\pmod p$ and $\omega=\zeta_p$. For $w=s,t$, the sequence satisfies $a_{n+w}=(-1)^{w/g}a_n$. The factor $X^w-c$ is permitted, with $c=1$ if $w/g$ is even and $c=p-1$ otherwise. By Theorem~\ref{thm:finite-det-concentration}, $R(\omega,a)$ lies in both width algebras at depth at most two.

At width $g$, $(X^g+1)(E)a=0$, and $X^g+1$ is coprime modulo $p$ to every allowed $X^g-c$, $1\le c\le2^g$, as well as to $X-2$. A determinant annihilator at width $g$ would therefore force $a$ to vanish on a tail by polynomial B\'ezout, a contradiction. The width-generated greatest lower bound is $\mathcal A_{\{g\}}$ by Theorem~\ref{thm:width-order}; hence the actual intersection is not width-generated.
\end{proof}
For example, $R(\zeta_5,(3^n)_n)$ belongs to the width-two and width-three algebras but not width one, by Corollary~\ref{cor:torsion-meet-fail} and Theorem~\ref{thm:finite-det-concentration}. Likewise $R(\zeta_5,((-1)^n)_n)$ belongs to a nonempty width set exactly when it contains a width greater than one: use $c=1$ for even widths, $c=4$ for odd widths at least three, and exclude width one by the same coprimality argument. These examples are not projective separations, by Theorem~\ref{thm:odd-depth-one-sharp}.
\begin{corollary}[Finite-order meets, scalars and sharp depth bounds]
\qsrpart{a}{cor:all-finite-meet}
For arbitrary nonempty $S,T$, put $J=\downarrow S\cap\downarrow T$. On the rank-one sector with a fixed finite phase order $q$, the exact meet identity holds if and only if $R_p(S)\cap R_p(T)=R_p(J)$ for every prime divisor $p$ of $q$, with $R_p$ as in Theorem~\ref{thm:det-meet-primes}. In particular the equality of exact rank-one intersections for \emph{all} finite phase orders is equivalent to the determinant-image meet criterion. If the latter fails, a common rank-one operator outside $\mathcal A_J$ exists, even for arbitrary mixed width sets. For finite $S,T$ this sufficient criterion for failure of the full exact operator meet is decidable. Conversely these rank-one results do not assert equality of the full operator intersection when the determinant criterion holds.

For odd $q$, the intersection of the projective rank-one sectors restricted to eventually periodic exponents is independent of $S,T$ and equals that of width one. Corollary~\ref{thm:nonperiodic-projective} disproves eventual-periodicity necessity for general odd-order projective realizations. Arbitrary projective operator meets remain open.
\qsrpart{b}{cor:scalar-periodic-mixed}
For odd $q\ge3$, $\omega=\zeta_q$ and $b_n\in\mathbb Z/q\mathbb Z$, the scalar family $\omega^{b_n}I_{2^n}$ belongs to $\mathcal A_S$ exactly when $a_n=2^n b_n$ satisfies the determinant criterion of Theorem~\ref{thm:finite-det-concentration}. In this case $b_n$ is eventually periodic. For any $q=2^m q'\ge2$ with $q'$ odd, restrict the rank-one phase sector to eventually periodic exponents. Then eventual projective membership is equivalent to exact membership of its $2$-primary component (with no condition when $m=0$). Consequently the projective gcd/downset intersection identity holds in this entire periodic finite-order rank-one sector, for arbitrary width sets. This includes all finite-order rank-one families that are exactly realizable at any widths.
\qsrpart{c}{cor:odd-exact-depth-scalar}
Fix odd $q\ge3$ and $\omega=\zeta_q$. For an exactly realizable $R(\omega,a)$ whose exponent is not eventually zero, its minimum exact depth is
\[
 \max\{1,d_{\rm sc}\},
\]
where $d_{\rm sc}$ is the minimum exact depth of the scalar $\omega^{2^{-n}a_n}I$. These minima are taken in the same allowed width algebra. If $d_{\det}$ is the minimum determinant-realization depth of $\omega^{a_n}$, the rank-one minimum lies between $\max(1,d_{\det})$ and $d_{\det}+1$. In particular determinant depth zero and nonzero tail imply exact rank-one depth one. For an eventually zero exponent all three minima are zero.
\qsrpart{d}{cor:odd-prime-sharp-sector}
Let $q\ge3$ be odd, $S$ finite and nonempty, and $a$ a sequence modulo $q$ not eventually zero, of finite determinant minimum depth $d$. Put $N_S(X)=\operatorname{lcm}_{s\in S}(X^s-2^s)$ over $\mathbb Z[X]$. If $a=a^{(1)}+a^{(2)}$ on a tail, where $M_S(E)^da^{(1)}=0$ and $(E-2)N_S(E)^da^{(2)}=0$ modulo $q$, then the exact rank-one minimum is $\max(1,d)$. In particular, for prime $q=p$, this holds if the minimal eventual polynomial divides
\[
 \operatorname{lcm}_{\mathbb F_p[X]}\bigl(\overline{M_S}^{\,d},(X-\bar 2)\overline{N_S}^{\,d}\bigr).
\]
The reductions are taken after the integer lcms.
\end{corollary}
\begin{proof}
\par\noindent\textup{(a)} By Theorem~\ref{thm:finite-det-concentration} exact rank-one membership is equivalent to the determinant exponent criterion. For sufficiency of the prime-root condition use the prime-power lifted B\'ezout argument of Theorem~\ref{thm:det-meet-primes} on each primary component of $\mathbb Z/q\mathbb Z$. For necessity, if a common nonzero root outside $R_p(J)$ exists, its monic irreducible polynomial over $\mathbb F_p$ divides allowed factors for $S,T$ and not any allowed factor for $J$. A nonzero recurrence sequence with this minimal polynomial exists over $\mathbb F_p$ (use a nonzero vector in its companion-matrix representation, or the recurrence with initial values not all zero; irreducibility forces minimality). The constant term is nonzero so the same holds eventually. Embed its additive exponent group into $\mathbb Z/q\mathbb Z$ by multiplying by $q/p$. This preserves the annihilator property and nonmembership. The concentration theorem now yields an actual common operator. The projective assertion follows directly from Theorem~\ref{thm:odd-depth-one-sharp} and width inclusion.
\par\noindent\textup{(b)} Necessity follows by the determinant. For sufficiency, Lemma~\ref{lem:odd-normalization} realizes $\omega^{2^{-n}a_n}I=\omega^{b_n}I$. Eventual periodicity follows from the recurrence for $a$ and periodic invertibility of $2^n$ modulo $q$. For the mixed-order assertion split the phase by Corollary~\ref{cor:finite-order-split}. Its odd component is projectively realizable at every nonempty width set by Theorem~\ref{thm:odd-depth-one-sharp}. The $2$-primary component has the exact/projective criterion of Theorem~\ref{thm:two-primary-rank-one}, and its meet identity is Corollary~\ref{cor:two-primary-rank-one-meet}. Combining the components proves the projective meet assertion. Finally exactly realizable rank-one families of finite order have eventually periodic exponents by the monic determinant recurrence, as above.
\par\noindent\textup{(c)} Exactly realizable rank-one families of finite order have eventually periodic exponents by Theorem~\ref{thm:finite-det-concentration}. By Theorem~\ref{thm:odd-depth-one-sharp} the normalized rank-one representative has depth at most one, so multiplication by it or its inverse transforms scalar realizations into rank-one realizations and conversely. The depth-zero obstruction in that theorem gives the equality of minima. Determinants give the lower bound. We prove the upper bound, taking finite $S_0\subseteq S$ realizing the determinant at depth $d=d_{\det}$.

Lemma~\ref{lem:odd-normalization} gives $d_{\rm sc}\le d+1$, and $d_{\rm sc}=0$ for $d=0$. This proves the bounds; an eventually zero exponent needs only a finite table.
\par\noindent\textup{(d)} The rank-one row of Lemma~\ref{lem:phase-difference} realizes the first component at depth at most $d$. For the second, $h_n=2^{-n}a_n^{(2)}$ satisfies $(E-1)M_S(E)^dh=0$ modulo $q$: substituting $2X$ in its original annihilator yields this polynomial up to a power of two, since $N_S(2X)=2^{\deg N_S}M_S(X)$. The last row of the lemma realizes the compensating scalar at depth at most $d$. The normalized rank-one part has depth at most one by Theorem~\ref{thm:odd-depth-one-sharp}, since the monic recurrence makes $a^{(2)}$ eventually periodic. Compose and correct finite exceptions. The determinant bound and the nonzero-tail depth-zero obstruction give equality. Over $\mathbb F_p$, the displayed lcm criterion gives the required decomposition by polynomial CRT, assigning each primary factor to one of the polynomials containing its full multiplicity.
\end{proof}

\begin{theorem}[Sharp finite-order exact depths at width one]\label{thm:finite-width-one-sharp}
Fix $q\ge2$ and $\omega=\zeta_q$, and let $a$ be a sequence modulo $q$. Write $d_{\det}$ for the minimum width-one depth of the determinant $\omega^{a_n}$, with value $+\infty$ if no realization exists. Then the minimum exact width-one rank-one depth is
\[
 d_{\rm ex}=\begin{cases}
 0,&a_n=0\text{ eventually},\\
 \max(1,d_{\det}),&\text{otherwise}.
 \end{cases}
\]
Here $d_{\det}$ is the least $d\ge0$ such that $(E-1)^d(E-2)^{d+1}a=0\pmod q$ eventually. For $q$ a power of two, the exact and eventual-projective depth minima coincide with $d_{\det}$. For an arbitrary finite $q=2^m q_o$ with $q_o$ odd and an eventually periodic exponent, the minimum eventual-projective width-one depth equals the maximum of the determinant minimum for its $2$-primary component and the number $\mathbf1_{\{a\bmod q_o\text{ is not eventually zero}\}}$ (the order-one component contributes zero). This includes nonrealizable cases with minimum $+\infty$.
\end{theorem}
\begin{proof}
At width one $L_{\{1\}}=(X-1)(X-2)$, giving the determinant criterion by Theorem~\ref{thm:det-full-depth}. By Corollary~\ref{cor:finite-order-split}, exact and projective minima are maxima over the odd and $2$-primary components, as is the determinant minimum by the modular recurrence criterion. It suffices to consider these two components.

For odd $q$, no finite determinant realization implies no exact realization; an eventually zero exponent needs only a finite table. Otherwise take $d=d_{\det}<\infty$. The polynomials $(X-1)^d$ and $(X-2)^{d+1}$ are comaximal over $\mathbb Z[X]$: $X-1$ and $X-2$ differ by one and powers of comaximal ideals are comaximal. B\'ezout projectors split $a$ into the two sectors of Corollary~\ref{cor:odd-prime-sharp-sector}, giving exact depth $\max(1,d)$. For eventually periodic exponents the projective minimum is zero or one by Theorem~\ref{thm:odd-depth-one-sharp}.

For $q=2^m$, Lemma~\ref{lem:two-primary-projective} gives the determinant lower bound for any projective realization, including when no finite determinant depth exists. For finite $d=d_{\det}$, $E-2$ is invertible on eventual sequences modulo $q$ (use the finite geometric series for $1-2E^{-1}$). Thus $(E-1)^da=0$ eventually and Lemma~\ref{lem:phase-difference} realizes the rank-one family at depth $d$. This includes $d=0$. Combining the components proves both formulas; nonperiodic odd-order projective membership beyond depth one is not classified here.
\end{proof}

\section{Projective intersections and alignment}\label{sec:alignment}
We study common exact representatives of projectively agreeing terms, not membership in the common-downset algebra.
\subsection{Scalar normalization}
\begin{theorem}[Scalar discrepancy normalization]
Let $U\in\mathcal A_S^{[d]}$ and $V\in\mathcal A_T^{[e]}$ satisfy $U_n=\lambda_n V_n$ on a tail. Let $K$ be the field of their initial entries and conjugates, $\Gamma$ the group generated by their initial determinants, and $H$ its saturation in $K^\times$. Fix a splitting of $H$ as in Lemma~\ref{lem:finite-phase-saturation}.
\qsrpart{a}{thm:free-alignment} Scalar corrections at ranks $d,e$ remove the free projection of $\lambda_n$, leaving bounded torsion, without any hypothesis on the original discrepancy.
\qsrpart{b}{cor:two-primary-alignment} Further scalar corrections at ranks at most $d+1,e+1$ remove the odd projection, leaving the two-primary projection of $\lambda_n$. If this projection is eventually trivial, a common exact representative exists at these bounds.
\qsrpart{c}{prop:projective-alignment} If the residual two-primary projection is eventually periodic, there is a common exact representative at ranks at most $d+1,e+1$. In particular this holds if the original discrepancy is eventually periodic and torsion.
\qsrpart{d}{cor:arbitrary-depth-one-alignment} If $d,e\le1$, there is a common exact representative in $\mathcal A_S^{[2]}\cap\mathcal A_T^{[2]}$ even for an arbitrary discrepancy. The bounds in (c) hold individually.
\end{theorem}
\begin{proof}
\textup{(a)} It suffices to use finite nonempty subsets of $S,T$ containing the actual widths. On the tail, $\lambda_n$ is a diagonal entry of $U_nV_n^{-1}$, hence is in $K$, and $\lambda_n^{2^n}=\det U_n/\det V_n\in\Gamma$ by determinant induction. Therefore $\lambda_n\in H$. Fix a basis $\gamma_1,\ldots,\gamma_r$ of the free factor of $H$ and let $v_i:H\to\mathbb Z$ be its coordinates. Put $a^U_i(n)=v_i(\det U_n)$, $a^V_i(n)=v_i(\det V_n)$ and $l_i(n)=v_i(\lambda_n)$. Then
\[
 a^U_i(n)-a^V_i(n)=2^n l_i(n).
\]
By determinant necessity, $a^U_i$ is annihilated by $P_{S,d}$ and $a^V_i$ by $P_{T,e}$ on tails. The rational CRT decomposition of Lemma~\ref{ws:slow} gives $a^U_i=u^U_i+2^n b^U_i$ and $a^V_i=u^V_i+2^n b^V_i$, with $u^U_i,u^V_i=o(2^n)$. With the original multiplicities retained, the fast components obey
\[
 (E-1)M_S(E)^d b^U_i=0,\qquad (E-1)M_T(E)^e b^V_i=0.
\]
Indeed the factors of $P_{S,d}$ at modulus two are precisely $(X-2)$ times the $d$th power of the squarefree lcm of $X^s-2^s$; replacing $E$ by $2E$ gives a nonzero constant times $(E-1)M_S(E)^d$. The same holds for $T$. Choose a single integer $D>0$ clearing the bounded denominators of all $b^U_i,b^V_i$ on a tail. Then
\[
 D(u^U_i-u^V_i)=2^nD(l_i-b^U_i+b^V_i)\in2^n\mathbb Z.
\]
The left side is $o(2^n)$, hence eventually zero, so $l_i=b_i^U-b_i^V$. Choose effective $\eta_i^D=\gamma_i$. The scalar row of Lemma~\ref{lem:phase-difference} realizes $\eta_i^{Db_i^U(n)}I$ and $\eta_i^{Db_i^V(n)}I$ at ranks $d,e$, with finite tables for early lengths. Multiply over $i$ (empty product identity) to obtain $C_U,C_V$. Their ratio is $\prod_i\gamma_i^{l_i}$, the free projection of $\lambda_n$; division leaves torsion. The new phases need not lie in $K$ or a fixed finite library.

\textup{(b)} Project the original determinants onto the odd factor $H_o$. The projected determinants obey the original annihilators; Lemma~\ref{lem:odd-normalization} realizes their unique $2^n$th roots in $H_o$ as scalar terms at ranks at most $d+1,e+1$. Their ratio is the odd projection of $\lambda_n$. Divide out these and the free corrections from (a). If the residual is trivial, finite tables align short lengths.

\textup{(c)} If the residual is periodic and has order dividing $2^b$ with $b\ge1$, its scalar matrix on the first $b$ bits is special unitary. Lemma~\ref{lem:periodic-local} and width grouping realize this schedule at rank one in either algebra. Divide it out on one side and correct short lengths. The rank bounds follow by taking maxima under composition. The trivial residual requires no such correction.

\textup{(d)} First apply (a) at ranks $d,e$. The remaining finite-valued scalar is the all-zero entry of the product of the corrected first representative with the inverse of the second. At rank at most one Theorem~\ref{thm:depth-one-transfer} gives a monic recurrence, so its finite-valued tail is periodic. Its two-primary projection is periodic as well, and (c) applies. If the original discrepancy is torsion at rank at most one, the same argument applies directly to it.
\end{proof}

At rank two, Lemma~\ref{lem:josephus-selector} gives $(-1)^{\mathbf1_{\{n=2^k:k\ge0\}}}I_{2^n}$: a nonperiodic scalar with determinant one. Thus determinant recurrences alone cannot establish periodicity; this is not an alignment obstruction.

\subsection{The group obstruction to common exact representatives}
\begin{theorem}[Two-primary abelian alignment obstruction]\label{thm:alignment-obstruction-group}\label{cor:odd-projective-alignment}
Work with the effective-angle unitary algebras modulo eventual exact equality, and let $G_S$ denote the resulting group. Write $Z_S$ for its subgroup of scalar families and $\pi$ for the quotient by eventual projective equivalence in the ambient group of unitary families. For nonempty $S,T$ define
\[
 P=\pi(G_S)\cap\pi(G_T),\qquad H_0=\pi(G_S\cap G_T).
\]
Then $H_0$ is normal in $P$, and $P/H_0$ is abelian and two-primary torsion: every element has order a power of two (including one). More precisely, there is a canonical injection
\[
 P/H_0\ \lhook\joinrel\longrightarrow\ Z_{S\cup T}/(Z_SZ_T),
\]
whose image consists exactly of the discrepancy classes $[UV^{-1}]$ of projectively agreeing $U\in G_S,V\in G_T$. Each image class has a representative taking values in a fixed finite group of power-of-two roots of unity on a tail.

If $p\in\pi(\mathcal A_S^{[d]})\cap\pi(\mathcal A_T^{[e]})$, some $2^m$th power of $p$ has a common exact representative at depths at most $d+1,e+1$. If $p,p'$ both have representatives with original bounds $d,e$, their commutator has a common exact representative already at bounds $d,e$. If the original representative terms have all their initial entries in one finitely generated field $K$ closed under conjugation, the exponent $2^m$ can be the order of the two-primary roots of unity in $K$, independently of the terms with such constants. The correcting terms may use other effective angles.
\end{theorem}
\begin{proof}
For $p\in P$ choose representatives $U\in G_S,V\in G_T$. The scalar $UV^{-1}\in Z_{S\cup T}$ is central, and changing representatives multiplies it by $Z_SZ_T$. Thus its coset defines a homomorphism, since $(UU')(VV')^{-1}=(UV^{-1})(U'V'^{-1})$. Its kernel is $H_0$: $UV^{-1}=ab$ with $a\in Z_S,b\in Z_T$ iff $a^{-1}U=bV$ gives a common representative. The first isomorphism theorem gives the injection, normality and abelianness.

By Theorem~\ref{cor:two-primary-alignment}, replace the original representatives by scalar-corrected terms $\widetilde U,\widetilde V$ at depths at most $d+1,e+1$ with $\widetilde U_n=\xi_n\widetilde V_n$ eventually and $\xi_n^{2^m}=1$ for some $m\ge0$. These corrections leave the coset unchanged. Their $2^m$th powers agree exactly on the tail and retain the depth bounds. Finite length tables make this a single all-length family without increasing either bound. Hence every obstruction element has two-power order. The residual phases belong to the two-primary factor of the saturation in the field generated by the original constants, by Theorem~\ref{thm:free-alignment} and Theorem~\ref{cor:two-primary-alignment}. When this field is contained in $K$, their orders divide the order of the two-primary roots of unity in $K$; finiteness follows from Lemma~\ref{lem:finite-phase-saturation}.

For the original representatives of $p,p'$ the scalar factors cancel in the commutator, giving $[U,U']=[V,V']$ eventually without any scalar correction. A finite table again supplies exact equality at short lengths. This proves the original-depth assertion directly.
\end{proof}

If $p\in P$ has finite odd order $q$, choose corrected representatives at depths $d+1,e+1$ as above with discrepancy of order dividing $2^m$. For $k>0$ with $k\equiv0\pmod{2^m}$ and $k\equiv1\pmod q$, their $k$th powers agree and represent $p$. The original depths suffice if the discrepancy already has bounded two-power order. Thus $H_0$ contains the subgroup generated by odd-order elements and commutators of $P$. For order $2^r q$, choose $k_o\equiv0\pmod{2^r}$, $k_o\equiv1\pmod q$; the odd-order factor $p^{k_o}$ lies in $H_0$, so only $p^{1-k_o}$ can obstruct alignment. This includes common odd-order rank-one projective families, without asserting common-downset membership.

\subsection{Nonperiodic scalar corrections available in every width algebra}
Let $\mathcal L$ be the Boolean algebra of subsets of the positive lengths generated by eventually periodic sets and the shifted powers-of-two sets $\{n:n-j=2^k,\ k\ge0\}$ for fixed integers $j\ge0$. All Boolean combinations here are finite.

\begin{lemma}[Boolean length selectors and fixed-block synthesis]\label{lem:boolean-length-selector}
For every $L\in\mathcal L$, every $b\ge1$ and every fixed algebraic $V\in SU(2^b)$, there is a width-one term of depth at most two acting on the first $b$ bits by $V$ if $n\in L$, by identity otherwise, at all $n\ge b$, and as identity at shorter lengths. It acts on the whole register, not just a subspace with initialized bits. It can be translated to every nonempty width set at the same depth.
\end{lemma}
\begin{proof}
First consider one-bit selectors for $D=\operatorname{diag}(e^{i\pi/6},e^{-i\pi/6})$. For the unshifted powers-of-two set use Lemma~\ref{lem:josephus-selector}. Apply $j$ identical-branch wrappers to shift the length argument by $j$ and then conjugate by a fixed wire permutation moving the selected bit to the first position; correct finitely many lengths $n\le j$ by a table. The wrappers do not raise depth. For eventually periodic sets use Lemma~\ref{lem:periodic-local} and finite tables. The complement of a selector $B$ is $DB^{-1}$, where $D$ acts on the first bit. To select the intersection of two sets, take the commutator of their selectors after conjugating one by a fixed algebraic rotation making the commutator noncentral whenever both are active. It is identity if either is inactive. The normal-generation construction of Lemma~\ref{lem:two-level-selection} converts that fixed noncentral selected value into $D$ without raising depth. Unions follow from complements and intersections. Induction on the finite Boolean expression proves the one-bit assertion with value $D$.

For $b>1$, on the first $b$ bits this selector acts as $D\otimes I_{2^{b-1}}$ or identity. Choose two prefix basis vectors from different eigenspaces of $D\otimes I$ and a fixed two-level algebraic special unitary $C$ mixing them. The commutator with $C$ is identity off their span (both operators preserve it and $C$ is identity off it), and is a fixed noncentral special unitary there when selected, by choosing a sufficiently small nonzero algebraic tilt as in Lemma~\ref{lem:two-level-selection}. On unselected lengths it is identity everywhere. Normal generation on that pair realizes any algebraic special unitary there, selected by the same $L$. Fixed permutations of the first $b$ bits move the pair to any two distinct basis positions. Algebraic Givens elimination expresses $V$ as a finite product of such special-unitary two-level gates. Compose the selected factors, using the predicate on the original length throughout; this avoids a length shift when controlling on a larger prefix. Fixed gates and inverses do not increase rank. Finite tables correct lengths below $b$, and width grouping in Theorem~\ref{thm:width-order} preserves depth. No clean ancillary qubit is required.
\end{proof}

\begin{corollary}[Two-primary scalar synthesis and alignment]
\qsrpart{a}{cor:universal-two-primary}
Let $a_n\in\mathbb Z/2^m\mathbb Z$, $m\ge1$, with each fiber $\{n:a_n=j\}$ in $\mathcal L$. Then $\zeta_{2^m}^{a_n}I_{2^n}$ has an exact realization of depth at most two in every nonempty width algebra. If $a_n$ is not eventually periodic, its minimum depth is exactly two in every such algebra.

In particular for any non-eventually-periodic $L\in\mathcal L$ and root of unity $\omega$ of order $q>1$, the scalar family $\omega^{\mathbf1_L(n)}I_{2^n}$ has minimum exact depth two if $q$ is a power of two, and has no finite exact realization in any width algebra if $q$ has an odd divisor greater than one. Consequently the possible orders of nonperiodic scalar phases at rank two are not bounded by rank alone.
\qsrpart{b}{cor:boolean-alignment}
Under the hypotheses of Theorem~\ref{thm:free-alignment}, suppose the residual discrepancy $\xi_n$ in Theorem~\ref{cor:two-primary-alignment} has its fibers in $\mathcal L$ on a tail. Then there is a common exact representative at depths at most $\max(d+1,2),e+1$, or alternatively $d+1,\max(e+1,2)$. If the original discrepancy already has this form with bounded two-power order, the bounds are $\max(d,2),e$ or $d,\max(e,2)$. For example the nonperiodic scalar from the preceding remark and all its higher two-power-order analogues can be removed at depth at most two, rather than constituting obstructions to alignment.
\end{corollary}
\begin{proof}
\par\noindent\textup{(a)} Take $b=m$ and for each exponent value $j$ use the selector for its fiber with $V=\zeta_{2^m}^j I_{2^b}$. This matrix is special unitary because $(\zeta_{2^m}^j)^{2^b}=1$. The product of the finitely many selected factors equals the required scalar on all lengths $n\ge b$; use a finite table at shorter lengths. A depth-one scalar has a monic recurrence for its all-zero diagonal entry by Theorem~\ref{thm:depth-one-transfer}; a finite-valued tail of such a recurrence is eventually periodic. This gives the lower bound even for arbitrary fixed constants. For the odd-divisor obstruction let $q_o>1$ be the odd part of $q$. A fixed power of a putative scalar realization yields a scalar with phase of order $q_o$. Its determinant is the corresponding primitive $q_o$th root raised to $2^n\mathbf1_L(n)$. Theorem~\ref{thm:det-full-depth} implies eventual periodicity modulo $q_o$. Inverting $2^n$ gives the contradiction. Taking powers of two for $L$ and arbitrary $m$ gives unbounded scalar orders, not an unbounded obstruction-group exponent.
\par\noindent\textup{(b)} Finite changes preserve $\mathcal L$. Corollary~\ref{cor:universal-two-primary} realizes the residual scalar in either algebra at rank at most two; divide it out and correct short lengths by a finite table. Composition takes the maximum rank.
\end{proof}

\subsection{Rank-preserving alignment on fixed local support}
\begin{theorem}[Finite local projective range]\label{thm:finite-local-alignment}
Fix $b\ge1$ and pairwise projectively distinct fixed algebraic unitaries $A_1,\ldots,A_k\in U(2^b)$, with $k\ge1$. Suppose a term $F\in\mathcal A_S^{[d]}$ satisfies, on a tail,
\[
 F_n=\lambda_n(A_{j(n)}\otimes I_{2^{n-b}}),\qquad |\lambda_n|=1,
\]
where $j(n)\in\{1,\ldots,k\}$ is arbitrary. For any fixed algebraic $V_1,\ldots,V_k\in SU(2^b)$ there is a term of depth at most $d$ realizing $V_{j(n)}\otimes I$ on that tail. In particular choose once and for all algebraic $\alpha_i$ with $\alpha_i^{2^b}\det A_i=1$ and set $V_i=\alpha_i A_i$. If $F$ and $G\in\mathcal A_T^{[e]}$ represent the same eventual projective family of this form, that family has a common exact representative at depths at most $d,e$. Its two scalar corrections on the common tail also have these respective depth bounds. No periodicity of the index function or original phases is assumed.
\end{theorem}
\begin{proof}
All constant gates below act on the first $b$ bits, with the identity on the suffix, and are available at depth zero by Lemma~\ref{lem:width-table}. Write $N=2^b$. Fix a target index $j$ and a coordinate pair in the $N$-dimensional prefix space. Start with a constant nonscalar special-unitary two-level gate $B$ on that pair. For each $i\ne j$, replace $B$ by
\[
 [B,C_i F A_i^{-1}C_i^{-1}],
\]
where $C_i$ is a fixed algebraic unitary to be chosen. On the tail the scalar in $F$ cancels in this commutator. When the index is $i$, its second operand is scalar, so the new gate is identity. Previously eliminated indices remain eliminated. On the target index the second operand, up to scalar, is a conjugate of $A_j A_i^{-1}$, a nonscalar algebraic unitary. Choose two eigenvectors with distinct eigenvalues, map their span to the designated pair and align their eigenbasis with that of the current target value of $B$. A sufficiently small nonzero algebraic tilt within the pair makes the two restrictions noncommuting and their commutator different from $-I_2$, by the same two-dimensional argument as in Lemma~\ref{lem:two-level-selection}. Outside the pair $B$ is identity and the second operand preserves the orthogonal complement. Thus the target value remains a fixed noncentral algebraic special unitary on the pair, identity off it. At other, not yet eliminated indices no preservation of the pair is required: their values remain unitaries on the prefix, and the later elimination step for each index uses a scalar second operand there. Algebraic eigenvectors, conjugating matrices and tilts exist as in Lemma~\ref{lem:two-level-selection}.

After finitely many eliminations, $B$ is identity on every nontarget index and a fixed noncentral $SU(2)$ gate on the designated pair at the target. The normal-generation part of Lemma~\ref{lem:two-level-selection} synthesizes any algebraic special unitary on this pair, selected by the same index. Fixed prefix permutations move the pair to any coordinate pair; two-level elimination synthesizes $V_j$ on the selected prefix. The case $k=1$ simply uses a constant gate. Compose these selected matrices over $j$. Each operation was composition, inverse, or a fixed gate, so no recursion rank is added. Short lengths can be set to identity by finite tables. For projectively agreeing $F,G$ the index is unique on the common tail. Apply the construction separately to each with the same choice $V_i=\alpha_i A_i$. The resulting exact families agree on the tail and can be made identical at earlier lengths by finite tables. Multiplying the original terms by the inverse of the common representative gives the claimed scalar corrections on the tail; replace their finitely many earlier matrices by identity using the finite length-table lemma to obtain scalar families at every length. This proves alignment and the factorization of the eventual scalar discrepancy.
\end{proof}
\begin{corollary}[A selector criterion for finite local projective membership]\label{cor:finite-local-selector}
Under the hypotheses on the fixed templates in Theorem~\ref{thm:finite-local-alignment}, fix a nonscalar algebraic $D\in SU(2)$ (for instance $D=\operatorname{diag}(i,-i)$). The family $A_{j(n)}\otimes I$ is eventually projectively realizable at depth at most $d$ if and only if for each $i$ the first-qubit selector $D^{\mathbf1_{\{j(n)=i\}}}\otimes I$ is realizable on a tail at depth at most $d$ in the same width algebra.
\end{corollary}
\begin{proof}
For necessity apply the theorem with targets $D\otimes I_{2^{b-1}}$ and identity. Conversely, for $b=1$ normal generation of nonscalar $D\in SU(2)$ gives each selected $\alpha_i A_i$. For $b>1$, choose a pair from distinct eigenspaces of $D\otimes I_{2^{b-1}}$ and take its commutator with a fixed special-unitary tilt on that pair. This leaves a noncentral two-level value at selected lengths and identity otherwise, as in Lemma~\ref{lem:boolean-length-selector}. Conjugate, synthesize the required two-level factors and compose over the indices. All operations preserve the depth bound. Finite tables supply early lengths if needed.
\end{proof}
The witnesses may use arbitrary effective gate constants; only the fixed templates are algebraic. This sector has trivial obstruction, and its minimum projective rank equals the exact rank of the chosen special-unitary normalization. No claim is made for arbitrary schedules or increasing support.

\subsection{A faithful exact encoding of all projective classes}
\begin{theorem}[Interleaved adjoint encoding]\label{thm:interleaved-adjoint}
For $S\ne\varnothing$ write $2S=\{2s:s\in S\}$. For any term $F\in\mathcal A_S^{[d]}$ there is a term $\mathfrak D(F)\in\mathcal A_{2S}^{[d]}$ defined at positive length $N$ as follows. Pair the first $2m$ bits, $m=\lfloor N/2\rfloor$, and identify the first and second positions of all pairs with two $m$-qubit registers. On these registers its action is
\[
 F_m\otimes\overline{F_m},
\]
with identity on a possible unpaired last bit, and identity at $N=1$. Here the bar means entrywise complex conjugation. This map is a homomorphism; two families have eventually equal encodings if and only if they are eventually projectively equal. Thus there is an injective homomorphism
\[
 \pi(\mathcal A_S^{[d]})\cap\pi(\mathcal A_T^{[e]})
 \longrightarrow
 \mathcal A_{2S}^{[d]}\cap\mathcal A_{2T}^{[e]}
 \quad\text{modulo eventual exact equality}.
\]
In particular the full projective intersection admits a faithful common exact encoding, with a factor two in register length and widths but no increase in rank. This is not a common exact representative on the original register.
\end{theorem}
\begin{proof}
Translate a term on one chosen lane, ignoring the other and any unpaired bit. Initial gates become guarded gates on complete pairs; composition translates componentwise. Binary branching at $m>1$ uses a two-bit branch tree testing only the chosen bit, with finite-table correction at $m\le1$. For width $s$ and threshold $b$, use width $2s$, threshold $2b+1$, translated operands and active prefixes whose chosen-lane word is active originally. Then $\lfloor(N-2s)/2\rfloor=m-s$ and the base regions coincide. Induction on syntax and length proves the action at both parities and preserves rank. Compose the two lane translations, conjugating the second term's constants (still effective and in the grammar).

The tensor action is the usual vectorized conjugation representation. It is multiplicative and phase invariant; conversely $U\otimes\bar U=V\otimes\bar V$ gives $W\otimes\bar W=I$ for $W=V^{-1}U$, so all eigenvalue ratios $w_i\bar w_j$ equal one and $W$ is scalar. This proves tail injectivity. Encodings agreeing on a tail can be aligned at shorter lengths by rank-zero tables.
\end{proof}

\subsection{Decoding on an invariant reference sector}
The interleaved encoding is faithful, but its inverse as a map to projective classes is not automatically a syntax translation. The following sufficient conditions give exact decoding without a rank increase.
\begin{definition}[Reference-ray syntax]\label{def:reference-ray-syntax}
Write $z_n=|0^n\rangle$. A reference-ray term is built from recursion-free terms fixing the ray $\mathbb C z_n$ at every positive length, composition of reference-ray terms, branching $\Br[A,B]$ with $A$ a reference-ray term and $B$ arbitrary, and the original recursion with all three earlier operands reference-ray terms. This is a restriction on a presentation, not on its denotation alone. In particular every term built from fixed diagonal prefix gates by composition, branching and recursion belongs to this syntax.
\end{definition}
\begin{theorem}[Rank-preserving reference-ray normalization]\label{thm:reference-ray-normalization}
For every reference-ray term $F\in\mathcal A_S^{[d]}$, its eigenphase on the reference ray,
\[
 F_n z_n=\alpha_n z_n,
\]
has a scalar realization $C(F)_n=\alpha_n I_{2^n}$ in $\mathcal A_S^{[d]}$. Consequently $N(F)=C(F)^{-1}F$ is an exact realization of the projective class fixing $z_n$ pointwise, at the original rank. If $F,G$ are reference-ray presentations in possibly different width algebras at ranks $d,e$ and agree projectively on a tail, their normalizations agree exactly there and yield a common exact representative at ranks $d,e$. No bound on local support, number of projective values, or multiplicative order of the phases is assumed.
\end{theorem}
\begin{proof}
A rank-zero operand is eventually $A\otimes I$ for fixed $A$ by induction on syntax. Its reference phase is eventually constant and effective at every length; realize a constant scalar and correct early lengths by Lemma~\ref{lem:width-table}. Multiply scalarizations under composition. For $\Br[A,B]$ use $\Br[C(A),C(A)]$, since the reference selects the zero branch; no condition on $B$ is needed. In a recursion of width $s$, threshold $b$ and active set $C$, the reference phase equals $\alpha_g(n)$ on the base and
\[
 \alpha_h(n)\alpha_p(n)\alpha_F(n-s)^{\mathbf1_{\{0^s\in C\}}}
 \quad(n>b).
\]
Recurse with the same width and threshold on the scalarized operands, activating all prefixes if $0^s\in C$ and none otherwise. Induction on length gives exactly the reference phase and preserves rank. If $F_n=\lambda_nG_n$, then $\alpha_F(n)=\lambda_n\alpha_G(n)$, so normalization cancels the discrepancy. Finite tables align early lengths.
\end{proof}
For example, $\operatorname{diag}(1,\beta)^{\otimes n}$ with an effective non-torsion phase $\beta$ has a diagonal reference-ray presentation of rank one at any permitted width $s$: use the fixed prefix product on $s$ bits, full active set, and the corresponding finite base. It has growing support and $n+1$ distinct eigenvalues, beyond the fixed local hypothesis of Theorem~\ref{thm:finite-local-alignment}. Reference-ray invariance of a denotation alone does not justify scalarizing its arbitrary syntax; two operands may leave the ray individually while their product fixes it.

\begin{theorem}[Paired reference-sector restriction]\label{thm:paired-sector-restriction}
At even length $2n$ let $J_n$ embed the first lane of interleaved pairs with the second lane initialized to $z_n$. Consider terms presented using fixed gates on a bounded number of complete leading pairs (identity if too short), composition, four-way branching on a complete leading pair (identity for $n=1$), and original recursions of widths $2s\in 2S$ with thresholds $B\ge2s$. Require each fixed gate to preserve the second-lane-zero subspace, each of the four-way branch operands selected by second-lane bit zero to have such a presentation (the other two may be arbitrary), and all three earlier recursion operands to have such a presentation. Each macro expands into the original grammar at the stated rank: a pair branch is two binary branch levels with the requisite finite-length guard. For every such term $K$ of rank at most $d$, the restriction $J_n^*K_{2n}J_n$ is unitary and has a realization in $\mathcal A_S^{[d]}$ at all positive lengths.

In particular, if an adjoint-encoded common exact family from Theorem~\ref{thm:interleaved-adjoint} admits such presentations over $2S$ and $2T$ at respective ranks $d,e$, then its original projective family has a common exact representative over $S,T$ at ranks $d,e$.
\end{theorem}
\begin{proof}
A fixed gate preserving the reference sector restricts to an effective unitary on the first lane (invariance implies surjectivity in finite dimension); use Lemma~\ref{lem:width-table}. Restrictions multiply, and a pair branch restricts to a binary branch on the second-lane-zero operands. For width $2s$ and threshold $B$, use width $s$, threshold $\lfloor B/2\rfloor\ge s$, restricted operands and active set $\{u:\operatorname{interleave}(u,0^s)\in C\}$. At even lengths the base, control and positive suffix match. Induction on syntax and length proves invariance and the restriction formula without increasing rank. If $K_{2n}=U_n\otimes\overline{U_n}$ on a tail, invariance and surjectivity of $U_n$ imply $U_n z_n=\alpha_n z_n$. Both presentations restrict to $\overline{\alpha_n}U_n$; finite tables align early lengths. Final-operator invariance alone does not justify restricting intermediate operands.
\end{proof}

\subsection{A limit on representation-independent decoding}
\begin{proposition}[Continuous adjoint-decoder obstruction]\label{prop:continuous-decoder-obstruction}
Let $N\ge2$ and $\mathcal D_N=\{U\otimes\bar U:U\in U(N)\}$. For every continuous map $L:\mathcal D_N\to U(N)$ there is a unitary $U$ with
\[
 \operatorname{tr}(U^*L(U\otimes\bar U))=0,
 \qquad
 \inf_{|\lambda|=1}\|L(U\otimes\bar U)-\lambda U\|_{\rm op}\ge\sqrt2.
\]
Thus no universal continuous phase-selection decoder exists, even with approximation error uniformly strictly below $\sqrt2$. The bound is sharp for $N=2$.
\end{proposition}
\begin{proof}
Let $f(U)=\operatorname{tr}(U^*L(U\otimes\bar U))$. Then $f(zU)=\bar z f(U)$ for $|z|=1$. If $f$ never vanished, $g=f/|f|$ would be a continuous map to the circle with winding $-1$ on the scalar loop $zI$. This loop is the product of the $N$ coordinate loops, which are based-homotopic by conjugation along paths to permutation matrices. Winding is additive on concatenation, and pointwise multiplication of based loops in a topological group is homotopic to concatenation. Hence its winding is divisible by $N$, a contradiction (cf.~\cite{Gavorova2024}).

At a zero of $f$, for every unit scalar $\lambda$ the squared Frobenius norm of the difference is $2N$ by expansion of the trace; the squared operator norm is at least its average squared singular value, namely two. At $N=2$ the constant decoder $L=I$ achieves projective distance at most $\sqrt2$ everywhere: place the two eigenvalues of $U$ symmetrically about one along their shorter circular arc by a scalar rotation, so their angular distances from one are at most $\pi/2$. This proves sharpness in dimension two only.
\end{proof}
At fixed $N=2^n$, effective rank-zero gates are dense in $U(N)$. The bound excludes a decoder continuous on the closure with uniform error $<\sqrt2$, not a syntax-dependent decoder or channel implementation. It does not prove the obstruction group nontrivial.

\noindent\textit{Remaining alignment problem.} Whether the two-primary obstruction vanishes remains open. A syntactic normal form for invariant reference sectors would suffice in that sector; common-downset membership is a separate question.

\section{Projection primitives and the Fourier obstruction}
\begin{lemma}[Projection erasure]\label{prop:projection-erasure}
Allow the original unnormalized projections $\operatorname{MEAS}[a]_n=|a\rangle\langle a|\otimes I_{2^{n-1}}$, $a\in\{0,1\}$, in the 2020 algebra. For any term $G$ let $G^\circ$ be obtained by replacing every projection by identity, preserving all widths. At every length and vector $v$ for which $\|Gv\|=\|v\|$, one has $G^\circ v=Gv$. In particular any term agreeing projectively with a unitary at sufficiently large lengths has a measurement-free term agreeing there. The width lower bounds and the exact-QFT impossibility thus hold even against the full original projection-allowing algebra.
\end{lemma}
\begin{proof}
Expand at fixed length into compositions and direct sums of contractions. Equality in $\|ABv\|\le\|Bv\|\le\|v\|$ forces equality at both actual inputs; for a direct sum, equality holds componentwise. An orthogonal projection preserving the norm fixes its input. Induction therefore permits erasing every projection in the expansion, equivalently in the original syntax. Apply this to every vector where $G_n$ is projectively unitary; the erased term satisfies the unitary width and field invariants.
\end{proof}

\subsection{Exact inclusion with projections}
The measurement-free order has a short-length refinement when unnormalized projections are included. Denote by $\mathcal P_S$ the algebra allowing the original $\operatorname{MEAS}[a]$ and the widths $S$, with the same well-defined positive-suffix convention $b\ge s$ for a width-$s$ recursion. An inactive recursion (all branches identity) is allowed by the original Scheme IV. The following result is relative to this convention; a different rule for empty suffixes or under-width inputs is not being classified.

\begin{lemma}[Finite projection-containing matrices and length localization]\label{lem:projection-table}
Let $V$ be the matrix of any fixed original term at a fixed positive length $j$. There is a recursion-free term $L_j(V)$ which is identity at lengths $n<j$ and $V\otimes I_{2^{n-j}}$ for $n\ge j$. If $j\ge a\in S$, there is a term of $\mathcal P_S$ which is $V$ at length $j$ and identity at every other length.
\end{lemma}
\begin{proof}
Apply the extension and inactive-recursion localization rules of Lemma~\ref{lem:guarded-eval}.
\end{proof}

\begin{theorem}[Full projection-algebra inclusion order]\label{thm:projection-order}
For nonempty $S,T\subseteq\mathbb N_{>0}$, under the positive-suffix convention,
\[
\boxed{\quad \mathcal P_S\subseteq\mathcal P_T\quad\Longleftrightarrow\quad
\min T\le\min S\ \text{ and }\ (\forall s\in S)(\exists t\in T)\ s\mid t.\quad}
\]
For eventual inclusion (allowing arbitrary finitely many disagreements), with or without length-dependent global phases, the minimum-width condition drops out and the divisibility condition alone is necessary and sufficient. The exact inclusion criterion is unchanged if equality is weakened to equality up to global phases at every length.
\end{theorem}
\begin{proof}
The unitary witnesses and projection erasure (Corollary~\ref{ws:main}, Lemma~\ref{prop:projection-erasure}) give the divisibility condition for all versions of inclusion.

For the remaining exact necessity let $s=\min S<a=\min T$. Use a width-$s$ recursion of threshold $s$ whose base deletes one basis vector at length $s$ (Lemma~\ref{lem:projection-table}), with identity exterior functions and all branches inactive. Its matrix is singular at $s$ and unitary at $s+1$. But at all lengths $1\le n\le a$, every $\mathcal P_T$ term reduces to the same recursion-free term by replacing each recursion by its base operand recursively. In any recursion-free term singularity is upward closed as a property of the positive length: it holds for initial gates and projections, composition of square matrices is singular iff at least one factor is singular, and a binary branch at length $n>1$ is singular iff at least one of its two suffix matrices at length $n-1$ is singular (at length one the branch is identity). Structural induction proves the property. Thus singularity at $s$ and nonsingularity at $s+1\le a$ are impossible, even after multiplication by nonzero constants.

For exact sufficiency put $a=\min T\le\min S$. Induct on the construction history as in Theorem~\ref{thm:width-order}; for a width-$s$ recursion with threshold $b\ge s$, choose $t=ms\in T$ and $B=b+(m-1)s$. The conditional direct-sum expansion is multiplicative for arbitrary square matrices, not just unitaries. Its exterior products are earlier translated terms. A dimension-preserving pre-operator, when that condition excludes annihilating vectors, is necessarily unitary: a fixed-length expansion is a product of unitaries and projections, and invertibility excludes any nonidentity projection. Such operators may be translated exactly by projection erasure and the unitary width theorem, so the pre-operator side condition is preserved. For the new base, first truncate the already translated original base $g$ by a width-$a$ inactive recursion with threshold $a$ and identity exterior operators. This gives $g$ up to length $a$, identity above $a$. Since $a\le s\le b$, it agrees with the desired $F$ at all lengths $\le a$. For every $a<j\le B$, compose the length-localized matrix $F_j$ from Lemma~\ref{lem:projection-table}. The resulting base agrees exactly with $F$ up to $B$. The grouping proof then gives equality at every larger length, without inverses of singular matrices.

For eventual sufficiency induct instead with earlier operands agreeing eventually. For a width-$s$ constructor use the same width $t=ms$, and choose $B$ sufficiently large that every exterior operand in the expansion agrees on all lengths used for $n>B$, that $B\ge b+(m-1)s$, and that $B-t+1\ge a$. If the pre-operator is required to be unitary, translate it exactly by the preceding unitary theorem. The new base may be identity except for the finite length-localized matrices $F_j$ for $B-t+1\le j\le B$. These lengths are all at least $a$. Induction on $n>B$ gives eventual equality, since the first recursive suffix $n-t$ lies in this terminal window or above it. This induction handles composition and branching by increasing the onset of agreement by a finite amount. No phase freedom is needed for sufficiency.
\end{proof}

By Theorems~\ref{thm:width-order} and~\ref{thm:projection-order}, $\mathcal A_{\{2\}}\subsetneq\mathcal A_{\{4\}}$, but $\mathcal P_{\{2\}}$ and $\mathcal P_{\{4\}}$ are incomparable; the former is contained in the latter only modulo finite-length disagreement. Projection algebras are equal exactly when their downsets and minimum widths agree.

\begin{corollary}[Projection-width poset need not be a lattice]\label{cor:projection-no-meet}
For nonempty width sets $S,T$ put $D=\downarrow S\cap\downarrow T$, $M=\max(\min S,\min T)$ and $J=\{j\in D:j\ge M\}$. Within the poset of nonempty-width-generated projection algebras, a common lower bound exists if and only if $J\ne\varnothing$. In that case the greatest lower bound is $\mathcal P_J$. The least upper bound always exists and is $\mathcal P_{S\cup T}$. In particular $\mathcal P_{\{2\}}$ and $\mathcal P_{\{3\}}$ have no lower bound in this poset, despite their nonempty set-theoretic intersection.
\end{corollary}
\begin{proof}
By Theorem~\ref{thm:projection-order}, any common lower bound $\mathcal P_V$ satisfies $\downarrow V\subseteq D$ and $\min V\ge M$, hence $V\subseteq J$. If $J$ is empty no such nonempty $V$ exists. Otherwise $\mathcal P_J$ is itself a lower bound, and any $V\subseteq J$ gives inclusion by the same theorem. A common upper bound must have downset containing $\downarrow S\cup\downarrow T$ and minimum width no greater than $\min(\min S,\min T)$; these are exactly the conditions for containing $\mathcal P_{S\cup T}$. For $S=\{2\}$ and $T=\{3\}$, $D=\{1\}$ and $M=3$. All recursion-free terms nevertheless lie in both algebras. Thus their set-theoretic intersection is not generated by any nonempty set of permitted widths.
\end{proof}

For a nonempty width set $S$, the operator family that deletes one basis vector at precisely one positive length $j$ and is identity at every other length belongs to $\mathcal P_S$ exactly when $j\ge\min S$. Sufficiency is Lemma~\ref{lem:projection-table}; necessity is the short-length upward-singularity argument of Theorem~\ref{thm:projection-order}. Hence for any two such algebras the single-length deletion families in their set-theoretic intersection are exactly those with $j\ge\max(\min S,\min T)$. This is a further exact intersection statement for a specified family, not a classification of all common contraction operators.

\begin{theorem}[Unbounded QFT is not a finite original term]\label{ws:qft}
Let $Q_n$ be the quantum Fourier transform (QFT), the standard Fourier matrix on $2^n$ basis vectors, with entries $2^{-n/2}\exp(2\pi i ab/2^n)$ in the computational basis indexed by $0\le a,b<2^n$. No finite term of the original measurement-free algebra, with any recursion widths, agrees projectively with $Q_n$ on a tail. The same obstruction applies if finitely many additional fixed-size gates are allowed. It also applies to unnormalized linear projection primitives, provided no new length-dependent constants are introduced.
\end{theorem}
\begin{proof}
Fix a term and let $K$ be the field over $\Qrat$ generated by its finitely many gate entries. Composition, branching and recursion preserve $K$. A scalar multiple of $Q_n$ has entry ratio $\zeta_n=\exp(2\pi i/2^n)$, so $\zeta_n\in K$. For a finite transcendence basis $t$ of $K/\Qrat$, put $d=[K:\Qrat(t)]<\infty$. Irreducible rational polynomials remain irreducible over $\Qrat(t)$, so every algebraic $\alpha\in K$ has degree $[\Qrat(\alpha):\Qrat]=[\Qrat(t,\alpha):\Qrat(t)]\le d$. But $[\Qrat(\zeta_n):\Qrat]=2^{n-1}$ is unbounded. This answers the exact-QFT question in~\cite[Sec.~6.1]{Yamakami2020}, including transcendental gate constants (cf.~\cite{FreedmanWang2007,GilesSelinger2013}).
\end{proof}

\section{Conclusion and open problems}
Open problems include vanishing of the alignment obstruction, common-downset intersections, invariant-sector normal forms, higher-rank transfer invariants and optimal finite-order ranks. For unchanged EQS, extensional Scheme-V elimination, admissible divide-and-conquer recursion and total-language quantum/probabilistic polylogarithmic-time separation remain open. General coherent normal forms, higher-type oracles and unbounded output expansion are not treated.

\appendix
\section{Native width-one compiler and clean transport}\label{app:compiler}
We use the notation, padding and recursion conventions of Section~\ref{sec:oneqrec}. All schematic recursions in this appendix have width and threshold one unless specified otherwise. The following construction proves Theorem~\ref{thm:oneqrec-fbqp}, including its exact machine interface.

\begin{definition}[Suffix stability]\label{def:suffix-stability}
For a formatted subspace $F_m\subseteq(\mathbb C^2)^{\otimes m}$, $m\ge1$, write $U:F_m\to F'_m$ if $U_mF_m\subseteq F'_m$. The action is suffix-stable on $F_m$ if, for every $k\ge0$,
\[
 U_{m+k}|_{F_m\otimes(\mathbb C^2)^{\otimes k}}
   =(U_m|_{F_m})\otimes I_{2^k}.
\]
A format may be specified by basis words and extended to their span. Its length and delimiters are part of the specification, not inferred from an arbitrary appended word. We use the same notation for a family of such subspaces.
\end{definition}
\begin{lemma}[Stability under routing and composition]\label{lem:suffix-stability}
Suffix-stable actions compose on successive formats; their inverses are stable on their images. Fixed local gates are stable on registers containing all their target wires. The update $D$ is stable on $H_j0W$, and $D^{-1}$ on $H_{j+1}W$, including $W$ empty. A fixed prefix control, $\Fx$, and $\Sk$ preserve stability on the respective formatted inputs whenever their operands act on nonempty designated payloads; $\It$ does so when each intermediate payload meets its operand's stability hypothesis.
\end{lemma}
\begin{proof}
Tensor the defining identities with the appended register; compose by substitution and invert on the image. Fixed gates and prefix controls use only specified wires. The $D$ identities hold for arbitrary suffixes. Skip and iteration reduce to operand stability, inducting on the count. Linearity includes entangled references.
\end{proof}

\begin{proposition}[Exact transport of descent clocks]\label{prop:native-clock-transport}
Let $V\in\mathcal A_{\{1\}}$ have recursion depth $e$. For any $r\ge1$, the fixed term $\It^{[r]}[V]$ (syntactic iteration of the constructor, not repeated execution) has depth at most $e+2r$. On a header stack with nonempty payload it acts as
\[
 \It^{[r]}[V](H_{j_1}\cdots H_{j_r}\,|\psi\rangle)
 =H_{j_1}\cdots H_{j_r}\,V^{j_1\cdots j_r}|\psi\rangle.
\]
In particular, encoding $j_i=D_{G_i,s_i}(n_i)$ transports any finite collection of independent reset clocks of Lemma~\ref{lem:clock-geometry} into the original width-one schematic algebra, with no additional quantum operations or allocated qubits. The headers are supplied; the identity holds on their formatted subspaces.
\end{proposition}
\begin{proof}
For $j=0$, $\Sk[V]$ crosses the delimiter and applies $V$; $\It[V]$ is identity. For $j>0$, recursion acts on $H_{j-1}W$, then $\Sk[V]$ adds one application to the same payload. Both preserve headers. Skip and iteration introduce one node each. Induction on $r$ multiplies the exponents and adds two per level; linearity extends to formatted subspaces.
\end{proof}

\begin{corollary}[Uniform preparation of descent headers]\label{cor:clock-preparation}
Let the $r$ arguments $n_i$ be available in binary, and let each fixed inverse $\rho_{G_i,s_i}$ be computable in time polynomial in the bit length and in linear bit space. Put $j_i=D_{G_i,s_i}(n_i)$ and $\ell_i=\lceil\log_2(n_i+2)\rceil$. The counted input of Proposition~\ref{prop:native-clock-transport} can be prepared by a uniform deterministic transducer in time
\[
 O\left(\sum_{i=1}^{r}(1+j_i)\operatorname{poly}(\ell_i)\right)
\]
and $O(\max_i\ell_i)$ additional work space, besides the $\sum_i(j_i+1)$ output header bits and the untouched payload. The transported operator then uses only the original unitary basis on that padded register.
\end{corollary}
\begin{proof}
For each $i$, copy $n_i$ to a private counter, emit a zero and replace the counter by $\rho_{G_i,s_i}$ until it reaches zero, then emit a delimiter one. Strict descent and the definition of $j_i$ give exactly $j_i$ zeros. The inverse routine and counter fit in $O(\ell_i)$ space; repeat sequentially for the fixed number of clocks. Attach the headers without accessing the payload. The bound charges header preparation only; it neither gives coherent internal preparation nor bounds the subsequent circuit time.
\end{proof}

\begin{lemma}[Cross, bubble, after, map, rotate]\label{oq:lem:lib}
For fixed widths $w,b\ge1$ the following are terms in $\A$.
\begin{align*}
\Cr_w(H_j u W)&=uH_jW,\\
\Bu_{w,b}(H_j uB_1\cdots B_jW)&=H_j B_1\cdots B_juW,\\
\Af_b[V](H_jB_1\cdots B_jW)&=H_jB_1\cdots B_jV(W),\\
\operatorname{Map}_b[V](H_j B_1\cdots B_jW)&=H_j V(B_1)\cdots V(B_j)W.
\end{align*}
Here $|u|=w$, $|B_i|=b$. In After, $W$ has positive length $m$ and $V$ is suffix-stable on its format; in Map, $V$ is a fixed $b$-bit gate extended by identity. Also $\Rt_b=\Br[\Bu_{b,b},\id]$ rotates the $j$ counted blocks left for $j\ge1$ and fixes the zero-block register. Cross, Bubble, Map and Rot are suffix-stable on their displayed formats; After is suffix-stable on its displayed format under the stated hypothesis.
\end{lemma}
\begin{proof}
For $\Cr_w$ use $\Q_1[\id,C,\id\mid \Cr_w,\id]$, where $C$ sends the leading $(w+1)$-bit word $au$ to $ua$. On the delimiter branch $1uW$ it produces $u1W$; on a zero branch it moves the restored zero past $u$. This proves the first identity. For the remaining identities let $\rho$ be the fixed swap of adjacent blocks of widths $w,b$. The exact recursion operands are
\begin{align*}
\Bu_{w,b}&=\Q_1[\id,\Br[\Sk[\Rp_b],\id],\Br[\Sk[\Rm_b\rho],\id]\mid \Bu_{w,b},\id],\\
\Af_b[V]&=\Q_1[\id,\Br[\Sk[\Rp_b],V],\Br[\Sk[\Rm_b],\id]\mid \Af_b[V],\id],\\
\operatorname{Map}_b[V]&=\Q_1[\id,\Br[\Sk[V\Rp_b],\id],\Br[\Sk[\Rm_b],\id]\mid \operatorname{Map}_b[V],\id].
\end{align*}
At a zero branch the pre-operation parks $B_1$ at the end (Bubble first exchanges $u$ and $B_1$); the post-operation restores the parked block after the smaller header (Map then applies the fixed gate to it). At the delimiter Bubble and Map do nothing. Induction gives their displayed identities for arbitrary $W$, hence stability. Rot is Bubble under the first zero with $u=B_1$.

For After, let $P$ be the permutation $B_1\cdots B_jW\mapsto WB_1\cdots B_j$ at the designated lengths. The recursion parks the blocks in order $B_1,\ldots,B_j$; its delimiter branch applies $V_{m+jb}$, and the unwind restores the blocks in reverse parking order. Identifying the fixed header sector with its data register, its exact action, even for a general $V$, is
\[
 \Af_b[V]|_{H_j}=P^{-1}V_{m+jb}P.
\]
Suffix stability gives $V_{m+jb}|_{F_m\otimes(\mathbb C^2)^{\otimes jb}}=(V_m|_{F_m})\otimes I$, yielding the third identity; for fixed $j$ this equality at $k=jb$ is also necessary. If a suffix of length $k$ is appended to $W$, the parked register has length $jb$ after that suffix; stability for $k+jb$ gives the same action on $W$ with both registers untouched. Hence After itself is stable. Merely leaving the parked bits unchanged would not suffice: $V_1=X$, $V_m=I$ for $m>1$ is a finite-table counterexample for $j=b=m=1$.
\end{proof}

Two routing conventions are useful and denote finite term compositions, not additional schemes. To act on $H_j q B^jW$ after the fixed $\ell$-bit register $q$ and the counted block register, use
\[\operatorname{Tail}_{\ell,b}[V]=\Cr_\ell^{-1}\Fx_\ell[\Af_b[V]]\Cr_\ell.\]
To carry a fixed-width register $u$ through a header, apply
\[\operatorname{CH}_w[V]=\Cr_w\Sk[V]\Cr_w^{-1}\quad\text{on }uH_j W.\]
It acts as $uH_jW\mapsto u'H_jW'$ if $V(uW)=u'W'$. To carry $u$ through $H_jB^j$, use
\[\operatorname{CB}_{w,b}[V]=\Cr_w\Bu_{w,b}^{-1}\Af_b[V]\Bu_{w,b}\Cr_w^{-1}.\]
It acts on $uW$ and preserves $H_jB^j$ provided $V$ is suffix-stable on the format of $uW$. Tail similarly requires stability on $W$. CH requires stability on $uW$ to preserve an arbitrary appended suffix; its unextended routing identity follows directly from Cross and Skip. Carrying past a fixed-width word uses a fixed swap, $\Fx$, and the inverse swap. Each wrapper preserves stability on its indicated format by Lemmas~\ref{lem:suffix-stability} and~\ref{oq:lem:lib}; their inverses preserve stability on the images. To act on a counted register following a fixed state under a header, conjugate $\Fx_\ell[V]$ by $\Cr_\ell$.

\begin{lemma}[Prefix deinterleaving]\label{oq:lem:deint}
For $1\le a<b$, the suffix-preserving term $\operatorname{Deint}_{a,b}$ constructed below has action \[H_j X_1Y_1\cdots X_jY_jW\longmapsto H_j X_1\cdots X_jY_1\cdots Y_jW,\] where $|X_i|=a$ and $|Y_i|=b-a$.
\end{lemma}
\begin{proof}
Use $p=\Br[\Sk[\Rm_b],\id]$ and the zero-branch-only recursion. Put
\[h=\Br[\Sk[\Rp_a]\Bu_{b-a,a}\Sk[\Rm_a]\Sk[\Rp_b],\id].\]
On unwind restore $X_1Y_1$, park $X_1$, bubble $Y_1$ through the $j-1$ already grouped $X$ blocks, restore $X_1$. On the delimiter branch do nothing. This proves the identity by induction.
\end{proof}

\subsection{Bounded machine evolution and reversible halting}
\paragraph{External machine premise.} For $f\in\FBQP$ and $0<\epsilon<1/2$, amplify a uniform circuit to produce $\widetilde{f(x)}$ zero-padded to polynomial capacity $M(n)\ge2|f(x)|+2$ with probability $\ge1-\epsilon/2$. Retain length and defer measurements using flags over a fixed algebraic universal basis. Let $\Gamma$ be the tape alphabet including $0,1,\blank$ and $\chi:\Gamma\to\{0,1\}$ the output-track bit ($\chi(\blank)=\chi(0)=0$, $\chi(1)=1$). Input and output tracks may coincide. By synchronization~\cite[Theorem~4.3]{BV1997}, prepare the gate list, capacity and scan bounds from length, preserving input. For each gate on $k$ bits, reversibly move them to the first $k$ cells of a blank scratch track, marking vacancies, and restore them afterward. These fixed-output-length permutations on framed tapes take polynomial scan time; the other data, list and addresses remain untouched (\cite[Sec.~7, p.~1449]{BV1997}). Routing and timing depend only on length.

Apply the stationary transformation machine~\cite[Definition~6.9 and Theorem~6.11]{BV1997} on the $k$-cell scratch track, leaving other tracks untouched~\cite[Lemmas~4.4--4.5]{BV1997}. It halts at a common time with transformed data tensor a fixed auxiliary basis string, independent of the data, hence also acts correctly with an entangled reference. For $L(n)$ gates choose error $\eta=\epsilon/(4\max(1,L(n)))$ per gate. Supply tolerance $\eta/4$ and a matrix description within $\min(1,c_k)\eta/8$ of the desired gate, where $c_k$ is the closeness factor of Theorem~6.11; two errors of $\eta/4$ leave room within $\eta$. Since $k$ is fixed, description length, time and tape are polynomial. Reversibly move retained auxiliary basis contents into fresh log segments before each invocation; total space is polynomial and the log is independent of quantum data. Telescoping bounds operator error by $\epsilon/4$ and probability error by $\epsilon/2$. Looping and dovetailing~\cite[Lemmas~4.8, 4.9, 4.13]{BV1997} compose these operations with stationary output routing to consecutive cells from the origin. The resulting single polynomial-time normal-form QTM has basis start and final states, head at the origin at halt and simultaneous halting on each input's branches; work may remain. The circuit is not supplied in the padded input.

Apply the five-step unidirection simulation~\cite[Lemma~5.5]{BV1997}. It preserves tape and head at simulated endpoints and basis start and final states (stage zero). The two intermediate direction subspaces are orthogonal by well-formedness. Choose fixed maximal independent subsets of effective spanning vectors, orthonormalize and extend in the complement; denominators are fixed nonzero constants, so no uniform zero test is needed. No intermediate stage signals halt. The normal-form swap of outgoing transitions at $(q_f,0)$ and $(q_0,4)$ preserves first-halt computation and outgoing-state displacement labels. The native interface also permits stationary displacement. Only a polynomial bound $T(n)$ and simultaneous halting on each input are needed.

\paragraph{Exact native simulation.} Let $q,q'$ range over the finite states, $a,a'\in\Gamma$, and $\delta(q,a;q',a',d)$ be the transition amplitude with head displacement $d\in\{-1,0,1\}$. Unidirectionality assigns a displacement $d(q')$ to each incoming state. The local matrix
\[W\ket{q,a}=\sum_{q',a'}\delta(q,a;q',a',d(q'))\ket{q',a'}\]
is square and unitary: its column inner products equal those of one-step images of configurations with the same head position and tape outside the scanned cell, since the output state determines the displacement. Well-formedness gives $W^*W=I$. Choose state width $\ell\ge1$ with start state $0^\ell$, symbol width $a\ge2$ with $2^{a-1}\ge |\Gamma|$, and codes blank $0^a$, $0=10^{a-1}$, $1=10^{a-2}1$. Encode each remaining symbol injectively with last bit $\chi$ of that symbol. There are enough distinct codes with each last bit even after reserving the three displayed codes. A full cell has width $b=a+1$: one origin-marker bit followed by the symbol. Extend $W$ by identity on the complement of the valid state-symbol subspace and by identity on the origin bit (using fixed wire permutations). Fix directions on invalid states arbitrarily, say zero.

Let $R=2T+3$. Encode $(q,h,\tau)$ by state $q$ and a head-centred ring: relative cell $r$ contains the origin marker $\mathbf1_{h+r\equiv0\pmod R}$ and the symbol at the unique position of $[-T,T]$ congruent to $h+r$, or blank if absent. Initially encode $x$ in the first $n\le T$ cells, with blanks elsewhere and the first origin marker set. All reachable heads and nonblank cells lie in $[-T,T]$, whose residues are distinct. The marker recovers $h$ and hence the tape, making the encoding isometric on the reachable span. Apply $W$ to state and head symbol, then rotate the ring left for $d(q')=+1$, right for $-1$, or leave it for zero. A transition writing $a'$ gives the encoding of $(q',h+d(q'),\tau[h\leftarrow a'])$ with the same amplitude. At stationary halt the ring starts at cell zero.

\begin{lemma}[Fresh-history stopping]\label{oq:lem:stop}
There is a single term $\mathcal S$ (depending only on the machine, not $T$ or $R$) acting on
\[H_T h_1\cdots h_T H_R qB_1\cdots B_R Z\]
for $T,R\ge1$, with the following action if $h_1=0$: rotate the history one place left, write the predicate ``the pre-step state is final'' into the last history bit, and apply the above machine step if that bit is zero; otherwise leave the machine unchanged. The indicated trailing word $Z$ is untouched. The term is globally unitary on every input, including malformed encodings.
\end{lemma}
\begin{proof}
Let $U$ act on $h H_R qB^R Z$, where $B^R=B_1\cdots B_R$. Move $h$ across $H_R$ by $\Cr_1^{-1}$; under $\Sk$ apply the local XOR of the final-state predicate into $h$, then the fixed unitary $W$ on $q,B_1$ controlled on $h=0$. Move the combined $(h,q)$ to before $H_R$ by $\Cr_{\ell+1}$. A finite branch tree on these $\ell+1$ bits selects $\Rt_b$, $\Rt_b^{-1}$ or identity on the counted ring: choose identity if $h=1$, otherwise the rotation determined by $q$. Restore $(h,q)$ by $\Cr_{\ell+1}^{-1}$ and move $h$ to before $H_R$ by $\Cr_1$. The local gates have all their targets within $h,q,B_1$ ($R\ge1$), and the selected rotations act on exactly the $R$ counted cells. Lemmas~\ref{lem:suffix-stability} and~\ref{oq:lem:lib} give stability of $U$ on $hH_RqB^R$, for arbitrary initial $h$ and valid or unused state codes. Thus After may park the other history bits after this payload. Now set
\[\mathcal S=\Br[\Af_1[U],\id]\Rt_1.\]
The first operation (rightmost) moves $h_1$ to the back of the history. The remaining header after its first zero is $H_{T-1}$, so After skips exactly the other $T-1$ bits and applies $U$ to $h_1$ and the machine. Global unitarity follows from the syntactic construction, not an assertion about malformed configurations.
\end{proof}
A separate outer $H_T$ drives $\It[\mathcal S]$. The following invariant proves that no absorbing unitary halt state or history erasure is needed.

\begin{lemma}[Exact stopped-evolution intertwining]\label{oq:lem:intertwine}
Fix $n,T$ with $T\ge n$ bounding the first stationary halting time $\tau(x)$ on each basis input $x$ of length $n$. Let $J_T$ encode the state and tape into the origin-marked head-centred ring of length $2T+3$. Let $\psi_i$ be the machine state after $i$ ordinary transitions for $0\le i\le\tau(x)$. On the formatted history and ring, the $j$-fold stopped term has the exact action
\[
 \mathcal S^j\bigl(H_T0^T H_R J_T\psi_0\bigr)
 =H_T0^{T-j}0^{\min(j,\tau(x))}1^{\max(0,j-\tau(x))}
 H_R J_T\psi_{\min(j,\tau(x))},\quad 0\le j\le T.
\]
Fixed surrounding registers and any untouched reference system are tensor factors of this identity. For superpositions of different basis inputs of the same length the identity is extended linearly (the history may depend on the input if their halt times differ).
\end{lemma}
\begin{proof}
The ring encoding $J_T$ is isometric and intertwines the ordinary transition on reachable configurations by the preceding construction. For $i<\tau(x)$ the state is supported on nonfinal states, so the fresh-history step implements that transition, preserving interference. At $i=\tau(x)$ the entire state is final; the step changes the next fresh zero to one and leaves the ring unchanged. At each $j<T$ the next history bit is unused. Induction gives the identity, including entangled references by linearity.
\end{proof}

\subsection{Native compilation on the original padded input}\label{oq:sec:init}
Choose the polynomial integer upper bound $T(n)$ to dominate $n$, the fixed binary output capacity $M(n)$ plus one, and a sufficiently large fixed constant. Fix $c=2b+6$, $d=\lceil\log_2c\rceil$, $p=cT$, $Z=11p+6$, $R=2T+3$, $z=(11c-2b-5)T-d-\ell-3b$, and $z'=z-n-1$. Write $B^R=B_1\cdots B_R$. The original input is $H_nH_pH_Zx$, with an arbitrary untouched suffix $S$ allowed. We build
\begin{equation}\label{oq:eq:layout}
H_nH_p0^dH_T H_T0^T H_R0^\ell B_1\cdots B_R H_{z'} H_n x S,
\end{equation}
where initially $B_i=0^b$ and later the first $n$ cells contain the coded input. All allocation in this display is a reversible rearrangement of existing bits.

\paragraph{Initializer.} Use the fixed wrappers
\begin{align*}
 A[V]&=\Sk[V], & F[V]&=\Fx_d[V], & B[V]&=\Af_1[V],\\
 Q[V]&=A[A[F[V]]], & L[V]&=A[F[A[B[V]]]], & C[V]&=L[\operatorname{Tail}_{\ell,b}[V]],\\
 \operatorname{Ring}[V]&=\Cr_\ell^{-1}\Fx_\ell[V]\Cr_\ell.
\end{align*}
Let $\operatorname{Sel}$ increment the $d$-bit phase code cyclically on its $c$ valid values, fixing the other values, then apply $D$ on the remaining word when the new code is zero. For loading, let $L_0$ be the fixed gate on a cell $u$ and one raw input bit: flip the first symbol bit of $u$ and swap its last bit with the raw bit. Set
\[
 V_{\rm load}=\operatorname{CH}_b[\Fx_b[\Rt_1]\operatorname{CH}_b[L_0]],\qquad
 K=\operatorname{Ring}[\Br[\Af_b[V_{\rm load}],\id]\Rt_b].
\]
Here $V_{\rm load}$ acts on $uH_{z'}H_nx$ for $n\ge1$: cross the residual header, cross the input header, apply $L_0$, and rotate the counted raw input one place left. The carrier wrappers restore both headers and the counted rotation fixes every appended word. Thus $V_{\rm load}$ is stable on this format. The After in $K$ can park the other $R-1$ cells; $K$ rotates the ring left and loads the old first cell, now last. The old raw bit becomes zero when the cell was blank.

Define $G$ by the following table, executing rows top to bottom and each row left to right. The unchanged prefix after the first row is $H_nH_p0^d$. The last column lists the remaining format, with $\psi_i,\tau(x)$ as in Lemma~\ref{oq:lem:intertwine} and $h$ its final $T$-bit history.
\begin{center}\small
\begin{tabular}{@{}>{\raggedright\arraybackslash}p{.12\linewidth}>{\raggedright\arraybackslash}p{.50\linewidth}p{.31\linewidth}@{}}
\toprule
Stage & Native terms in execution order & Resulting format\\\midrule
Short clock & $A[A[F[X]]],\ A[\It[\operatorname{Sel}]]$ & $H_TH_{Z-d-T-1}xS$\\
History & $Q[A[X]],\ Q[\It[D]]$ & $H_TH_T0^T H_{Z-d-3T-2}xS$\\
Ring clock & $Q[A[B[X]]],\ Q[\It[B[D]]],\ Q[\It[B[D]]],\ Q[A[B[D^3]]]$ & $H_TH_T0^TH_R0^\ell B^RH_zxS$\\
Input clock & $A[C[D^{-1}]],\ \It[C[D^{-1}]],\ A[C[A[X]]],\ \It[C[A[D]]]$ & $H_TH_T0^TH_R0^\ell B^R H_{z'}H_nxS$\\
Load/mark & $\It[L[K]],\ \It[L[\operatorname{Ring}[\Rt_b^{-1}]]],\ A[L[A[\Fx_\ell[X]]]]$ & $H_TH_T0^TH_R J_T\psi_0 H_{z'}H_n0^nS$\\
Run & $Q[\It[\mathcal S]]$ & $H_TH_T h H_R J_T\psi_{\tau(x)}H_{z'}H_n0^nS$\\\bottomrule
\end{tabular}
\end{center}

\paragraph{Verification of the formats.} In the first row $d$ zeros hold the phase code, and the next zero becomes $H_0$. Each of the $T=p/c$ complete phase cycles consumes one further zero via $D$; the phase returns to zero. The second row seeds a second header and consumes $T$ zeros, after which the next $T$ zeros constitute the history. The third row applies $X$ and then $2T+3$ copies of $D$ past this history, creating $H_R$. The next $\ell+bR$ zeros are the state and ring. The total number of consumed zeros is
\[
 d+2(T+1)+T+(R+1)+\ell+bR=(2b+5)T+d+\ell+3b+6.
\]
The residual pool is therefore $H_z$. Choose the fixed lower bound on $T$ so that $z\ge n+1$. The fourth row shortens $H_z$ by $n+1$ using $D^{-1}$, then converts the resulting $n+1$ zeros into $H_n$ by $X$ and $n$ copies of $D$. Every $X,D,D^{-1}$ in these four rows is applied with its required nonempty format and enough remaining pool zeros. They are stable there by Lemma~\ref{lem:suffix-stability}; the fixed wrappers, including $B$ and Tail, preserve stability by Lemma~\ref{oq:lem:lib}. In particular their operands do not depend on any parked history, ring cells or trailing bits. The first four rows leave $xS$ untouched.

In the fifth row $n$ applications of $K$ fill the first $n$ original cells in order and zero the raw register. The subsequent $n$ right rotations restore the ring position and $X$ marks its first cell. When $n=0$ both counted loops are empty. Stability of $K$ follows from that of $V_{\rm load}$ and the routing lemma; the marking and rotation operations are stable on the displayed formats. All payloads of After are nonempty (the loading operand includes one cell and both headers). The last row executes the stable stopped step $T$ times. The history and configuration identity follows from Lemma~\ref{oq:lem:intertwine}. This proves that $G$ is suffix-stable on $\phi^p(x)$ and its inverse on the image, including arbitrary superpositions of valid inputs of a fixed length. No additional qubits are allocated.

\subsection{Output extraction and soundness}
In each full cell move the last symbol bit to the first position by a fixed wire permutation. By construction this bit is $\chi$ of the represented tape symbol, including symbols with additional work-track information. On the counted ring, apply $\operatorname{Map}_b$ of this permutation and use Lemma~\ref{oq:lem:deint} with extracted width $1$. Use the conjugation through the fixed state register and the fixed prefix wrappers of Section~\ref{oq:sec:init}. Since the machine writes the already encoded fixed-capacity output, the leading $R$ extracted bits start with $\widetilde{f(x)}$ on successful outcomes; the remaining cell bits and subsequent cells may be arbitrary.

To move these bits to the very front, define $P=\Q_1[\id,\Rm_1,\id\mid P,\id]$. Induction gives $P(H_jW)=W1 0^j$; the last-bit delimiter is not discarded. This avoids the noninjective map that would put an unreversed variable header after arbitrary data. Apply $P$ to peel $H_n,H_p$, move the $d$ phase bits to the end by $\Rm_d$, and peel the first $H_T$. Under the second $H_T$ apply $\It[\Rm_1]$ to move its $T$ history bits to the end; then peel this header. Under $H_R$ apply $\Sk[\Rm_\ell]$ to move the state bits to the end; peel $H_R$. Explicitly, with $M_b$ the fixed permutation moving the last cell bit first, $\operatorname{Deint}_{1,b}$ the term in Lemma~\ref{oq:lem:deint}, and the wrappers $Q,B,\operatorname{Ring}$ above, the initial extraction term is $Q[A[B[\operatorname{Ring}[\operatorname{Deint}_{1,b}\operatorname{Map}_b[M_b]]]]]$. Follow it, in execution order, by $P,P,\Rm_d,P,\It[\Rm_1],P,\Sk[\Rm_\ell],P$. These final garbage-routing operations are used on the full original input, not claimed to preserve arbitrary appended suffixes. Each is a native unitary.

The constructor rules give the following sufficient rank upper bounds, not minimum ranks (operand rank $r$):
\[
\begin{array}{c|c@{\qquad}c|c}
\text{Macro} & \text{Rank} & \text{Construction} & \text{Rank}\\ \hline
\Cr_w & 1 & \operatorname{Deint}_{a,b} & 4\\
\Bu,\Rt,\operatorname{Map}\text{ (fixed gate)} &3&\text{stopped step},V_{\rm load}&4\\
\Af_b[V]&1+\max(2,r)&K&5\\
\Sk[V],\It[V]&r+1,\ r+2&\text{initializer rows 1--6}&4,5,7,8,10,8\\
\Fx_w\text{ (fixed }w\text{)}&r& E_{\rm out}&8
\end{array}
\]
The row maxima use the displayed wrappers $A,F,B,Q,L,C$; guarded deinterleaving dominates the extraction peeling ranks. Composition gives rank at most ten, independent of the machine and padding.

\begin{theorem}[Native machine compilation interface]\label{thm:native-machine-interface}
Fix a finite state set and tape alphabet with effective transition constants, a basis start state and a designated basis final state. Suppose the local state--symbol matrix $W$ is unitary, and the head displacement is a fixed function of the outgoing state, with values in $\{-1,0,1\}$. Use a fixed binary input coding and an output-track map $\chi$ as above. Suppose each length-$n$ basis input first halts simultaneously on all its branches at time $\tau(x)\le T(n)$ with head at the origin. Let the fixed-capacity output occupy at most $M(n)\le T(n)-1$ consecutive cells from the origin, and let $T$ be an integer-valued polynomial upper bound chosen as in Section~\ref{oq:sec:init}.

There is one width-one term $G$ of rank at most ten, independent of $n$ and $x$ (but depending on the finite machine and coding), such that on the original padded input $\phi^p(x)$, with $p=cT$, its exact final state is the last-row format in the initializer table: the simulated configuration is $J_T\psi_{\tau(x)}$, with the fresh-history queue from Lemma~\ref{oq:lem:intertwine}. The identity holds with an untouched appended payload, and by linearity with an arbitrary reference on the formatted subspace of each fixed input length. There is a rank-at-most-eight extraction permutation $E_{\rm out}$ on the entire padded register for which, for every binary prefix $y$ of length at most $M(n)$, the probability of observing $y$ on the leading wires of $E_{\rm out}G\ket{\phi^p(x)}$ equals its output-track probability on the halted machine state. No additional qubits are allocated by either term. Neither statement assumes that the halted state is a basis configuration.
\end{theorem}
\begin{proof}
Lemma~\ref{lem:width-table} supplies the fixed local matrix, and Lemmas~\ref{oq:lem:stop},~\ref{oq:lem:intertwine} prove stopped evolution with fresh history. The six initializer rows construct the starting format and perform $T$ transitions, with the stability and ranks established above. The extraction permutation pulls each leading-prefix projector back to the output-track projector on encoded halted configurations, first on basis configurations and hence on their span. Isometry preserves the projected norm. QTM normalization is used only to supply machines satisfying the hypotheses, not in this simulation proof.
\end{proof}

Apply Theorem~\ref{thm:native-machine-interface} to the stationary machine above. Extraction preserves its output probability; a fixed SWAP gives direct one-bit readout. For soundness, expand the witness with $\epsilon=1/6$ by Lemma~\ref{lem:circuit-expansion} on the computable polynomial-length padded input. Approximate to operator error at most $1/12$ up to phase on data and clean auxiliaries. For unit vectors $v,w$ and a projector $P$, $|\|Pv\|^2-\|Pw\|^2|\le2\|v-w\|$; hence success remains at least $2/3$. Distinct prefix-free codes have orthogonal prefix projectors. Measure and parse aligned pairs $00,01$ until $11$, rejecting $10$ or a missing delimiter. This proves Theorem~\ref{thm:oneqrec-fbqp}, also for decisions.

\noindent\textit{Rank, padding and running time.} At input length $n$, padded length is $N=2n+12p(n)+9$. Lemma~\ref{lem:circuit-expansion} bounds rank-ten gate count by $O((N+1)^{11})$ before universal approximation. For $p(n)=O((n+1)^k)$, $k\ge1$, this is $O((n+1)^{11k})$ with term-dependent constants. Neither $k$ nor the padding polynomial is uniform over machines.

\subsection{Clean realization of SR evaluations}\label{sec:clean-clock}
The compiler also prepares bounded unary clocks with clean work and input-length-dependent capacity.

\begin{lemma}[Counted coherent readout]\label{lem:clean-readout}
Let $G$ be the initializer and stopped simulator of Section~\ref{oq:sec:init}, before the final garbage routing. Let $m\le R$ be positive. There is a fixed width-one term $C$ acting on
\[
 H_{m+1}H_{m+1}u\,\mathsf{Conf}\,|\psi\rangle,
 \qquad |u|=m,
\]
where $\mathsf{Conf}$ has the layout (\ref{oq:eq:layout}) with its history, state and cell registers arbitrary and with the displayed header lengths, such that it XORs the last symbol bits of the first $m$ ring cells into $u$ in order and preserves the entire simulated register and $|\psi\rangle$. It acts linearly on superpositions of all these data, including an external reference.
\end{lemma}
\begin{proof}
Use the carrier wrappers $\operatorname{CH}_w,\operatorname{CB}_{w,b}$ and $\operatorname{Ring}$ already proved in Lemma~\ref{oq:lem:lib} and its following paragraph. For a bit $v$ immediately preceding a formatted simulated register, let $L_0$ be the fixed gate XORing the last bit of the first ring cell into $v$ on $v q B_1\cdots$. Define
\[
 U_0=\Fx_1[\operatorname{Ring}[\Rt_b]]\operatorname{CH}_1[L_0].
\]
This acts on $vH_RqB^R\cdots$ by the desired XOR followed by a left rotation of the ring. Let $\operatorname{CF}_d[V]$ be the fixed carrier wrapper moving $v$ across $d$ bits, applying $V$ to $v$ and the subsequent payload, then restoring the $d$ bits (a fixed block swap, $\Fx_d[V]$, and the inverse swap). The complete carrier operation, on $v\mathsf{Conf}$, is
\[
 U=\operatorname{CH}_1[\operatorname{CH}_1[\operatorname{CF}_d[
 \operatorname{CH}_1[\operatorname{CB}_{1,1}[U_0]]]]].
\]
Its successive wrappers bypass $H_n,H_p,0^d,H_T$ and then $H_T$ with its $T$ history bits. The local gate $L_0$ has all targets within $v,q,B_1$ and $\operatorname{Ring}[\Rt_b]$ acts on exactly the counted ring; thus $U_0$ is stable on $vH_RqB^R$. In particular CB may park the $T$ history bits after that format. The other carrier wrappers preserve stability on the successive formats. Hence $U$ is stable on $v\mathsf{Conf}$, allowing After in $K_c$ to park the other buffer bits. The effective count of either leading $H_{m+1}$ is $m$ under its first zero. Set
\begin{align*}
 K_c&=\Br[\Br[\Af_1[U],\id]\Rt_1,\id],\\
 C&=\Br[\It[\Br[\Af_1[Q[A[B[\operatorname{Ring}[\Rt_b^{-1}]]]]],\id]],\id]\,
     \Br[\It[K_c],\id],
\end{align*}
where $A,B,Q$ are the fixed wrappers of the initializer. Each iteration of $K_c$ rotates the $m$ buffer bits once left, skips the first $m-1$ bits and acts on the parked old first bit by $U$. After $m$ iterations the buffer is in its original order with the desired XOR, while the ring is rotated left by $m$. The first (leftmost) factor performs $m$ counted inverse ring rotations: $\operatorname{Ring}[\Rt_b^{-1}]$ is stable on $H_RqB^R$, so its enclosing $B$, $A$, $Q$ wrappers are stable on the corresponding simulation format. After may therefore skip the buffer. This restores the ring without changing any copied bit. The displayed formulas are compositions of earlier terms, and their actions on arbitrary formatted basis states give the claimed linear identity. No clean-state assumption on the buffer is needed for this identity.
\end{proof}

\begin{proposition}[Clean reversible realization of SR evaluations]\label{prop:clean-sr-lift}
Let $h:\{0,1\}^*\to\{0,1\}^*$ be a deterministic polynomial-time function with fixed positive polynomial output length $m(n)$ at input length $n$ (a shorter bounded output may be encoded and padded). With a suitable polynomial $p$ there is a fixed width-one term of rank at most twelve acting exactly by
\[
 H_{m+1}H_{m+1}u\,\phi^p(x)|\psi\rangle
 \longmapsto H_{m+1}H_{m+1}(u\oplus h(x))\,\phi^p(x)|\psi\rangle,
 \qquad |u|=m=m(n).
\]
All work is restored on coherent $x,u$ and entangled payloads. Given the evaluator and explicit polynomial time/output bounds, translation and padding are effective. Thus any fixed SR evaluation with polynomial-time binary evaluation admits this clean lift, without injectivity. For a single bounded recursion, polynomial-bit arguments and values, polynomially many descent stages and polynomial-time inverse, base and update suffice: store descent and evaluate updates in reverse order. Nested terms require polynomial total evaluation work. No bound for the full classical SR hierarchy is asserted.
\end{proposition}
\begin{proof}
Use stationary reversible simulation~\cite[Theorem~4.3]{BV1997}, retaining input/work and routing the $m$ output bits to the output track; bound time by at least $m$. Unidirectional normalization~\cite[Lemma~5.5]{BV1997} and Section~\ref{oq:sec:init} yield native $G$ halting with $h(x)$. Put $\mathcal G=\Sk[\Br[\Af_1[G],\id]]$ and apply $\mathcal G^{-1}C\mathcal G$ with Lemma~\ref{lem:clean-readout}'s $C$. Stability preserves skipped bits and holds on the inverse image; copying by XOR leaves the simulated state unchanged, so inversion restores it. Linearity gives coherence. Readout rank is at most ten, the two wrappers around $G$ give at most twelve; inverse and composition preserve maxima.
\end{proof}

\begin{corollary}[Computable padding and clean coherent clocks]
\qsrpart{a}{cor:computable-padding}
Every total computable binary string function $h$ has an exact clean width-one XOR realization with computable, input-length-dependent padding, and a fixed prefix-free encoding of its output. In particular every total computable member of any SR row admits this lift. If a deterministic evaluator takes at most $t(n)$ steps on length-$n$ inputs, the padded input and output buffers may have total size polynomial in $n+t(n)+1$, and the native term has polynomial circuit size in that padded length. The term is fixed for the evaluator, not for individual inputs or their answers. This is a resource-parametric simulation, not preservation of the weak-basis inclusion order or construction rank of the source.
\qsrpart{b}{cor:internal-clock}
Fix an increasing generator $G_0$ and stride $s$ whose inverse is computable in polynomial time in the argument bit length. Suppose a fixed polynomial $m(n)\ge2$ satisfies $D_{G_0,s}(y)\le m(n)-2$ for every $n$-bit integer $y$ (leading zeros allowed). There is a polynomial $p$ and a single width-one schematic unitary term of rank at most twelve which, for every $n$-bit word $x$ representing $y$, performs exactly
\begin{multline*}
 H_{m+1}H_{m+1}0^m\phi^p(x)|\psi\rangle\\
 \longmapsto H_{m+1}H_{m+1}H_jH_{m-j-2}\phi^p(x)|\psi\rangle,
 \qquad j=D_{G_0,s}(y),\quad m=m(n).
\end{multline*}
The map extends coherently to superpositions of $n$-bit inputs, even with $j$ varying across basis components and with arbitrary entanglement to the payload or a reference. All work is restored and the register length is unchanged. The initial capacity and padding depend only on $n$, not $j$.
\qsrpart{c}{cor:coherent-clock-fold}
Under the hypotheses of Corollary~\ref{cor:internal-clock}, fix a unitary $V$ on $w$ qubits, where $w$ and its effective matrix entries are fixed. Let $L(n)=|\phi^p(x)|$, depending only on $n$. With the additional unary length header $H_L$, one width-one term of rank at most thirteen acts on the formatted subspace as
\begin{multline*}
 H_{m+1}H_{m+1}0^mH_L\phi^p(x)|\psi\rangle\\
 \longmapsto H_{m+1}H_{m+1}0^mH_L\phi^p(x)V^{D_{G_0,s}(y)}|\psi\rangle,
\end{multline*}
where $V$ acts on the first $w$ bits of a payload of length at least $w$. In particular all supplied clock buffers and simulator work are returned exactly, including for coherent binary clock arguments.
\end{corollary}
\begin{proof}
\par\noindent\textup{(a)} Compute $t(n)\ge n+1$ by running the evaluator on all $2^n$ inputs. Encode $z=H_nxH_t$ of length $N=2n+t+2$. A fixed polynomial-time routine checks format, simulates for $N$ steps (up to a fixed tape-model constant), and returns the prefix-free output padded to $2N+2$, or a default code if invalid or timed out. Apply Proposition~\ref{prop:clean-sr-lift}. Capacity depends only on $n$; circuit expansion is polynomial in padded length, with $t$ preparation charged separately. This is compute--copy--uncompute~\cite{Bennett1973}.
\par\noindent\textup{(b)} Iterating the inverse computes $j$ in polynomial time in $n$ under the capacity and efficiency hypotheses. The fixed-length output $h(x)=H_jH_{m-j-2}$ has exactly $m$ bits. Apply Proposition~\ref{prop:clean-sr-lift} with the zero buffer. Its clean transport term $P_c=\mathcal G^{-1}C\mathcal G$ gives the asserted identity. The construction uses supplied polynomial capacity and length metadata; it neither allocates qubits nor prepares exponentially long unary headers in polynomial binary-input space.
\par\noindent\textup{(c)} Modify the clean preparation $P_c$ of Corollary~\ref{cor:internal-clock} by crossing the unchanged $H_L$ before applying the simulator: replace $G$ inside $\mathcal G$ by $\Sk[G]$. In the copy term replace $U$ by $\operatorname{CH}_1[U]$, and in its inverse-rotation factor replace $Q[A[B[\operatorname{Ring}[\Rt_b^{-1}]]]]$ by its $\Sk$ wrapper. These changes preserve $H_L$ and give the same clean output header pair before $H_L\phi^p(x)$. The term
\[
 P_c^{-1}\,\Sk[\Sk[\It[\Sk[\Af_1[V]]]]]\,P_c
\]
acts as required: $\It$ reads $H_j$, skips $H_{m-j-2}$ and uses $\Af_1$ to bypass the $L$ simulator bits. The fixed local $V$ is stable on payloads of length at least $w$, so parked data and headers survive. Carrier wrappers make $P_c$ stable and its inverse stable on the prepared format; inversion restores the clock after the payload changes. Linearity covers entanglement. Extra Skip and carrier wrappers increase preparation rank by at most one; the counted local update has rank at most eight. Thus rank is at most thirteen with polynomial padding determined by $n$.
\end{proof}
The fixed-local restriction is essential: arbitrary length-dependent operators may act on parked registers.

\begin{proposition}[What the two grammars preserve]\label{prop:grammar-interface}
For a fixed target term $V$ of rank $e$, let $j_i=D_{G_i,s_i}(n_i)$, with $r$ fixed. On the formatted subspace with supplied headers $H_{j_1}\cdots H_{j_r}$, a nest of $r$ fixed-workspace reset clocks with update $V$ has a width-one realization of rank at most $e+2r$. Its encoding preserves the ordered product of repeated updates, inverses (when the update is unitary), all payload inner products, and the action on a reference. Replacing a stride by a different stride replaces its count by the exact ceiling formula of Lemma~\ref{lem:clock-geometry}. If the count is instead supplied by a coherent binary argument with the polynomial capacity and inverse-computation hypotheses of Corollary~\ref{cor:internal-clock}, the single-clock fixed-local fold has a clean rank-at-most-thirteen realization with all clock and simulator work restored.
\end{proposition}
\begin{proof}
Proposition~\ref{prop:native-clock-transport} intertwines the repeated update with the encoded update via the header-prepending isometry. Its invariant image gives the assertions on composition and inversion, and tensoring with identity handles references. The coherent case is Corollary~\ref{cor:coherent-clock-fold}; the count formula is Lemma~\ref{lem:clock-geometry}.
\end{proof}
No inclusion-reflecting embedding of the classical SR hierarchy is asserted: preservation of minimal basis, source rank and global pointwise majorant is not established.

\section{Output and guard semantics}\label{sec:historical-model}
We examine output specifications and guarded linear operations motivated by schematic quantum recursion~\cite{Yamakami2020,Yamakami2024}. The results concern the explicit rules below (source versions are identified in the bibliography), not alternative encodings, invariant-sector semantics or general quantum channels. For vacuum-extension semantics see~\cite{Barsse2026}.
Put $H_j=0^j1$, $\phi^p(x)=H_{|x|}H_{p(|x|)}H_{11p(|x|)+6}x$, and $\widetilde{0^m}=0^{2m}11$. Consider two zero-error output tests, with $m=|f(x)|\le p(|x|)$: (i) the output on $\phi^p(x)$ begins with $f(x)$; (ii) the output on $\widetilde{0^m}H_{m+1}\phi^p(x)$ is $f(x)H_{m+1}H_{m+1}\phi^p(x)$. In each test a single fixed term must work for every $x$.
\begin{theorem}[Insufficiency of ambiguous output tests]\label{thm:raw-prefix-counterexample}
There is a polynomially bounded noncomputable binary string function satisfying both tests with constant padding and measurement-free recursion-free witnesses.
\end{theorem}
\begin{proof} Choose a nonrecursive subset $A\subseteq\{2,3,\ldots\}$. Set $f(\lambda)=1$, $f(x)=00$ if $|x|\in A$, and $f(x)=0$ otherwise for nonempty $x$. Its output length decides $A$, so $f\notin\FBQP$ (effective bounded-error finite-time quantum computations are computable by approximating their probabilities). Put $p=2$. The padded input $\phi^p(x)$ starts, for $|x|\ge2$, with $00$, for $|x|=1$ with $0$, and for $x=\lambda$ with $1$. Thus the identity satisfies (i).

For (ii), the supplied header depends on the desired output length $m$. With $m=1,2$ the input header $\widetilde{0^m}H_{m+1}$ is respectively $0011001$ and $0000110001$. The required output header $f(x)H_{m+1}H_{m+1}$, when $f(x)=0^m$, is respectively $0001001$ and $0000010001$. Implement the following three fixed reversible bit flips on the first eight wires (when the length is sufficient): flip bit 3 when bits 1,2 are zero and bit 4 is one; flip bit 5 when bits 1--4 are zero and bit 6 is one; then flip bit 1 when bits 2--7 equal $001001$ and bit 8 is one. Each is a controlled NOT with a predicate on unchanged wires, hence a reversible finite-prefix operation in the recursion-free schematic basis by fixed swaps and branching. The first two flips produce the two zero-output headers just displayed, leaving $\phi^p(x)$ unchanged. The third flip is inactive on the second header, and on the first is active exactly when $x=\lambda$, since bit 8 there is the first bit of $H_{|x|}$. For the empty input, the first header is $0011001$ and the first bit of $\phi^p(\lambda)$ is one. The first two flips yield $0001001$ on the header, and the third yields $1001001=1H_2H_2$. Thus the same gate sequence handles the empty input without an extra promise. The complete prescribed target is obtained exactly. Thus (ii) also accepts the noncomputable $f$.
\end{proof}

The prefix test allows ambiguous output lengths; the clean-output test supplies the desired length in the input header. To formulate the usual bounded-error FBQP problem, use a prefix-free output (or an effectively recoverable length) and padding determined only by the input length. These conventions remove the information ambiguity.
\begin{lemma}[Bilinear XOR: Walsh criterion]\label{thm:xor-fourier}
On a fixed decoded $m$-bit target, let $a=\sum_y a_y|y\rangle$ be a normalized state. Define $C_a|x\rangle=\sum_y a_y|x\oplus y\rangle$, a bilinear XOR translation. Set $\widehat a(t)=\sum_y(-1)^{t\cdot y}a_y$. Then $C_a$ is unitary if and only if $|\widehat a(t)|=1$ for every $t$; it is a unitary involution if and only if all $\widehat a(t)\in\{-1,1\}$. In particular, among normalized states with nonnegative real coefficients only computational-basis states give unitary translations.
\end{lemma}
\begin{proof}
Walsh diagonalization of $C_a=\sum_y a_y X^y$ gives eigenvalues $\widehat a(t)$, proving both criteria. For $a_y\ge0$, $\widehat a(0)=1$ and $\sum_y a_y^2=1$ imply $\sum_{y<z}a_ya_z=0$; hence exactly one coefficient is one. Basis translations are unitary involutions.
\end{proof}

For $a=(|0\rangle+|1\rangle)/\sqrt2$, the eigenvalues are $\sqrt2,0$ and self-XOR is $|0\rangle+|1\rangle$, so the proposed general rule is not unitary. A unitary length-preserving oracle $O$ instead admits the reversible interface $|x\rangle|y\rangle\mapsto |x\rangle O|x\oplus y\rangle$, which prepares $O|x\rangle$ on a zero target but does not assert bilinear XOR. The XOR interface for deterministic binary functions remains valid. The criterion concerns fixed-length translations, not arbitrary higher-type quantum functionals.

Agreement of computational-basis statistics does not imply coherent approximation: $|+\rangle$ and $|-\rangle$ have identical basis statistics but their density matrices differ by trace norm $2$.
\begin{theorem}[Exact guard criterion and distance to unitarity]\label{thm:v12-guard}
Let $P$ be an orthogonal projection on a finite-dimensional Hilbert space $\mathcal H$, $Q=I-P$, and $U$ unitary. Put $T=UP+Q$ and $s=\|QUP\|$. Then
\[
 \|T\|=\sqrt{1+s},\qquad
 \inf_{V\text{ unitary}}\|T-V\|=1-\sqrt{1-s}.
\]
In particular, $T$ is a contraction if and only if it is unitary, if and only if $UP=PU$. More generally, for an orthogonal resolution $\sum_jP_j=I$ and unitaries $U_j$, the piecewise map $T=\sum_jU_jP_j$ is a contraction if and only if it is unitary, if and only if $P_iU_i^\dagger U_jP_j=0$ for all $i\ne j$.
\end{theorem}
\begin{proof}
Relative to $P\mathcal H\oplus Q\mathcal H$, put $B=QUP$. Then $T^\dagger T=\left(\begin{smallmatrix}I&B^\dagger\\B&I\end{smallmatrix}\right)$ has eigenvalues $1\pm\sigma_i(B)$ and remaining ones. Thus the extreme singular values are $\sqrt{1\pm s}$ (also when a sector is zero). A smallest-singular-value vector gives the distance lower bound. A unitary polar extension attains the maximum endpoint deviation~\cite{Higham1986}, equal to $1-\sqrt{1-s}$ by concavity. For $s=0$, $U$ preserves $P\mathcal H$ and, by unitarity, its complement. In the general partition, diagonal blocks of $T^\dagger T$ are identity, so its trace equals the dimension. Contractivity forces every eigenvalue to be one. The off-diagonal blocks are the stated cross terms.
\end{proof}

\begin{corollary}[Guarded iteration criterion]\label{cor:v12-scheme-v}
At a fixed total register length, let $P_j$ be the mutually orthogonal projections associated with a disjoint parse and iteration-count convention (including a count-zero inactive sector). For a globally unitary ground operation $G$, the piecewise iteration $T=\sum_jG^{k_j}P_j$ is unitary exactly when
\[
 P_iG^{k_j-k_i}P_j=0\quad(i\ne j).
\]
\end{corollary}
\begin{proof}
Apply Theorem~\ref{thm:v12-guard} with $U_j=G^{k_j}$. Sector invariance is sufficient but is not required by the exact condition. A different, always unitary repair retains an explicit orthogonal sector label and acts by the controlled block diagonal $\sum_j\ket{j}\!\langle j|\otimes G^{k_j}$. Erasure of such a label is not automatic. Neither repair alone establishes definability in a restricted recursion syntax.
\end{proof}

\begin{proposition}[Mixed-parse obstruction]\label{prop:v-no-unitary}
Parse a binary word as $xr_0y$ at the first aligned delimiter $r_0$. Let $k=\lceil\log_2|y|\rceil$ for nonempty $y$. The code-controlled NOT flips the first bit of $xr_0$ when $|x|>2$, $|y|>1$ and $|x|\le |r_0|k$, and fixes other words. Guarded iteration applies this operation $k$ times when $x$ is nonempty, $|y|>1$ and $|x|\le |r_0|k$, and otherwise fixes the word. Extend linearly on each fixed-length register. Neither the code-controlled NOT nor this iteration is a contraction on the entire Hilbert space. In particular no globally norm-preserving replacement realizes either action on every input, even with clean ancillary registers, or approximates its linear vector action uniformly on unit inputs with error strictly less than one.
\end{proposition}
\begin{proof}
Take the one-bit delimiter $r_0=1$, $a=|0001\,0^8\rangle$ and $b=|1001\,0^8\rangle$. The code of $a$ has length three and its payload length is eight. Both guards hold: the code length exceeds two and is at most $|r_0|\lceil\log_2 8\rceil=3$. NOT on its code sends $a$ to $b$. The first delimiter of $b$ is its first bit, so its empty code causes the identity guard; $b$ is fixed. This yields a collision already in the code-controlled NOT. For guarded iteration the initial iteration count on $a$ is three. If the code-controlled operand is reparsed after each application, the first sends $a$ to $b$ and the rest fix $b$. If the raw NOT is applied to the original code three times, the odd count again sends $a$ to $b$. In both readings the iteration fixes $b$ by its empty-code guard.
For either collision, linear extension maps $(a+b)/\sqrt2$ to $\sqrt2 b$ and $(a-b)/\sqrt2$ to zero. Its operator norm is at least $\sqrt2$, and an isometry has error one on the normalized difference. A clean ancillary realization remains an isometry. If mixed-parse superpositions are excluded instead, no global linear operator has been specified. The obstruction concerns unrestricted global linear semantics, not an invariant-sector interpretation, and does not preclude a modified computationally equivalent syntax.
\end{proof}

\begin{theorem}[SIZE collision and reversible repair]\label{thm:v12-size}
The following unary-header size map has no global contractive linear realization: for $m\ge1$, $n=|x|\ge1$ and $k=\lceil\log_2 n\rceil$, send $H_m x$ to $H_kH_{m-k-1}x$ if $n\le2^{m-1}$, and fix inactive words. Its prescribed active action has a globally involutive length-preserving permutation extension, but no injective extension agreeing with the prescribed inactive identity on all its active images.
\end{theorem}
\begin{proof}
Take $a=H_3 00=000100$ and $b=H_1H_1 00=010100$. The active rule on $a$ uses $k=\lceil\log_2 2\rceil=1$ and yields $b$. The parse of $b$ as $H_1 0100$ is inactive, since its payload length four exceeds $2^{1-1}$. Thus both inputs map to $b$. The normalized sum of these inputs has output norm $\sqrt2$ and their normalized difference has output zero. This contradicts contractivity, not merely unitarity.
The active map is injective (the two headers recover $m$, the payload is unchanged). Its image is disjoint from the active domain: for $k=0$ the first header is zero; otherwise the new payload length $n+m-k>n>2^{k-1}$. Swap each disjoint input-image pair and fix other words, giving a length-preserving involution. Zero-length payload conventions may be handled separately; the six-qubit collision persists. Definability in a restricted fast-recursion syntax is not asserted.
\end{proof}

\section*{Declaration of generative AI and AI-assisted technologies in the manuscript preparation process}
During the preparation of this work, the author used ChatGPT (OpenAI) for copyediting, grammatical correction, and language polishing of the manuscript. The author is responsible for reviewing and editing the resulting text, verifying the mathematical claims and references, and the full content of the article.


\begin{thebibliography}{99}
\setlength{\itemsep}{2pt}
\bibitem{MilneANT} J. S. Milne, \emph{Algebraic Number Theory}, v3.07 (2017), Sec.~5 (Dirichlet unit theorem). \url{https://www.jmilne.org/math/CourseNotes/ANT.pdf}.
\bibitem{Yamakami2020} T. Yamakami, \emph{A schematic definition of quantum polynomial time computability}, J. Symbolic Logic \textbf{85} (2020), no.~4, 1546--1587. \url{https://doi.org/10.1017/jsl.2020.45} (author version consulted: \url{https://arxiv.org/abs/1802.02336v3}).
\bibitem{Yamakami2024} T. Yamakami, \emph{Elementary quantum recursion schemes that capture quantum polylogarithmic time computability of quantum functions}, Math. Struct. Comput. Sci. \textbf{34} (2024), no.~7, 710--745. \url{https://doi.org/10.1017/S0960129524000264} (author version consulted: \url{https://arxiv.org/abs/2311.15884v3}).

\bibitem{SR} K. Osipov, \emph{Step Recursion: A Three-Parameter Refinement of the Grzegorczyk Hierarchy}, arXiv:2608.04871 (2026).
\bibitem{BV1997} E. Bernstein and U. Vazirani, \emph{Quantum complexity theory}, SIAM J. Comput. \textbf{26} (1997), no.~5, 1411--1473.
\bibitem{Yao1993} A. C.-C. Yao, \emph{Quantum circuit complexity}, Proc. 34th FOCS, IEEE (1993), 352--361.
\bibitem{Watrous1999} J. Watrous, \emph{Space-bounded quantum complexity}, J. Comput. System Sci. \textbf{59} (1999), no.~2, 281--326.
\bibitem{FeffermanRemscrim2021} B. Fefferman and Z. Remscrim, \emph{Eliminating intermediate measurements in space-bounded quantum computation}, Proc. 53rd Annual ACM SIGACT Symposium on Theory of Computing (STOC 2021), ACM (2021), 1343--1356.
\bibitem{DawsonNielsen2006} C. M. Dawson and M. A. Nielsen, \emph{The Solovay--Kitaev algorithm}, Quantum Inf. Comput. \textbf{6} (2006), no.~1, 81--95. \url{https://arxiv.org/abs/quant-ph/0505030}.
\bibitem{Ferrari2025} F. Ferrari, E. Hainry, R. P\'echoux, and M. Silva, \emph{Quantum programming in polylogarithmic time}, MFCS 2025, LIPIcs \textbf{345}, 47:1--47:18. \url{https://doi.org/10.4230/LIPIcs.MFCS.2025.47}.
\bibitem{BBCMW2001} R. Beals, H. Buhrman, R. Cleve, M. Mosca, and R. de Wolf, \emph{Quantum lower bounds by polynomials}, J. ACM \textbf{48} (2001), no.~4, 778--797.
\bibitem{Yamakami2025QFO} T. Yamakami, \emph{Quantum first-order logics that capture logarithmic-time/space quantum computability}, in \emph{CiE 2024}, LNCS \textbf{14773}, Springer (2024), 311--323. \url{https://doi.org/10.1007/978-3-031-64309-5_25} (expanded version arXiv:2501.12007, 2025).
\bibitem{DLMZ2010} U. Dal Lago, A. Masini, and M. Zorzi, \emph{Quantum implicit computational complexity}, Theor. Comput. Sci. \textbf{411} (2010), 377--409. \url{https://doi.org/10.1016/j.tcs.2009.07.045}.
\bibitem{Ying2014} M. Ying, \emph{Quantum recursion and second quantisation}, arXiv:1405.4443 (2014).
\bibitem{HPS2023} E. Hainry, R. P\'echoux, and M. Silva, \emph{A programming language characterizing quantum polynomial time}, FoSSaCS 2023. \url{https://doi.org/10.1007/978-3-031-30829-1_8}.
\bibitem{HPS2025} E. Hainry, R. P\'echoux, and M. Silva, \emph{Branch sequentialization in quantum polytime}, FSCD 2025, LIPIcs, 22. \url{https://doi.org/10.4230/LIPIcs.FSCD.2025.22}.
\bibitem{GRZ2021} U. Girish, R. Raz, and W. Zhan, \emph{Quantum logspace algorithm for powering matrices with bounded norm}, ICALP 2021, LIPIcs, 73. \url{https://doi.org/10.4230/LIPIcs.ICALP.2021.73}.
\bibitem{Barenco1995} A. Barenco et al., \emph{Elementary gates for quantum computation}, Phys. Rev. A \textbf{52} (1995), 3457--3467. \url{https://doi.org/10.1103/PhysRevA.52.3457}.
\bibitem{CrosswhiteBacon2008} G. M. Crosswhite and D. Bacon, \emph{Finite automata for caching in matrix product algorithms}, Phys. Rev. A \textbf{78} (2008), 012356. \url{https://doi.org/10.1103/PhysRevA.78.012356}.
\bibitem{Chuklanov2026} P. Chuklanov, \emph{Parameterized equivalence checking of quantum circuit generators via weighted automata}, arXiv:2611.12049 (2026).
\bibitem{Droste2009} M. Droste, W. Kuich, and H. Vogler (eds.), \emph{Handbook of Weighted Automata}, Springer (2009). \url{https://doi.org/10.1007/978-3-642-01492-5}.
\bibitem{FreedmanWang2007} M. H. Freedman and Z. Wang, \emph{Large quantum Fourier transforms are never exactly realized by braiding conformal blocks}, Phys. Rev. A \textbf{75} (2007), 032322. \url{https://doi.org/10.1103/PhysRevA.75.032322}.
\bibitem{GilesSelinger2013} B. Giles and P. Selinger, \emph{Exact synthesis of multiqubit Clifford+T circuits}, Phys. Rev. A \textbf{87} (2013), 032332. \url{https://doi.org/10.1103/PhysRevA.87.032332}.
\bibitem{Higham1986} N. J. Higham, \emph{Computing the polar decomposition---with applications}, SIAM J. Sci. Stat. Comput. \textbf{7} (1986), 1160--1174. \url{https://doi.org/10.1137/0907079}.
\bibitem{Bennett1973} C. H. Bennett, \emph{Logical reversibility of computation}, IBM J. Res. Develop. \textbf{17} (1973), 525--532. \url{https://doi.org/10.1147/rd.176.0525}.
\bibitem{SG2026} M. Silva, A. Guernut, and M. Lewis, \emph{pfoqcompiler: Efficient compilation of quantum control in recursive first-order quantum programs}, arXiv:2611.08503 (2026).
\bibitem{Gavorova2024} Z. Gavorov\'{a}, M. Seidel, and Y. Touati, \emph{Topological obstructions to quantum computation with unitary oracles}, Phys. Rev. A \textbf{109} (2024), 032625. \url{https://doi.org/10.1103/PhysRevA.109.032625}.
\bibitem{Barsse2026} K. Barsse, R. P\'echoux, and S. Perdrix, \emph{Quantum control and general recursion beyond the unitary case}, LICS 2026, LIPIcs, 14:1--14:26. \url{https://doi.org/10.4230/LIPIcs.LICS.2026.14}.
\bibitem{Barloy2020} C. Barloy, N. Fijalkow, N. Lhote, and F. Mazowiecki, \emph{A robust class of linear recurrence sequences}, CSL 2020, LIPIcs \textbf{152}, 9:1--9:16. \url{https://doi.org/10.4230/LIPIcs.CSL.2020.9}.
\bibitem{Amy2024} M. Amy, A. N. Glaudell, S. Kelso, W. Maxwell, S. S. Mendelson, and N. J. Ross, \emph{Exact synthesis of multiqubit Clifford-cyclotomic circuits}, arXiv:2311.07741v2 (2024).
\bibitem{Deng2024} H. Deng, R. Tao, Y. Peng, and X. Wu, \emph{A case for synthesis of recursive quantum unitary programs}, Proc. ACM Program. Lang. \textbf{8} (POPL, 2024), art. 59. \url{https://doi.org/10.1145/3632901}.
\bibitem{CHPV2026} K. Chardonnet, E. Hainry, R. P\'echoux, and T. Vinet, \emph{Resource-aware quantum programming with general recursion and quantum control}, FSCD 2026, LIPIcs \textbf{378}, 12. \url{https://doi.org/10.4230/LIPIcs.FSCD.2026.12}.
\end{thebibliography}
\end{document}